\documentclass[12pt]{article}

\usepackage{amsmath}
\usepackage{epsfig}
\usepackage{rotating}
\usepackage{latexsym}

\def\stackunder#1#2{\mathrel{\mathop{#2}\limits_{#1}}}

\def\inbox#1{\begin{quotation} \noindent {\it #1} \end{quotation}}

\def\outbox#1#2{
\vskip 1cm
\begin{flushright}
\parbox{7cm}{\it #1}
\end{flushright}
\begin{flushright}
{\footnotesize (#2)}
\end{flushright}
\vskip 1cm}

\begin{document}

\pagenumbering{roman}

\title{\LARGE
From Monte Carlo Integration \\
to Lattice Quantum Chromo Dynamics
\large \\ 
An introduction}
\author{Massimo Di Pierro \\ \\
Fermilab, Batavia, IL 60510, USA \\
Email: {\tt mdp@fnal.gov}}
\maketitle

\vspace{-3.8in}
\begin{flushright}
FERMILAB-FN-699 \\ 
hep-lat/0009001
\end{flushright}
\vspace{2.9in}

\begin{abstract}
Lectures on Lattice Field Theory and Lattice QCD given at the Graduate Students 
Association (GSA) Summer School (Fermilab).

In these lectures we provide a short introduction to the Monte Carlo integration method 
and its applications.
We show how the origin of ultraviolet divergences if Field Theories 
is in the undefined formal product of 
distributions and how one can define the Path Integral in terms of regularized 
distributions in order to cancel these divergences. This technique provides the only non 
perturbative regularization procedure of continuum Field Theories and, at the same time, 
provides a practical method to compute correlation (Green) functions (using Monte Carlo 
integration for the regularized path integrals). We then apply these tools to 
formulate QCD on a lattice. Some of the examples are accompanied by complete computer 
programs. \\ \\
Freely download libraries and examples from:\\ \\
{\tt http://latticeqcd.fnal.gov/software/fermiqcd/}
\end{abstract}

\newpage \phantom{blank page} \newpage

\section*{Introduction}

\noindent Two are the main tasks a physicist has to confront with:
\begin{itemize}
\item {\bf [Induction]}: given the symmetries of the measured observables,
build the underlying theory (i.e. write down an action).
\item {\bf [Deduction]}: given the action, $\mathcal S$, compute correlation functions 
and, from them, physical observables (to test the theory and to make predictions).
\end{itemize}
In this notes we will focus the second task.

In the first section we wil see how the perturbative expansion of the
Path Integral does not provide a satisfactory definition of the latter and 
a non-pertubartive regularization is necessary in order to define it properly.

For us the word ``non-perturbative'' means ``exact up to a given precison that 
can be arbitrarily small''.

In the second section we will show how to compute numerically $K$-dimensional
integrals (for large integer $K$) using Monte Carlo techniques. 

In the third section we will define the regularized Path Integral in terms of regularized 
distributions. This is equivalent to discretize the the space-time on which the quantum 
fields are defined. The lattice spacing $a$ will play the role of
an ultraviolet cut-off. We will then define the continuum Path Integral
in terms of $K$-dimensional integrals in the limit $a \rightarrow 0$. 
We will show how this limit is the origin of ultraviolet divergences and how one 
can renormalize the theory by giving an $a$ dependence to the coupling constants that
appear in the action.

Finally in the third section we will see how one can approximate a continuum Path 
Integral of QCD with a $K$-dimensional integral (for a finite $K$) and compute it 
numerically using Monte Carlo integration. We will discuss the sources of numerical 
errors and we will present, as an example, one full Lattice QCD application (the computation 
of $f_B\sqrt{m_B}$).

Some more examples of typical Lattice QCD computations are given in the Appendix.

The emphasis in these lectures will be given to three aspects that we consider crucial 
and make of lattice a privileged tool in respect to other model independent methods:
\begin{itemize}
\item The lattice regularization provides the only non-perturbative definition of Path Integral 
and, therefore, of Field Theories.
\item It is possible to quantify with precision the error committed in the numerical 
approximation of the integrals.
\item It is, in principle, possible to reduce arbitrarily this error by approaching 
the continuum limit (reducing the lattice spacing) and increasing the statistical samples 
in the Monte Carlo integration.
\end{itemize}

This lectures are intended to be an introductory tutorial and they are not meant to be 
complete and/or exhaustive on the subject of lattice QCD. For more complete introductory 
reviews on the subject see \cite{creutz} and \cite{rothe}. For a more complete 
and formal review see \cite{montvay}.

\vskip 5mm

I wish to thank G. Chiodini, B. Dobrescu, E. Eichten, J. Juge, A. Kronfeld, 
P. Mackenzie and J. Simone for helpful comments and suggestions regarding these notes.

This work was performed at Fermilab, a U.S. Department of Energy Lab (operated by the 
University Research Association, Inc.), under contract DE-AC02-76CHO3000.

 \clearpage\newpage

\tableofcontents \newpage

\pagenumbering{arabic}

\section{Correlation functions, masses and matrix elements}

\label{chap0}

\outbox{He who loves practice without theory is like the sailor who boards 
ship without a rudder and compass and never knows where he
may cast}{Leonardo da Vinci}

The symbol of Path Integral~\cite{peskin}
\begin{equation}
\langle 0| T\{\phi (x_{1})...\phi (x_{n})\}| 0 \rangle
\stackrel{def}{=} \int [\text{d}\phi ]\phi (x_{1})...\phi (x_{n})e^{-{\mathcal{S}}_{\text{E}}[\phi ]}
\label{PI}
\end{equation}
provides a definition of the most general Euclidean correlation
function (the left hand side) in terms of an infinite-dimensional integral
(the path integral at the right hand side). ${\mathcal S}_{\text{E}}[\phi ]$ 
is the Euclidean action of the system and $\phi (x)$ represents the degrees
of freedom of the system as function of the space-time Euclidean coordinates.

Without loss of generality we will only deal with Path Integrals in the 
Euclidean space since Minkowskian correlation functions can be obtained by 
analytic continuation of the Euclidean ones. 
In particular we are interested in extracting particle masses and matrix 
elements which do not have an explicit time dependence. Therefore, 
up to a phase, they are the same in the Minkowskian and in the Euclidean 
formulation of the theory.

In the case of the $\phi^4$ scalar model the perturbative expansion, 
after renormalization, is always convergent therefore it provides us 
with a definition of the Path Integral. Unfortunately...

\inbox{In general, the perturbation expansion does not
provide a satisfactory definition of the Path Integral.}

Let's consider, for example, the case of QCD. The perturbative
expansion of Path Integrals of QCD is only convergent at high energy ($%
p >> 200MeV$), i.e. at short distance ($a \equiv 1/p << 1$fm ). 
In fact the expansion parameter $\alpha _{s}=\frac{g(a)^2}{4 \pi}$ is scale 
dependent and it becomes big at low energy,
i.e. large distance~\cite{qcdreview, marciano, afreedom}. Moreover usual 
regularization prescriptions (such as Pauli-Villars and
dimensional regularization), which are necessary to give sense to the Path
Integral, are only defined in a perturbative way.

One could argue that, even if the perturbative expansion of eq.(\ref{PI}) is
only valid at high energy, it still defines eq.(\ref{PI}) because one can
analytically continue the correlation functions from high to low 
energy and make somehow sense out of them. This is not correct! 

First, the perturbative series is an ``asymptotic series'' and it does not
necessarily converge. Second (the most important point), the perturbative
expansion is performed around the classical minimum of the action. The
Euclidean action of QCD has many different minima that correspond to
multi-instanton configurations. These are tunneling
transitions between different vacua of QCD. These 
effects are not taken into account in the perturbative expansion but,
nevertheless, they play a crucial role at
low energy. They are believed to be responsible for the phenomenon of
confinement~\cite{confinement}.

\inbox{The lattice discretization of the space-time provides us with the only non-perturbative regularization of the Path Integral and, therefore, with an EXACT definition of the latter.}

\inbox{For a small and finite cut-off (the lattice spacing $a$) the regularized theory can be seen as an effective theory of the continuum one (corresponding to the limit $a \rightarrow 0$). Such an effective theory is finite and free of ultraviolet divergences therefore it can be simulated numerically!}

\inbox{Moreover, in the case of QCD and gauge theories in general, 
lattice regularization has the nice feature to preserve gauge 
invariance even for every finite $a$.}

We now introduce two examples in which we require a 
non-perturbative definition of the Path Integral to compute some
important phenomenological quantities.

\subsection{2-point correlation functions}

QCD is the theory of strong interactions, hence it must predict the
masses of mesons and baryons from first principles. Let's consider here,
for example, a B meson.

The following current
\begin{equation}
J^\mu(x)=\bar h(x)\gamma ^{\mu }\gamma ^{5}q(x)
\end{equation}
(where $x_{\mu }=(x_{0}\equiv t_x,x_{1},x_{2},x_{3})$, $t_x$ is the Wick rotated
time and $h$ ($q$) is the fermionic field representing the heavy (light) 
quark) has the same flavor quantum numbers of a B meson. If we apply $J^0$
to the vacuum, it must create some linear combination of a static B meson 
and its excited states
\begin{equation}
J^{0 }| 0 \rangle =\varepsilon _{0}| B \rangle
+\varepsilon _{1}| B^{(1)} \rangle +\varepsilon _{2}|
B^{(2)} \rangle +...
\end{equation}
(in general there is a continuum of states but, for notation, we write them here as a sum.)
From now on all our states are normalized as $\langle B^{(n)} | B^{(n)} \rangle = 2 m_{B^{(n)}}$.

We define the following (zero momentum Fourier transform of the) two point
correlation function
\begin{eqnarray}
C_{2}(t_x) &=&\int {\text{d}^3\mathbf{x}} \langle 0| J^0
(x)J^{0 \dagger}(0)| 0 \rangle   \nonumber \\
&=&\int [\text{d}A_{\mu }][\text{d}q_{i}][\text{d}\overline{q}_{i}]\left(
\int {\text{d}^3\mathbf{x}}J^0(x)J^{0\dagger }(0)\right) 
e^{-{\mathcal{S}}_{\text{E}}^{\text{QCD}}}  \label{c2}
\end{eqnarray}

One can insert in the correlator a complete set of states...
\begin{eqnarray}
\int {\text{d}^3\mathbf{x}}\langle 0| J^{0 }(x)J^{0 \dagger }(0)|
0\rangle  &=&\sum_{n} \int {\text{d}^3\mathbf{x}} \langle 0| J^{0
}(x) \frac{| B^{(n)}\rangle \langle B^{(n)}|} 
{2m_{B^{(n)}}} J^{0 \dagger
}(0)| 0 \rangle  \nonumber \\
&=&\sum_{n} \langle 0| J^{0
}(0)e^{-H t_x}\frac{| B^{(n)} \rangle \langle B^{(n)}|}
{2m_{B^{(n)}}} J^{0 \dagger }(0)| 0 \rangle  \nonumber\\
&=&\sum_{n}\frac{| \langle 0| J^{0 }(0)|
B^{(n)} \rangle | ^{2}}{2 m_{B^{(n)}}} e^{-E^{(n)}t_x} \nonumber\\
&=&| Z _{0}| ^{2} e^{-E^{(0)} t_x}+| Z
_{1}| ^{2}e^{-E^{(1)} t_x}+... \label{fit1}
\end{eqnarray}
where $e^{-H t_x}$ is the Euclidean translation operator and 
\begin{equation}
H| B^{(n)} \rangle =E^{(n)}| B^{(n)} \rangle 
\end{equation}

The long distance behavior of $C_{2}(t_x)$ is dominated by the 
exponential associated with the lightest state $| B^{(0)}\rangle $ (with energy 
$E^{(0)}=m_{B}$)
\begin{equation}
C_{2}(t_x)\stackunder{t_x \rightarrow \infty }{\simeq }| Z_{0}
| ^{2}e^{-m_{B} t_x}  \label{c2mb}
\end{equation}

Therefore by definition%
\footnote{
$\langle 0| J^\mu | B \rangle = f_B p^\mu$
}
\begin{equation}
Z_{0} =  \frac{\langle 0| J^{0 }(0)|
B^{(0)} \rangle}{\sqrt{2m_B}} = \frac{1}{\sqrt{2}}f_B \sqrt{ m_B }
\end{equation}

Eq.(\ref{c2}) is not defined (yet) in the limit $t_x \rightarrow
\infty $ because this limit is clearly large distance and therefore
non-perturbative. 

\inbox{If we had a non perturbative definition of eq.(\ref
{c2}), we could compute $C_{2}(t_x)$ and extract both $m_B$ and $f_B$
by fitting the result with eq.(\ref{c2mb}).}

\begin{figure}
\begin{center} 
\epsfxsize=12cm
\epsfysize=12cm
\epsfbox{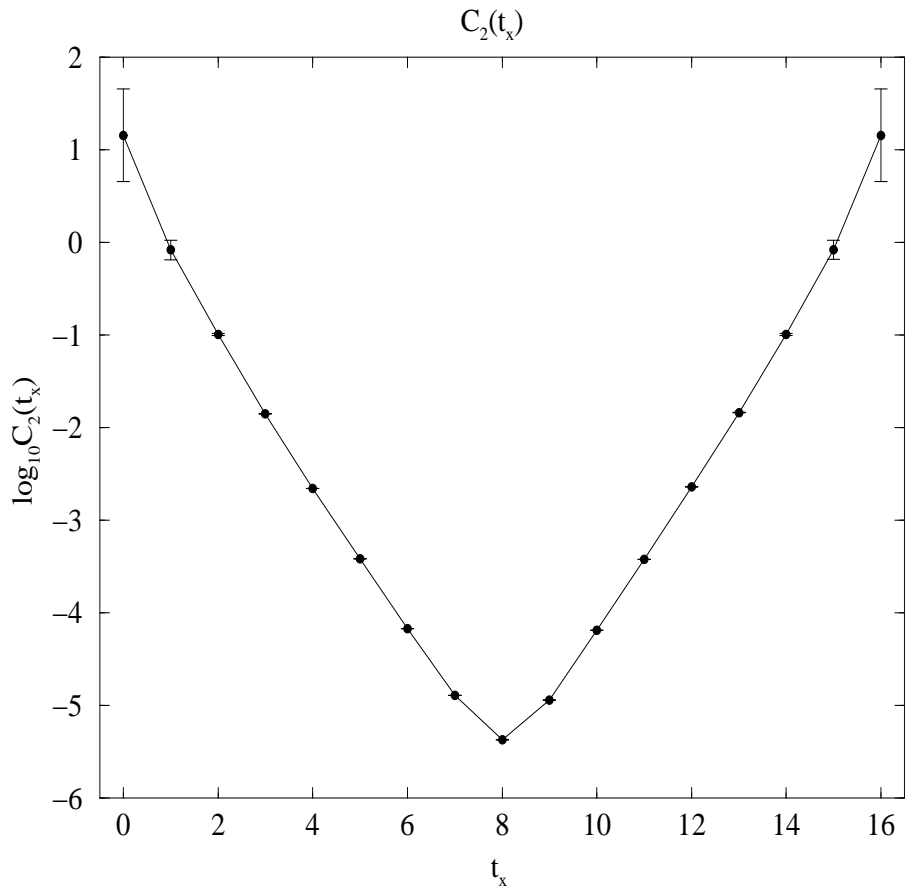} 
\end{center}
\caption{Example of a numerical two-point correlation function for a $B$ meson 
for some (arbitrary) input values of the light and heavy quark masses. 
The plot is symetric arout $t_x=8$ because of periodic boundary conditions.
The curve (for $t_x<8$) can be fitted with $C_2(t)= \frac12 f_B^2 m_B e^{-m_B t}$
to determine $m_B$ and $f_B$. \label{c2a}}
\end{figure}

An example of two-point correlation function computed on a lattice 
(using program {\tt C2.C} in the Appendix) is shown in fig.~\ref{c2a}.

\subsection{3-point correlation functions}

$V_{cb}$ is a very important parameter of the Standard Model%
\footnote{$V_{cb}$ is one of the elements of the 
Cabibbo-Kobayashi-Maskawa matrix $V$ and it is known 
with very poor precision.
A non unitary CKM matrix would be a clear signal of 
physics beyond the Standard Model.}. 
It must be determined by comparing QCD contributions with experiment. 
By definition
~\cite{donoghue, ckm}:
\begin{equation}
| V_{cb}| ^{2}=\frac{f_B^2 m_B}
{\left\langle B| \bar h \gamma ^{\mu }Lq\bar h
\gamma ^{\mu }L q| \bar B\right\rangle } \text{[perturbative factor][experiment]}
\end{equation}
From $C_2(t_x)$ we have the coefficients in the numerator but we also 
need a QCD prediction for the matrix element in the
denominator. To reach this goal we define a three-point correlation function
\begin{eqnarray}
C_{3}(t_x,t_y) &=&\int \text{d}^3{\mathbf{x}} \int {\text{d}^3\mathbf{y}} 
\left\langle 0| J^{0}(x) {\mathcal O} J^{0}(-y)| 0\right\rangle 
\nonumber \\
&=&\int [\text{d}A_{\mu }][\text{d}q_{i}][\text{d}\overline{q}_{i}]\left(
\int {\text{d}^3\mathbf{x}} \int {\text{d}^3\mathbf{y}} 
J^{0 }(x){\mathcal O}J^{0}(-y)\right) e^{-{\mathcal{S}}_{\text{E}}^{\text{QCD}}}
\label{c3}
\end{eqnarray}
where
\begin{equation}
{\mathcal O}=\overline{h}(0)\gamma ^{\mu }Lq(0)\overline{h}(0)\gamma ^{\mu }L q(0)
\end{equation}
(L, defined in the appendix, is the left-handed projector)
We play the same trick as before (inserting two complete sets of states) and
we obtain
\begin{eqnarray}
C_{3}(t_x, t_y) &=&\sum_{n,m}\int \text{d}^3{\mathbf{x}} \int {\text{d}^3\mathbf{y}}
 \langle 0| J^{0
}(x) \frac{| B^{(n)} \rangle \langle B%
^{(n)}|}{2 m_{B^{(n)}}} {\mathcal O} \frac{| \bar B^{(m)} \rangle \langle \bar B^{(m)}|}
{2 m_{B^{(m)}}}
J^{0}(-y)| 0 \rangle  \nonumber \\
&=&\sum_{n,m} Z ^{(n)} Z ^{(m)} \frac{\langle B%
^{(n)}| {\mathcal O}| \bar B^{(m)}\rangle}{\sqrt{2 m_{B^{(n)}} 2 m_{B^{(m)}}}}
 e^{-E^{(n)} t_x}e^{-E^{(m)} t_y}
\end{eqnarray}
which for $t_x \rightarrow \infty $ is dominated by
\begin{equation}
C_{3}(t_x, t_y)\stackunder{t_x,t_y\rightarrow \infty }{\simeq }| Z
_{0}| ^{2}e^{-m_{B} (t_x+t_y)}\frac{\langle B| {\mathcal O}|
\bar B \rangle}{2 m_B}   \label{fitc3}
\end{equation}

Again...
\inbox{If we had a non-perturbative definition of eq.(\ref{c3}), 
we could compute $C_{3}(t_x, t_y)$ and extract 
matrix elements (such as $\langle B | {\mathcal O}|
\bar B\rangle$) by fitting the result with eq.(\ref{fitc3})\footnote{%
This is what one actually does to determine $V_{cb}$. 
At present the main error on $V_{cb}$ is due to theoretical 
uncertanties.}}.

We will see in the next sections how the lattice provides both 
a definition and a practical way of computing the right-hand side 
of eq.(\ref{PI}), (\ref{c2}) and (\ref{c3}).

 \clearpage\newpage

\section{Monte Carlo Integration}

\label{chap1}

\outbox{Although this may seem a paradox, all exact science is dominated by the 
idea of approximation}{Bertrand Russell}

In this section it will be shown how to compute numerically 
multidimensional integrals using a
statistical method (the Monte Carlo integration).

\subsection{Riemann integrable functions}

Let's consider the simplest of the integrals
\begin{equation}
\int_{\alpha }^{\beta }f(x)\textrm{d}x  \label{int1}
\end{equation}
Riemann gave us a definition of this integral:

To check if the function $f(x)$ in the domain 
$\mathcal{D}=[\alpha ,\beta ]$ is Riemann integrable we divide the 
domain in $N$ small intervals 
\begin{equation} 
[x_{0}=\alpha,x_{1}],[x_{1},x_{2}],...,[x_{N-2},x_{N-1}],
[x_{N-1},x_{N}=\beta ]
\end{equation}
and compute 
\begin{equation}
\varepsilon (N) = a\sum_{i=0}^{N-1}\stackunder{x\in [x_{i},x_{i+1}]}{\max }%
f(x)-a\sum_{i=0}^{N-1}\min_{x\in [x_{i},x_{i+1}]}f(x) 
\end{equation}
where $a \stackrel{def}{=} \frac{\beta -\alpha }{N}$

\inbox{$f(t)$ is Riemann integrable if $\varepsilon (N)$ goes to
zero for $N\rightarrow \infty$, i.e. $a \rightarrow 0$.}

\begin{figure}
\begin{center} 
\epsfxsize=12cm
\epsfysize=12cm
\epsfbox{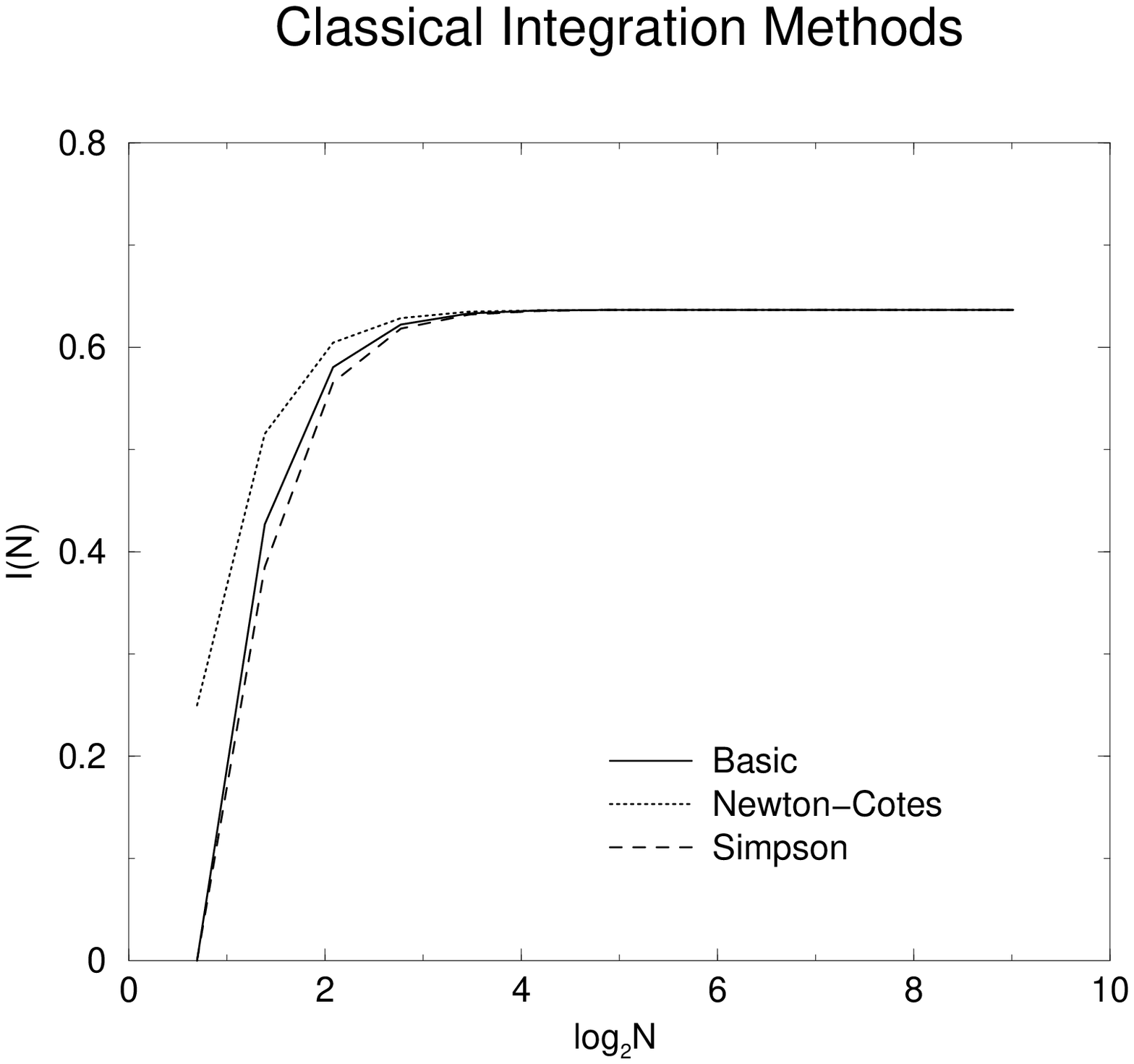} 
\end{center}
\caption{Convergent behavior of different numerical 
discretizations for the integral of eq.(\ref{int1}).
\label{program1}}
\end{figure}

This definition of integration is also very practical because it gives
a method of performing numerical integrations: 
\begin{equation}
\int_{\alpha }^{\beta }f(x)\text{d}x=\sum_{i=0}^{N-1}a\cdot
f(x_{i})+O(f^{^{\prime }}/N^{2})  \nonumber
\end{equation}

There are other possible ways of approximating a continuum 
integral with a discrete sum\footnote{%
for example using trapezoids instead of rectangles (Newton-Cotes method) 
\begin{equation}
\int_{\alpha }^{\beta }f(x)\text{d}x=\sum_{i=0}^{N-1}a\cdot \frac{%
f(x_{i})+f(x_{i+1})}{2}+O(f^{\prime \prime }/N^{3}) 
\end{equation}
or using interpolating polynomia (Simpson's method) 
\begin{equation}
\int_{\alpha }^{\beta }f(x)\text{d}x=\sum_{i=0}^{N-2}a\cdot \frac{%
f(x_{i})+4f(x_{i+1})+f(x_{i+2})}{6}+O(f^{\prime \prime \prime \prime
}/N^{5}) 
\end{equation}
See the Numerical Recipes for more details.} and, if they converge, they all
converge to the same number. 

\inbox{The difference between the different integration
methods resides in the behavior of the numerical integral as function of $N$: 
some methods converge faster than others. For practical
purposes this is a crucial issue.} 

\texttt{program1.c} is a {\tt C} program that computes the integral 
\begin{equation}
I=\int_{0}^{1}\sin (\pi x)\text{d}x\simeq 0.\,63662
\label{int10}
\end{equation}
using three different simple methods. The convergent behavior of the
different methods is shown in fig.1. The error done when one stops at a
finite $N$ is known as \emph{disctretization error}.

For multidimensional integrals this method turns out to be too slow for
practical applications. Therefore we need a new tool, Monte Carlo
integration.

\subsection{Basic Monte Carlo}

Monte Carlo integration is another numerical method for integration 
and it has a statistical foundation. The algorithm is the following:

\begin{itemize}
\item  Generate a set of $N$ random points $\{x^{[i]}\}$ with uniform
distribution in the integration domain $\mathcal{D}$

\item  For each point $x^{[i]}$ compute $f(x^{[i]})$

\item  Compute the average $I(N)=\frac{\beta -\alpha }{N}%
\sum_{i=0}^{N-1}f(x^{[i]})$
\end{itemize}

\inbox{If $f(x)$ is Riemann integrable, 
the function $I(N)$ converges to the integral of $f(x)$ when $%
N\rightarrow \infty$.}

\texttt{program2.c} is a {\tt C} program that computes 
numerically, using Monte Carlo integration, the same integral 
of eq.~(\ref{int10}).
Its convergent behavior is shown in fig.2 and compared with the 
standard methods showed in section 2.1.

\begin{figure}
\begin{center} 
\epsfxsize=12cm
\epsfysize=12cm
\epsfbox{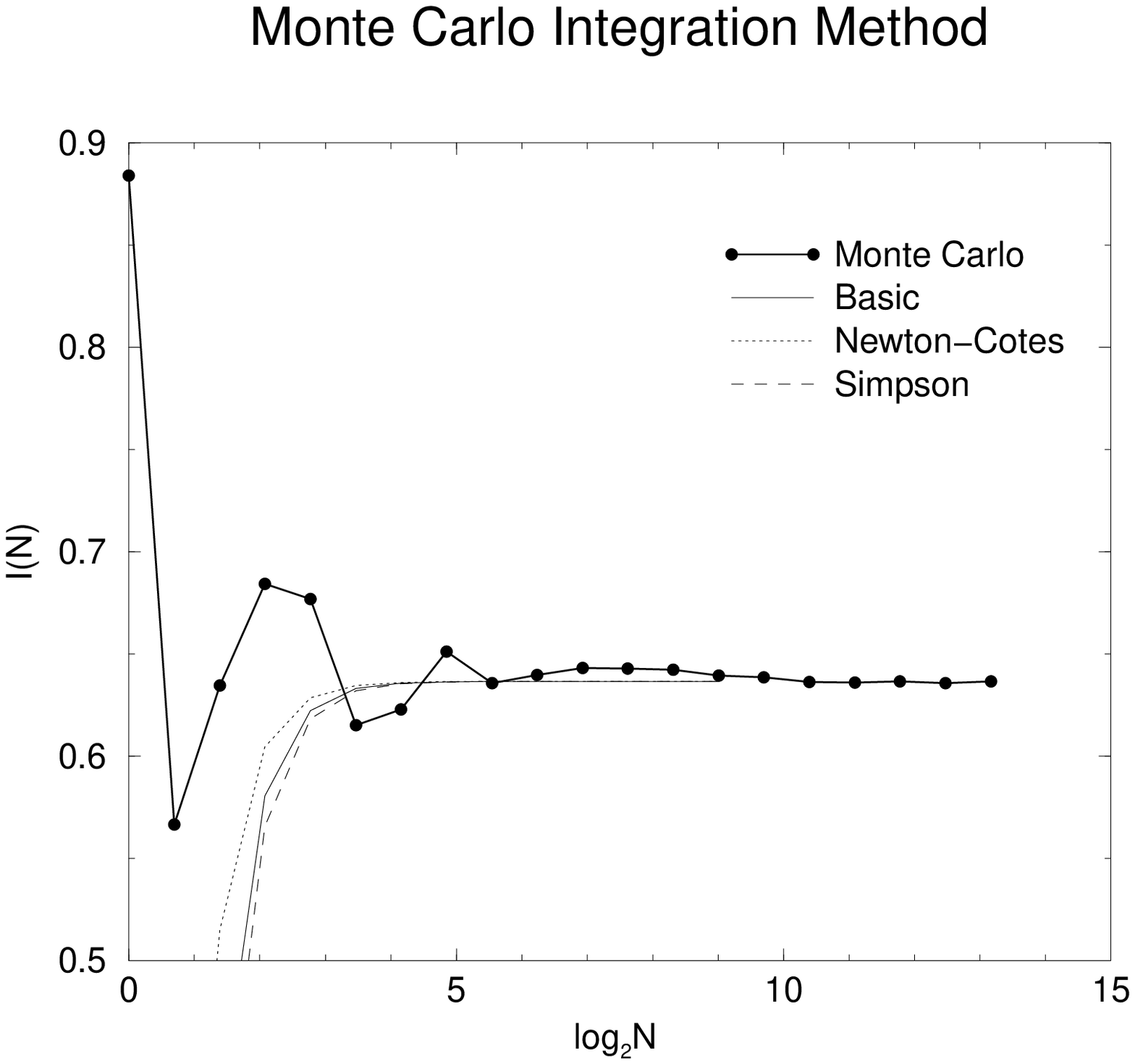} 
\end{center}
\caption{Convergent behavior of Monte Carlo Integration of eq.(\ref{int1}).
\label{program2}}
\end{figure}

This algorithm can easily be extended to arbitrary dimensions. For example 
\texttt{program3.c} computes the following integral 
\begin{equation}
I=\int_{0}^{1}dx_{0}\int_{0}^{1}dx_{1}%
\int_{0}^{1}dx_{2}(3x_{0}^{2}x_{1}+2x_{2}^{3})=1
\label{int2}
\end{equation}
and its convergent behavior is shown in fig.3.
Monte Carlo integration is not very efficient for 1D integrals but
it becomes more and more efficient (when compared with conventional
integration methods) for higher dimensional integrals.

\begin{figure}
\begin{center} 
\epsfxsize=12cm
\epsfysize=12cm
\epsfbox{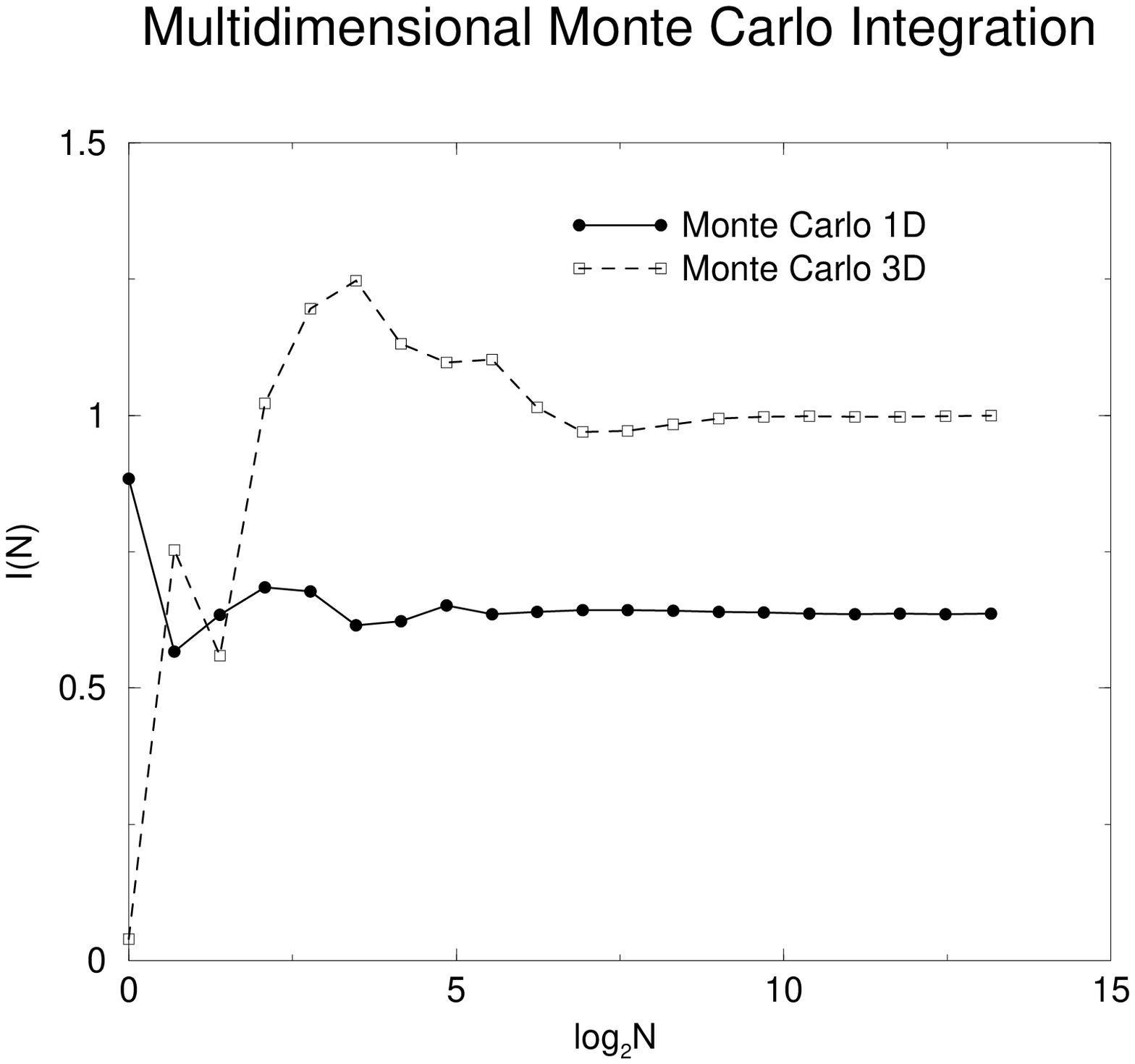} 
\end{center}
\caption{Convergent behavior Monte Carlo Integration of eq.(\ref{int2}).
\label{program3}}
\end{figure}

From now on we will use the upper index $\mathbf{x}^{[i]}$ to label the
Monte Carlo points and a lower index $x_{j}$ to label a particular coordinate of
a given point. For example $x_{2}^{[3]}$ is the second coordinate of
the 3rd point.

\subsection{Metropolis Monte Carlo}

In the next section we will deal with $K$-dimensional integrals of the form 
\begin{equation}
I_{k}=\int_{0}^{1}dx_{0}\int_{0}^{1}dx_{1}...\int_{0}^{1}dx_{K-1}f(%
{\mathbf x })P({\mathbf x})  \label{int4}
\end{equation}
\begin{equation}
\text{where }{\mathbf x }=(x_{0},x_{1},...,x_{K-1})
\end{equation}
and we have to compute them for different functions $f({\mathbf x})$ but the
same $P({\mathbf x})$~\footnote{%
think for example to the case of many Euclidean green functions $G_{k}$
defined as 
\begin{equation}
G_{k}=\int [\text{d}{\mathbf x}]f_{k}({\mathbf x})P({\mathbf x}) 
\end{equation}
where $P({\mathbf x})\propto e^{-S_{E}({\mathbf x})}$ and $S_{E}({\mathbf x})$
is the Euclidean action of a system in the configuration ${\mathbf x}$.}. The
clever trick we play is to modify our algorithm in the following way:

\begin{itemize}
\item  Generate a set of $N$ random points $\{{\mathbf x}^{[i]}\}$ in the
integration domain $\mathcal{D}$ using $P({\mathbf x}^{[i]})$ for the
probability distribution of the point ${\mathbf x}%
^{[i]}=(x_{0}^{[i]},x_{1}^{[i]},...,x_{K-1}^{[i]})$

\item  For each point ${\mathbf x}^{[i]}$ compute $f({\mathbf x}^{[i]})$

\item  Compute the average
\begin{equation}
I(N)=\frac{\text{Vol}(\mathcal{D})}{N}
\sum_{i=0}^{N-1}f({\mathbf x}^{[i]})
\label{montecarlo}
\end{equation}
where Vol$(\mathcal{D})$ is the volume
of the integration domain.
\end{itemize}

\inbox{If $f({\mathbf x})P({\mathbf x})$ is Riemann integrable then 
the function $I(N)$ converges to the integral eq.(\ref{int4}) when $%
N\rightarrow \infty .$ The factor $P({\mathbf x})$ in the integrand has been
absorbed into the probability distribution of generating the random points.}

We are now left with the problem of generating random points ${\mathbf x}%
^{[i]}$ with a given distribution $P({\mathbf x}^{[i]})$. There are many
algorithms that do this job. The simplest one is the Metropolis algorithm
~\cite{metropolis}.

\begin{enumerate}
\item  Start with $i=0$ and a point ${\mathbf x}^{[i]}$ chosen at random in
the integration domain.

\item  Generate another random point $\mathbf{y}$ in the integration
domain and a random number $\alpha $ in the interval $[0,1).$

\item  If $P({\mathbf y})/P({\mathbf x}^{[i]})>\alpha $ then 
${\mathbf x}^{[i+1]}=\mathbf{y}$ else ${\mathbf x}^{[i+1]}=\mathbf{x}^{[i]}$.

\item  Increase $i$ by $1$ and repeat steps 2,3,4.

\end{enumerate}

\begin{figure}
\begin{center} 
\epsfxsize=12cm
\epsfysize=12cm
\epsfbox{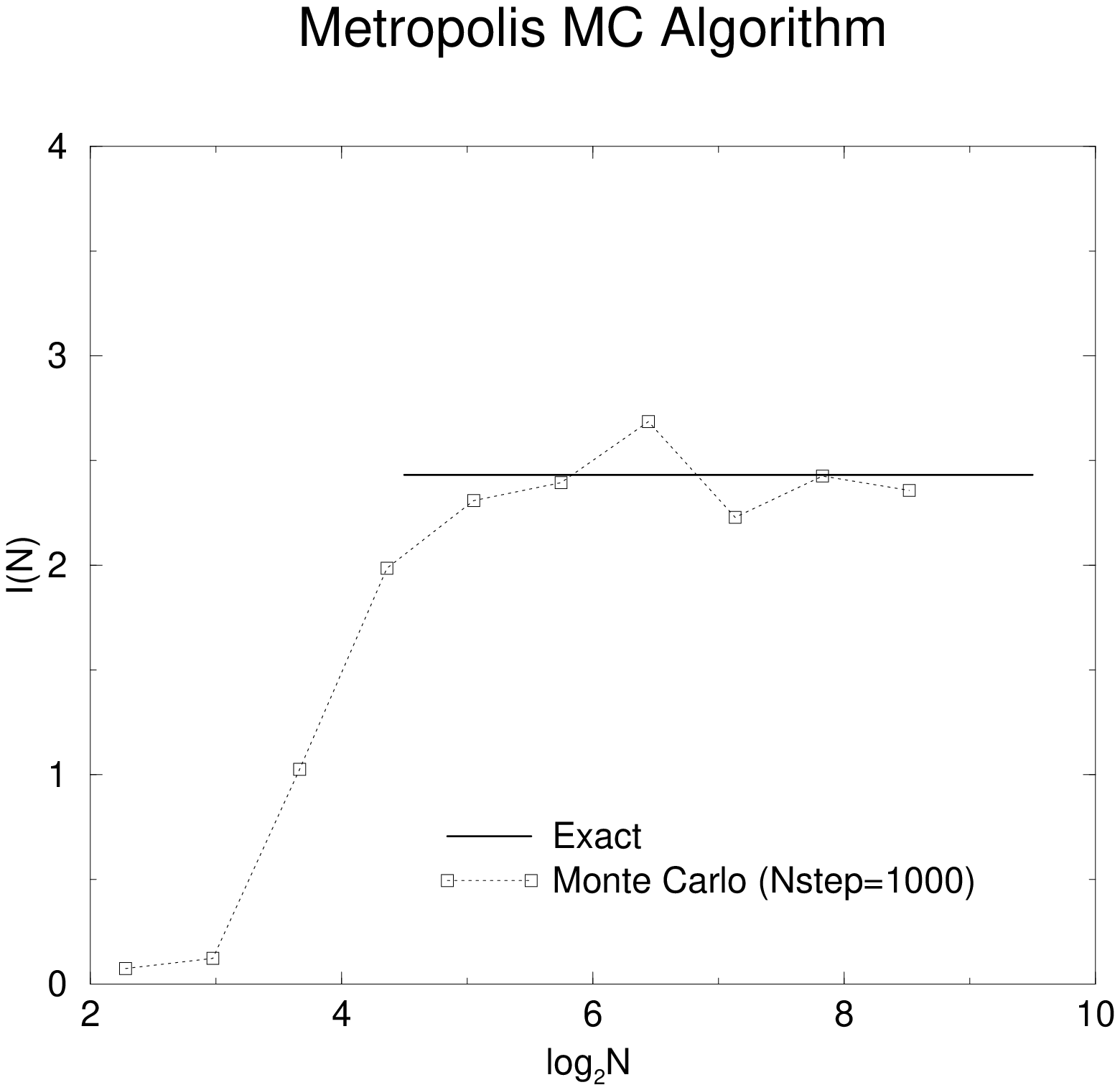} 
\end{center}
\caption{Convergent behavior of the Metropolis algorithm.\label{program4}}
\end{figure}

The succession of points $\{{\mathbf x}^{[i]}\}$ generated by this algorithm
has the required probability $P({\mathbf x}^{[i]})$. This kind of
succession is also known as Markov chain.

Each point of the Markov chain is called a {\it configuration}.

The Metropolis Monte Carlo only works with a real probability function, 
$P(x)$, but the functions $f(x)$, and the integration variables $\mathbf{x}$, 
can be complex, vectors or tensors.

\texttt{program4.c} computes the following 3D integral\footnote{%
it can easily be computed analytically by factorization and one finds $%
I=2.43131$ } 
\begin{equation}
I=\int_{0}^{1}dx_{0}\int_{0}^{1}dx_{1}\int_{0}^{1}dx_{2}f({\mathbf x})P(%
{\mathbf x})
\end{equation}
\begin{eqnarray*}
\text{where }P({\mathbf x}) &=&e^{-(x_{0}^{2}+x_{1}^{2}+x_{2}^{2})} \\
\text{and } f({\mathbf x}) &=&128x_{0}^{3}x_{1}^{2}x_{2}
\end{eqnarray*}

Its convergent behavior is shown in fig.4.

\subsection{Bootstrap errors}

Monte Carlo integration does not just provide us with a way to compute
multidimensional integrals. There are two techniques, known and Jackknife
and Bootstrap~\cite{errors}, that permit us to evaluate the statistical error we commit
when we truncate the Markov chain and we compute the numerical integral on a
finite number of configurations, $N$. We consider here the Bootstrap
method.

Let's assume we have a finite number, $N$, of configurations that constitute
the Markov chain $\{{\mathbf x}^{[i]}\}$ and the integral $I$ which is
computed numerically as 
\begin{equation}
I \simeq I(N)\stackrel{def}{=} \frac{\text{Vol}({\mathcal D})}N\sum_{i=0}^{N-1}f({\mathbf x}^{[i]})
\label{iii}
\end{equation}

\inbox{Since in general the $f({\mathbf x}^{[i]})$ are not Gaussian distributed,
standard error analysis is not applicable.} 

The Bootstrap algorithm consists of the following steps:

\begin{itemize}
\item  Construct a table of $N\times M$ integer random numbers $k_{ij}$
where $i\in \{0,1,...,N-1\}$, $j\in \{0,1,...,M-1\}$ and $k_{ij}\in
\{0,1,...,N-1\}$ for each couple $(i,j)$ ($M$ is an input parameter that we
choose to be equal to $100$).

\item  For each $j$ compute
\begin{equation}
\overline{I}_{j}\stackrel{def}{=} \frac{\text{Vol}({\mathcal D})}N \sum_{i=0}^{N-1}f({\mathbf x}^{[k_{ij}]})
\end{equation}

\item  Reorder the set $\{\overline{I}_{j}\}$ so that, for each $j$, 
$\overline{I} _{j}< \overline{I} _{j+1}$
\end{itemize}

\inbox{The result for the integral $I$ lies between 
$\overline{I}_{33}$ and $\overline{I}_{66}$ at 65\% confidence level.}

\begin{figure}
\begin{center} 
\epsfxsize=12cm
\epsfysize=12cm
\epsfbox{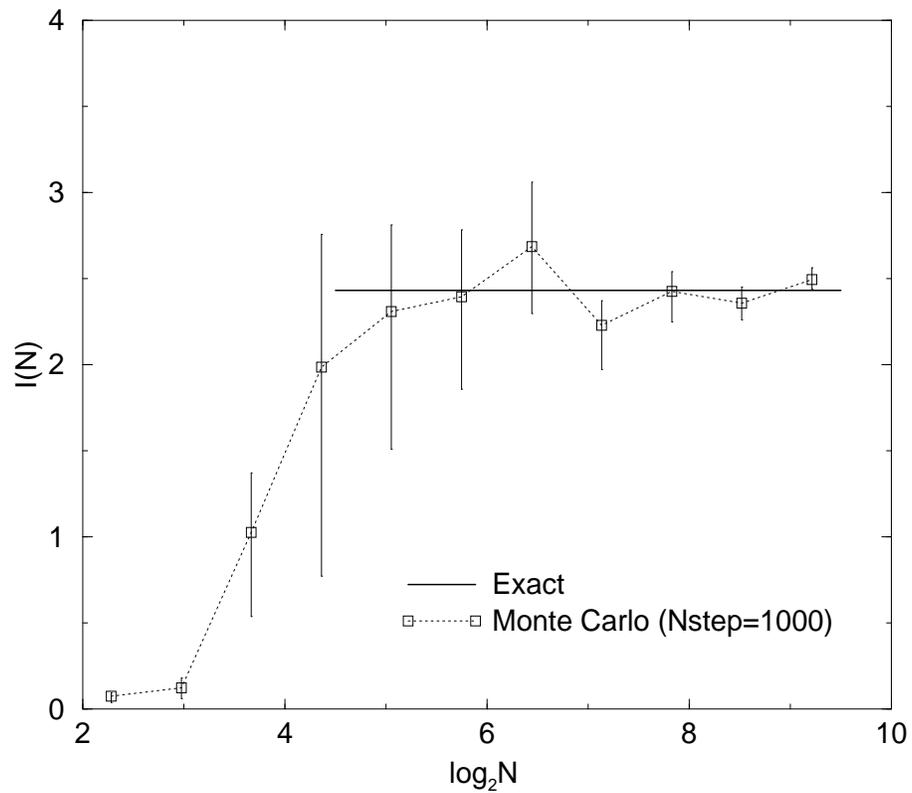} 
\end{center}
\caption{Convergent behavior of the Metropolis and Bootstrap algorithm. \label{program5}}
\end{figure}

The idea behind the Bootstrap algorithm is that, for each $j$, and for $N>>1$%
, the $j$-th Markov chain $\{{\mathbf x}^{[k_{ij}]}\}$ (chain in $i$) has the
same probability distribution of the original chain $\{{\mathbf x}^{[i]}\}$.
Therefore the different $\overline{I}_{j}$ (the numerical integral computed
using the derived chain $j$) have the same probability distribution as that
of the derived chain $j$ associated to $\overline{I}_{j}$.

\texttt{program5.c} implements the Bootstrap algorithm as complement of the
Metropolis algorithm. Fig.5 shows the same data of fig.4 including the
Bootstrap error. The kind of error one does when truncates the Markov chain
is known as \emph{statistical error}.
 \clearpage \newpage

\section{Definition of Path Integrals}

\outbox{Calculus required continuity, and continuity was supposed 
to require the infinitely little; but nobody could discover what the
infinitely little might be}{Bertrand Russell}

The concepts of Regularization and Renormalization play a fundamental
role in the definition of the Path Integral (and in physics in general) 

There is a physical reason for it:
\begin{itemize}
\item One never measures the value of a field (associated to a particle)
in every point in space-time but one measures its integral over the test
function of the physical detectors, which have a finite
extension. Therefore there are mathematical reasons to require that  
the fields are defined in the space of distributions.
\item One wants to model the unknown short distance physics by
introducing a Lagrangian density which contains only local (contact)
interactions. 
\end{itemize}

If one tries to combine the previous statements in a Quantum Field
Theory, one encounters the problem of divergences and must find a way of
dealing with them.  The mathematical origin of these divergences
is the presence of
undefined formal products of distributions in the Lagrangian from which
the path integral is computed. Regularization and Renormalization are in fact,
in mathematical terms, the solution to the problem of defining these
products of distributions%
\footnote{For general reviews on the subject see~\cite{rabin,jaffe}. 
For an application to Quantum Mechanics see~\cite{lepage}}.

We analyze here, as
an explanatory example, the problem of defining the product of
$\delta$ functions
by regularizing them by a sequence of smooth functions that
become more localized at zero. We then show how 
an arbitrary quantum
field can be expanded and regularized using delta functions and how
the presence of a finite spatial cut-off in the regularized
distributions is equivalent to a finite cut-off in the momentum
expansion of the field itself.

This is not intended to be an introduction to Renormalization,
but it is presented as an alternative view of its meaning 
having in mind the lattice as typical regulator.

\subsection{Toy example: Regularizing distributions}

The $\delta$ function is a distribution which is defined as
\begin{equation}
\int \delta (x-\overline{x})F(x)\text{d}x=F(\overline{x})
\end{equation}
for any smooth test function $F(x)$. The same $\delta $ function can be
thought of as the limit of an ordinary function $\delta (a,x)$%
\begin{equation}
\delta (x)=\stackunder{a\rightarrow 0}{\lim }\delta (a,x)
\end{equation}
where $\delta(a,x)$ must be enough regular, localized in $x$ within a
precision $a$ and its integral normalized to one.
\index{regularization}
This procedure is called {\it regularization}. Some possible
regularization schemes are%
\footnote{The $\theta(x)$ function is defined to be $0$ for $x<0$ and
$1$ for $x>0$. It is discontinuous in $x=0$.}
\begin{eqnarray}
\delta (a,x) &=&\frac 1a\left[ \theta (x+a/2)-\theta (x-a/2)\right] 
\label{lattice1} \\
\delta (a,x) &=&\left( \pi a^2\right) ^{-\frac 12}\exp (-x^2/a^2)
\label{lettice3} \\
\delta (a,x) &=&\frac{\sin(\pi x/a)}{\pi x} \label{lattice4} 
\end{eqnarray}

\begin{figure}
\begin{center} 
\begin{tabular}{ccc}
\epsfxsize=3cm
\epsfysize=3cm
\epsfbox{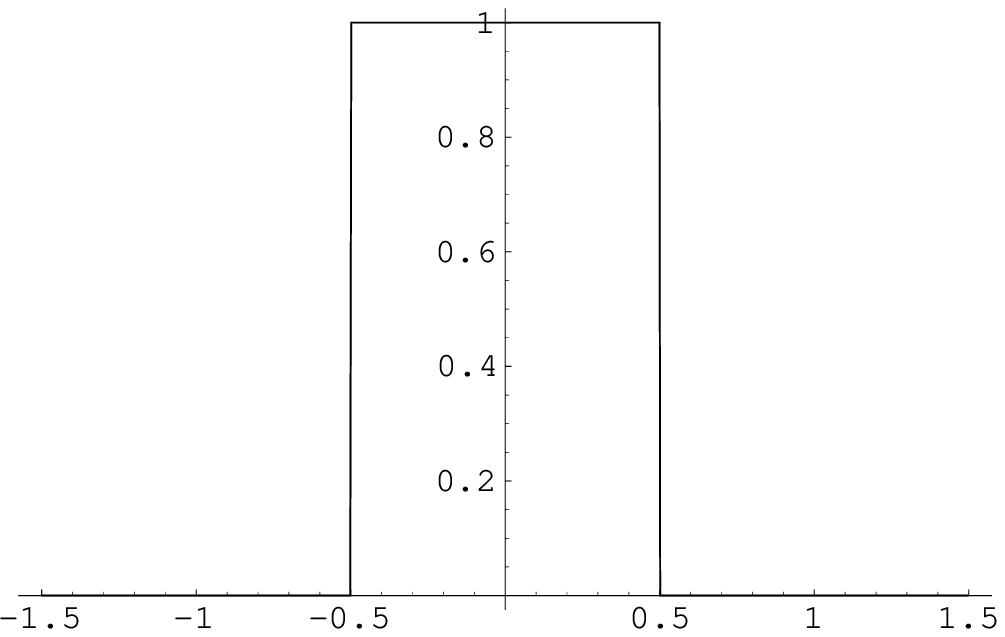} &
\epsfxsize=3cm
\epsfysize=3cm
\epsfbox{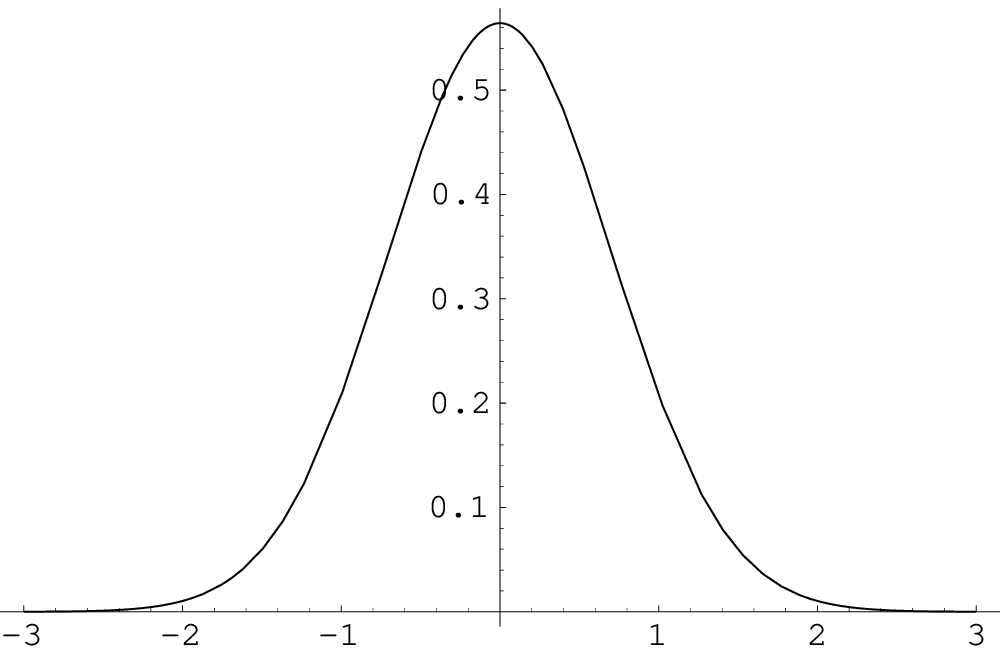} &
\epsfxsize=3cm
\epsfysize=3cm
\epsfbox{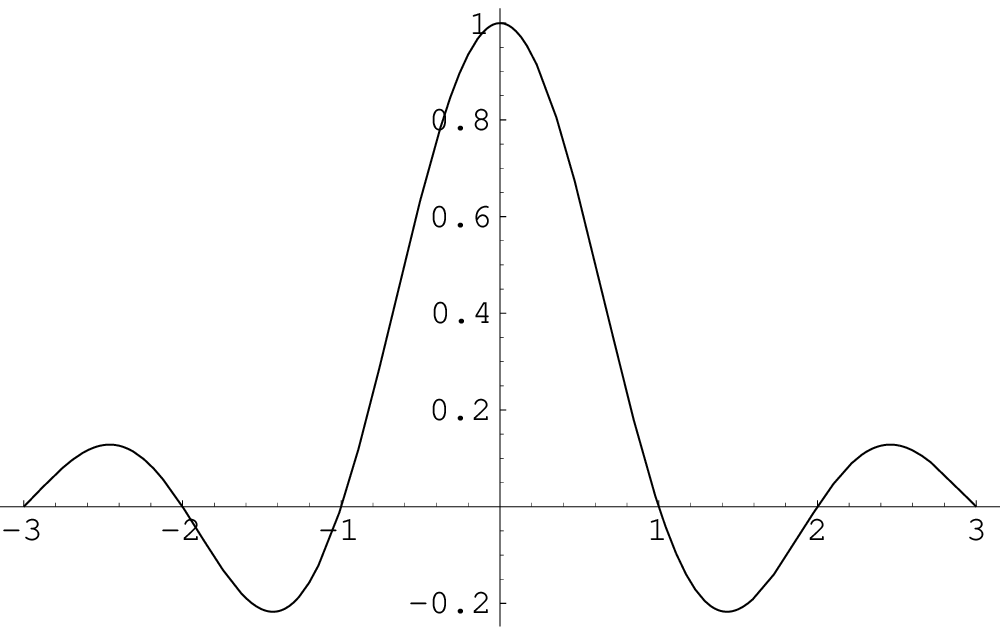} 
\end{tabular}
\end{center}
\caption{Examples of possible regularizations for a delta
function. The $x$ axis is in units of $a$, the $y$ axis is in units of
$a^{-1}$. \label{deltas}}
\end{figure}

They are sketched in figure \ref{deltas}. 
The different schemes are equivalent in
the sense that they give the same result for the following limit 
\begin{equation}
\stackunder{a\rightarrow 0}{\lim }\int \delta (a,x-\overline{x})
F(x)\text{d}x=F(\overline{x})
\end{equation}
but they do not give a well defined limit in expressions of the form 
\begin{equation}
\stackunder{a\rightarrow 0}{\lim }\int [\delta (a,x-\overline{x})]^{n+1}
F(x)\text{d}x=?
\label{questionmark}
\end{equation}
In fact eq.~(\ref{questionmark}) is divergent as $a^{-n}$. 
Suppose one wants to give a well defined meaning to the limit in
eq.~(\ref{questionmark}) by removing somehow the divergence that
occurs. One way of doing it is by making $F(x)$ dependent
on $a$, according with the prescription for $\delta (a,x)$, 
\begin{equation}
F(x)\rightarrow F_R(a,x) = Z_R^{-n}(a)F(x)
\end{equation}
where $Z_R^{-1}(a) = a+O(a^2)$. \index{renormalization}
In other words the divergence of the integral is absorbed in the
normalization of the function $F(x)$. 
This procedure is called {\it renormalization} and it depends on which
regularization has been chosen. From now on the scheme of
eq.~(\ref{lattice1}) will be considered in particular.
After renormalization
\begin{equation}
\int [\delta (a,x-\overline{x})]^{n+1}Z_R^{-n}(a)
F(x)\text{d}x = \text{const.} + O(a)
\label{questionmark2}
\end{equation}
and its limit for $a \rightarrow 0$ becomes well defined.
Therefore, up to order $a$ terms one
can redefine the integral of eq.~(\ref{questionmark}) in the following
way
\begin{equation}
\int [\delta (x-\overline{x})]^{n+1} F(x)\text{d}x \stackrel{def}{=} 
\stackunder{a\rightarrow 0}{\lim } \int 
[\delta (a,x-\overline{x})]^{n+1} Z_R^{-n}(a)F(x)\text{d}x
\end{equation}
The situation can be even more complicated if $F(x)$ itself is defined
in terms of delta functions. For example one can consider the case when
$F(x) = \exp[g\delta^2(x)]$. In this case it is not sufficient to
regularize $\delta$ and renormalize $F$ to get rid of the
divergence, one is forced to renormalize $g$ as well.

It is a general statement that, if the function $F(x,g)$ depends on some
constant $g$, one has to renormalize the constant 
\begin{equation}
	g \rightarrow g_R(a)
\end{equation}
by imposing a constraint 
\begin{equation}
\int [\delta (a,x-\overline{x})]^{n+1} Z_R^{-n}(a)F(x,g_R(a))\text{d}x=
\text{const.}  \label{rgequation}
\end{equation}

Eq.~(\ref{rgequation}) fixes the behavior of $g_R(a)$ as function of
$a$. Its solution, \index{RGE}\index{running}
$g_R(a)$, can have a non-trivial behavior in $a$. Eq.(\ref
{rgequation}) is a particular case of what is generally known as the 
{\it Renormalization Group Equation} (RGE)~\cite{peskin}.
The behavior of $g_R(a)$ versus $a$ is called {\it running}. 
Another common way of writing the renormalization group equation is
\begin{equation}
	\frac{\text{d}}{\text{d}\log a}\int [\delta (a,x-\overline{x})]^{n+1}
	Z_R^{-n}(a)F(x,g_R(a))\text{d}x = 0
\end{equation}
or explicitly
\begin{equation}
\left(a\frac{\partial}{\partial a} -
\beta(g_R)\frac{\partial}{\partial g_R} + n\gamma(g_R) \right) \int
[\delta (a,x-\overline{x})]^{n+1}Z_R^{-n} F(x,g_R)\text{d}x = 0
\label{rgequation_2}
\end{equation}
where 
\begin{align}
&\beta(g_R) \stackrel{def}{=} -\left. \frac{\partial}{\partial \log a} g_R(a)
\right|_{g_R} \\
&\gamma(g_R) \stackrel{def}{=} \left. \frac{\partial}{\partial \log a} Z_R(a)
\right|_{g_R}
\end{align}
If the original constant $g$ is dimensionless, $g_R(a)$ must also depend on
some other scale, say $\Lambda $, to cancel the dimension of $a$. In other
words $g_R$ must be a function of $a\Lambda $, an adimensional
quantity. 
This simple example shows how the renormalization procedure may force
one to introduce a second scale $\Lambda $ of which the renormalized
constant is a function. \index{dimensional transmutation} 
This phenomenon is called {\it dimensional transmutation}. 

Dimensional transmutation is also a characteristic of QCD (and gauge 
theories in general).
In typical physical problems, $a$ is a free parameter and it can be chosen
(the physics does not depend on it providing it is small enough).
$\Lambda$, on the other side, characterizes the typical scale of 
the physics one is describing. 
If one measures $g_R(a\Lambda)$ at some
arbitrary physical scale $a=\bar a$ one can uniquely determine 
the value of $\Lambda$.
Once $\Lambda$ is known one can predict the value of $g_R$ at any other scale $a$.

\subsection{Defining the Path Integral}

We now go back to the most general correlation function, defined in terms of the Path Integral. 
We rewrite it as
\begin{equation}
\langle 0 | 
T\{\phi (x_{1}) ... \phi (x_{n})\}
| 0 \rangle \stackrel{def}{=}
 \int [ {\textrm d} \phi ] 
F[\phi(x), g] 
\label{PI1}
\end{equation}
where $F[\phi(x),g]$ is the integrand
\begin{equation}
F[\phi(x),g] \stackrel{def}{=} \phi (x_1) ... \phi (x_n) 
e^{-{\mathcal{S}}_{\text{E}}[\phi,g]}
\label{PI2}
\end{equation}
and $g$ is the coupling constant that appears in the action.
For the moment we simply consider a one-dimensional scalar field theory $\phi(x)$ defined in the interval $[0,L]$. The specific form of the action, $S_E[\phi,g]$ is unimportant but we assume that the action contains an interaction term of the form
\begin{equation}
g \int \phi^n(x) \text{d}x
\label{interactionterms}
\end{equation}
with $n>2$. This makes $F[\phi,g]$ a non-trivial functional of $\phi(x)$.

\begin{figure}
\begin{center} 
\begin{tabular}{ccccc}
\epsfxsize=3cm
\epsfysize=3cm
\epsfbox{delta1.eps} & \hfill &
\epsfxsize=3cm
\epsfysize=3cm
\epsfbox{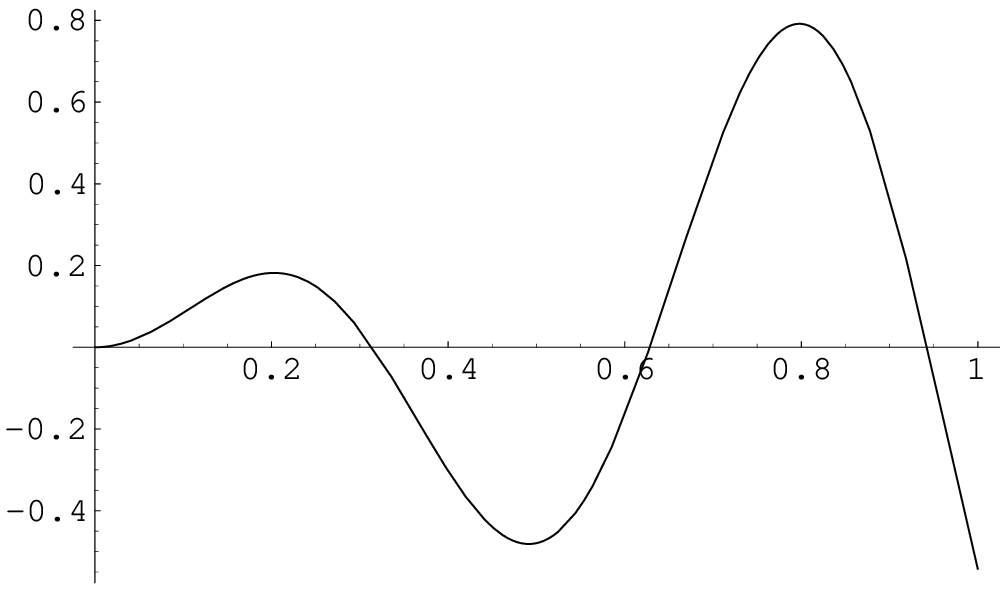} & \raisebox{1.2cm}{$\rightarrow$} &
\epsfxsize=3cm
\epsfysize=3cm
\epsfbox{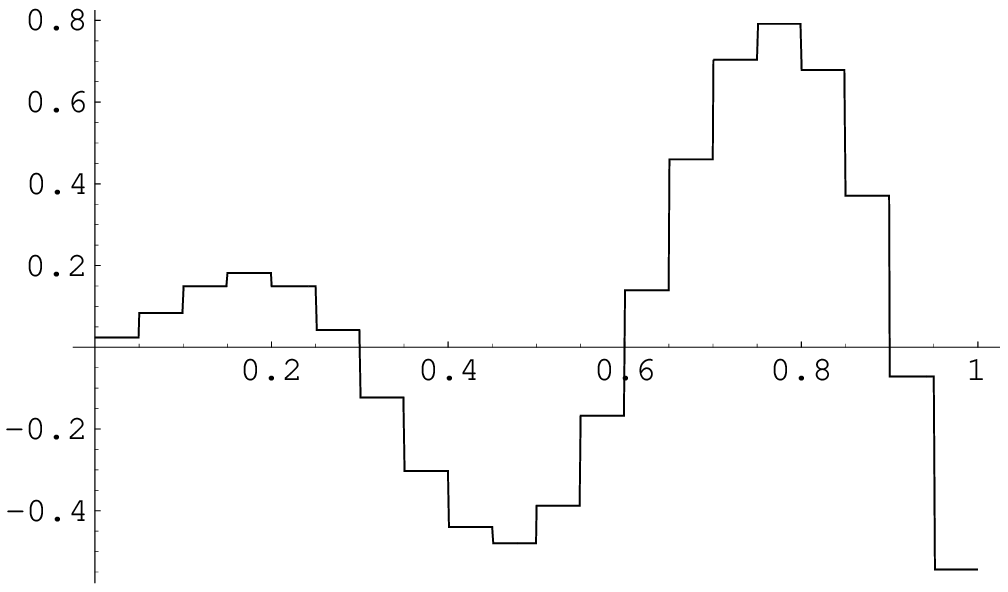} 
\end{tabular}
\end{center}
\caption{[left]:Example of a possible regularizations for the delta
function. The $x$ axis is in units of $a$, the $y$ axis is in units of
$a^{-1}$. [center-right]:Example of a continuum field configuration $\phi$ and its approximation with a linear combinations of regularized delta functions. \label{delta1}}
\end{figure}

\def\imgfar{\raisebox{-5mm}{
\epsfxsize=2.5cm
\epsfysize=1.5cm
\epsfbox{fa00.eps} }
}
\def\imgfaq{\raisebox{-5mm}{
\epsfxsize=2.5cm
\epsfysize=1.5cm
\epsfbox{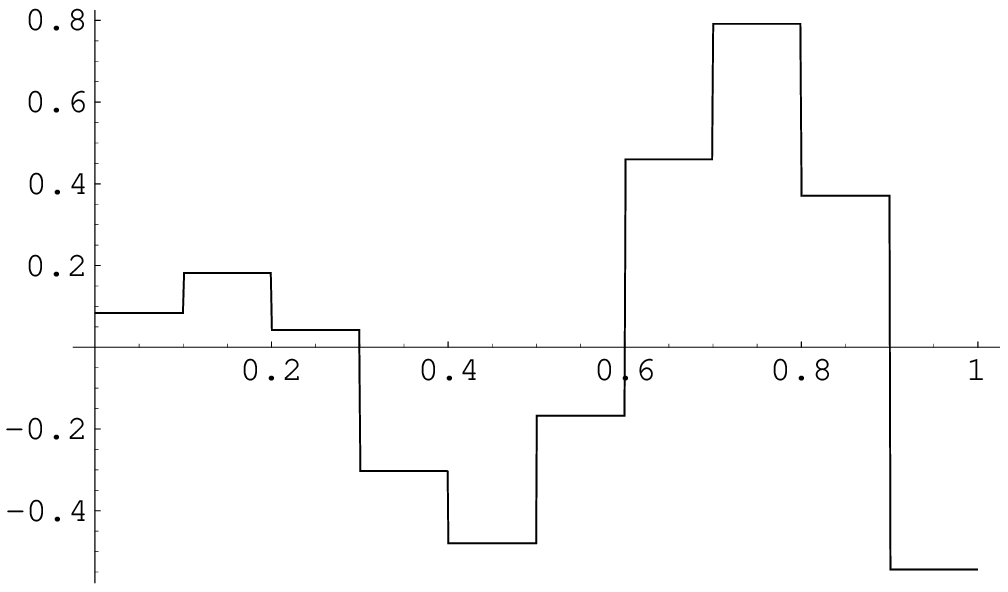} }
}
\def\imgfaw{\raisebox{-5mm}{
\epsfxsize=2.5cm
\epsfysize=1.5cm
\epsfbox{fa20.eps} }
}
\def\imgfae{\raisebox{-5mm}{
\epsfxsize=2.5cm
\epsfysize=1.5cm
\epsfbox{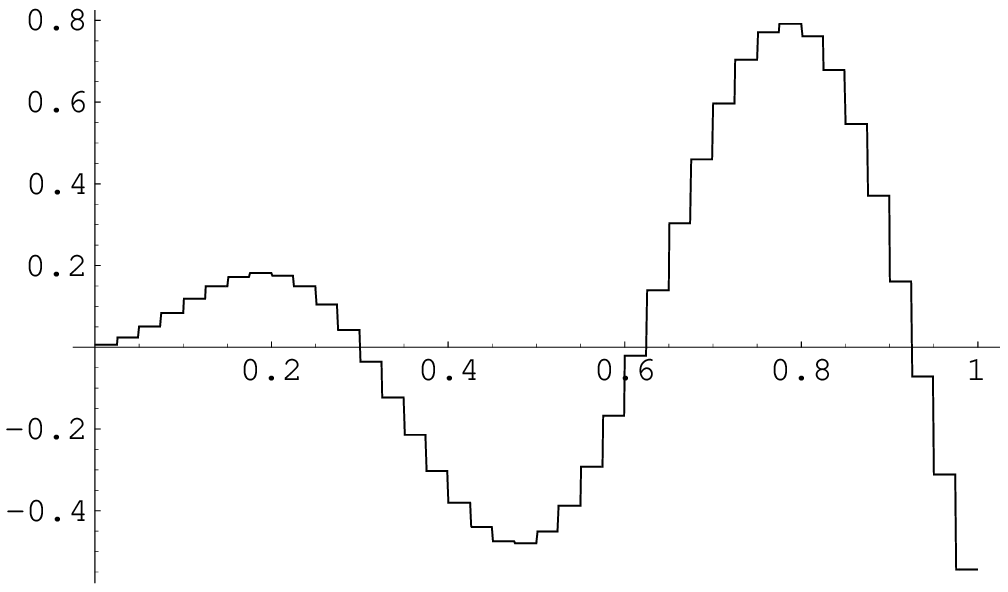} }
}
\def\imgfbr{\raisebox{-5mm}{
\epsfxsize=2.5cm
\epsfysize=1.5cm
\epsfbox{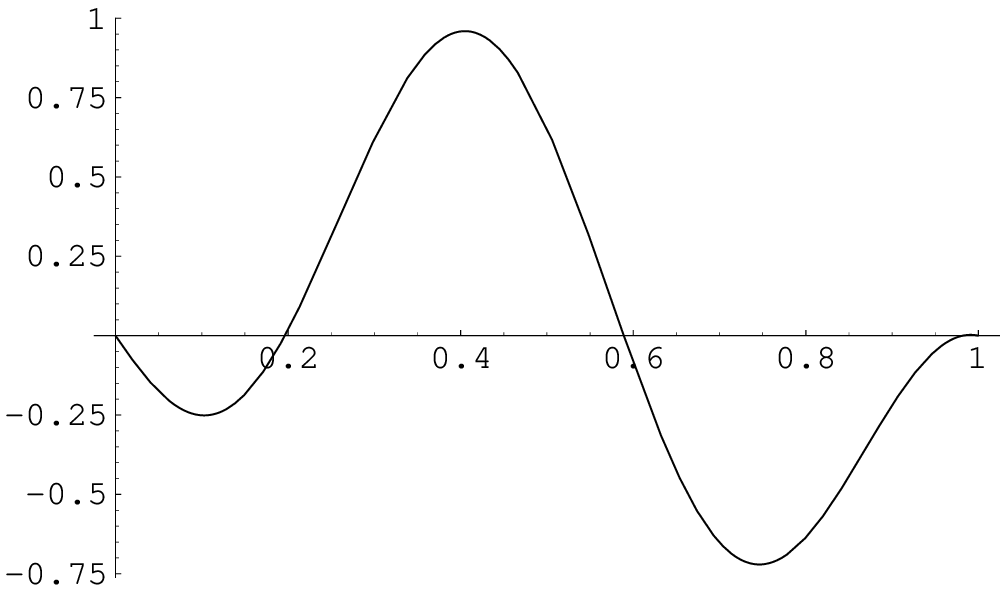} }
}
\def\imgfbq{\raisebox{-5mm}{
\epsfxsize=2.5cm
\epsfysize=1.5cm
\epsfbox{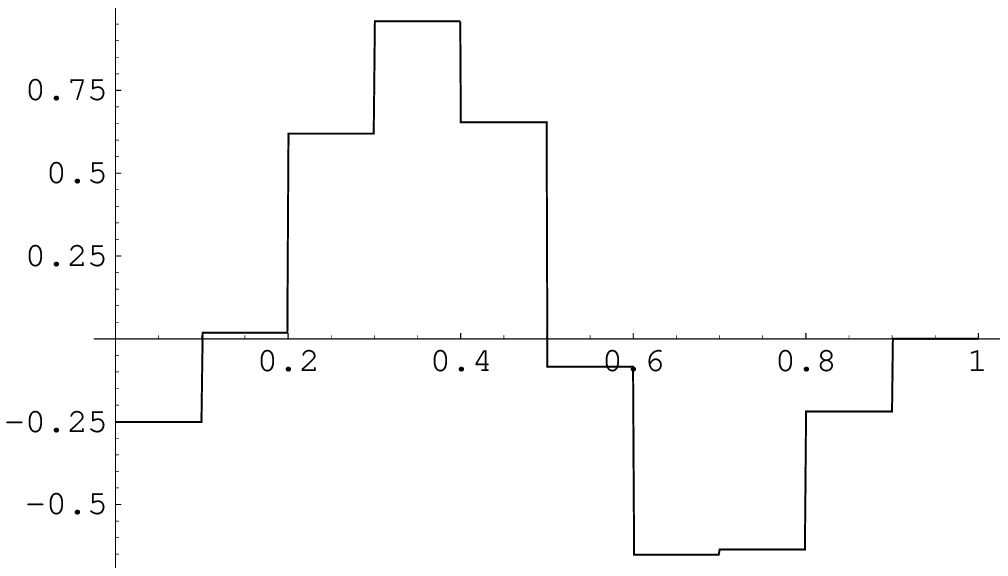} }
}
\def\imgfbw{\raisebox{-5mm}{
\epsfxsize=2.5cm
\epsfysize=1.5cm
\epsfbox{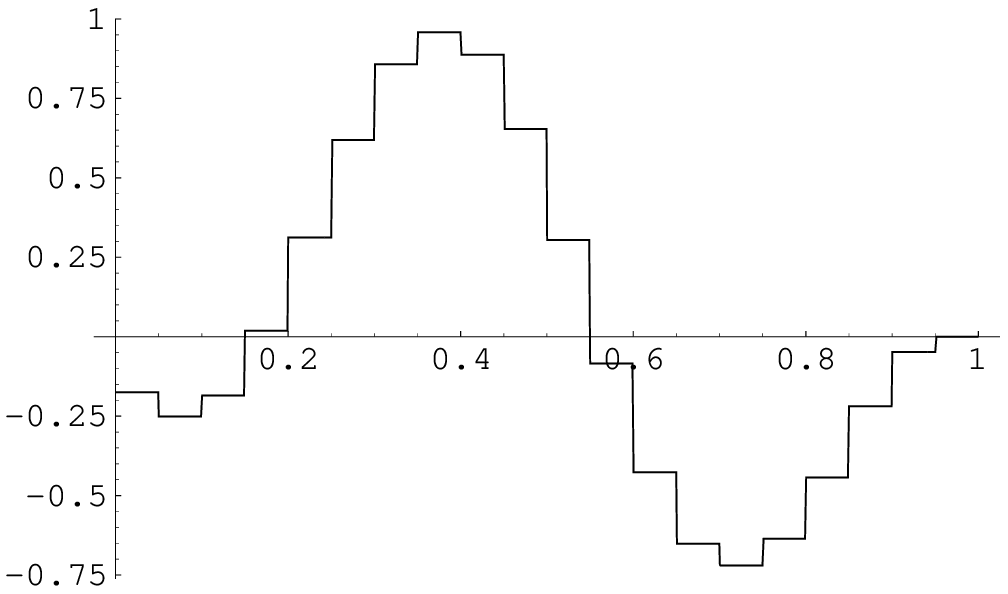} }
}
\def\imgfbe{\raisebox{-5mm}{
\epsfxsize=2.5cm
\epsfysize=1.5cm
\epsfbox{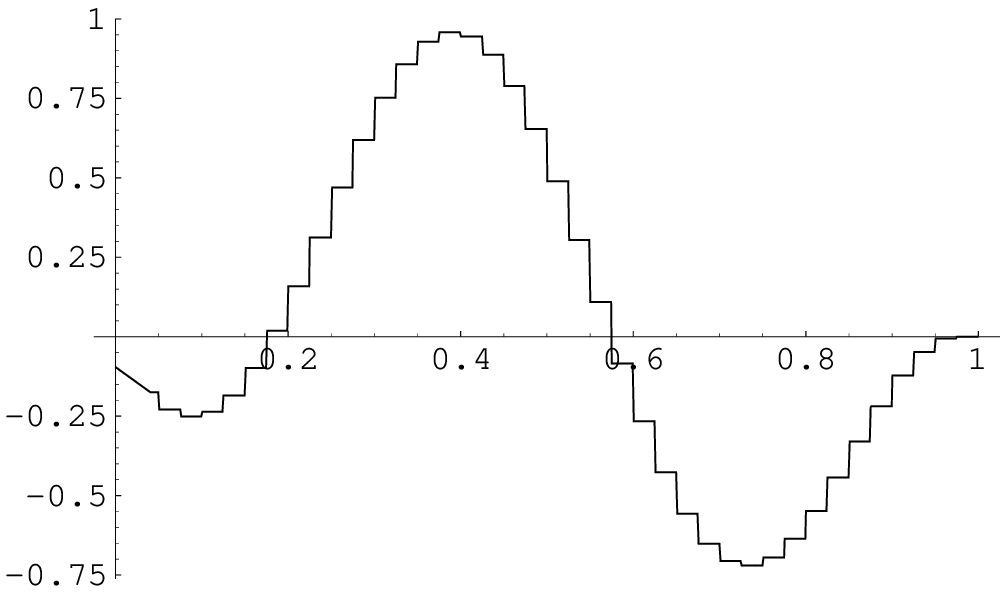} }
}
\def\imgftr{\raisebox{-5mm}{
\epsfxsize=2.5cm
\epsfysize=1.5cm
\epsfbox{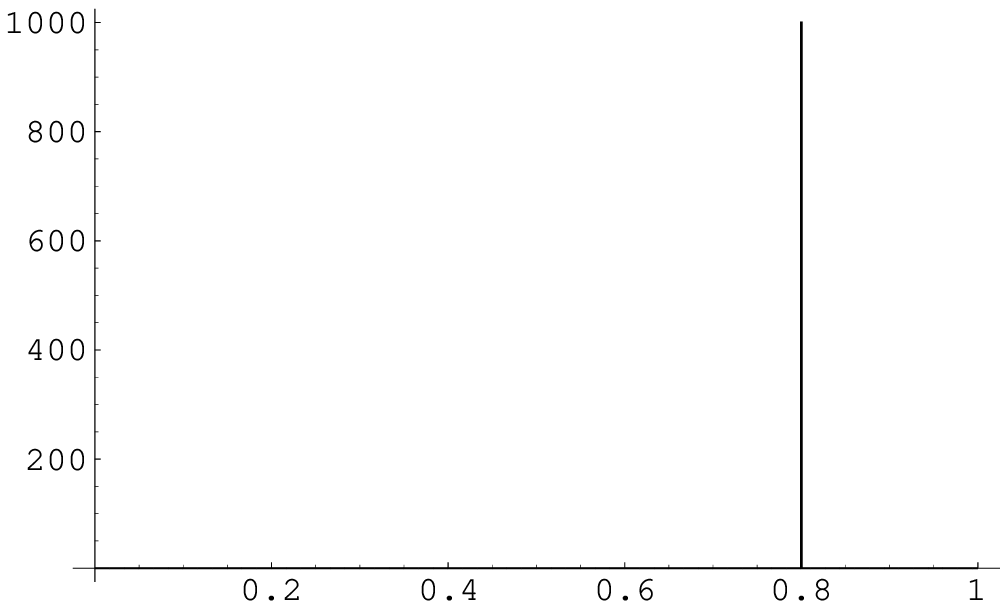} }
}
\def\imgftq{\raisebox{-5mm}{
\epsfxsize=2.5cm
\epsfysize=1.5cm
\epsfbox{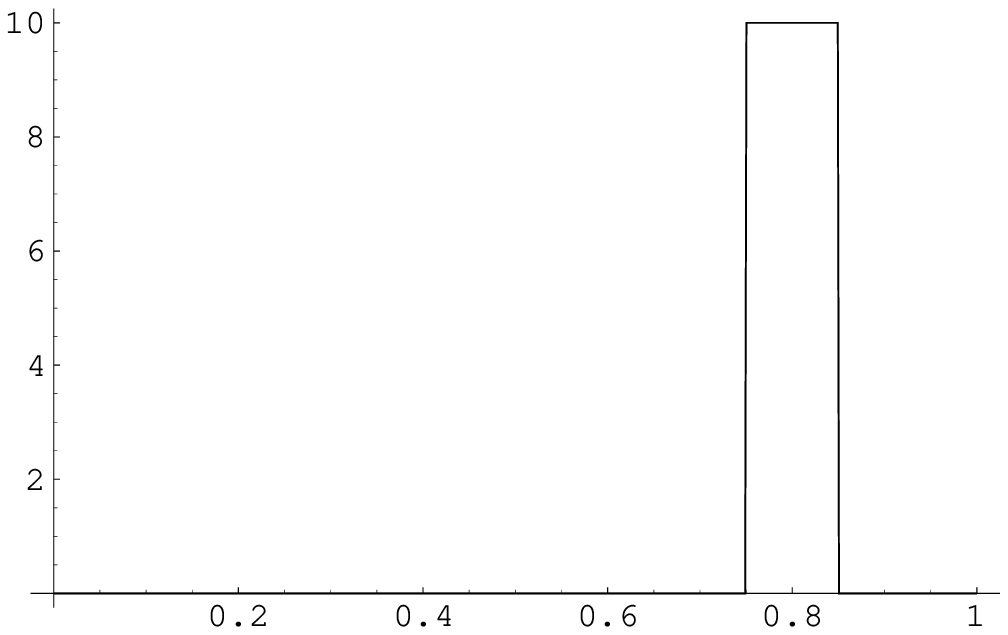} }
}
\def\imgftw{\raisebox{-5mm}{
\epsfxsize=2.5cm
\epsfysize=1.5cm
\epsfbox{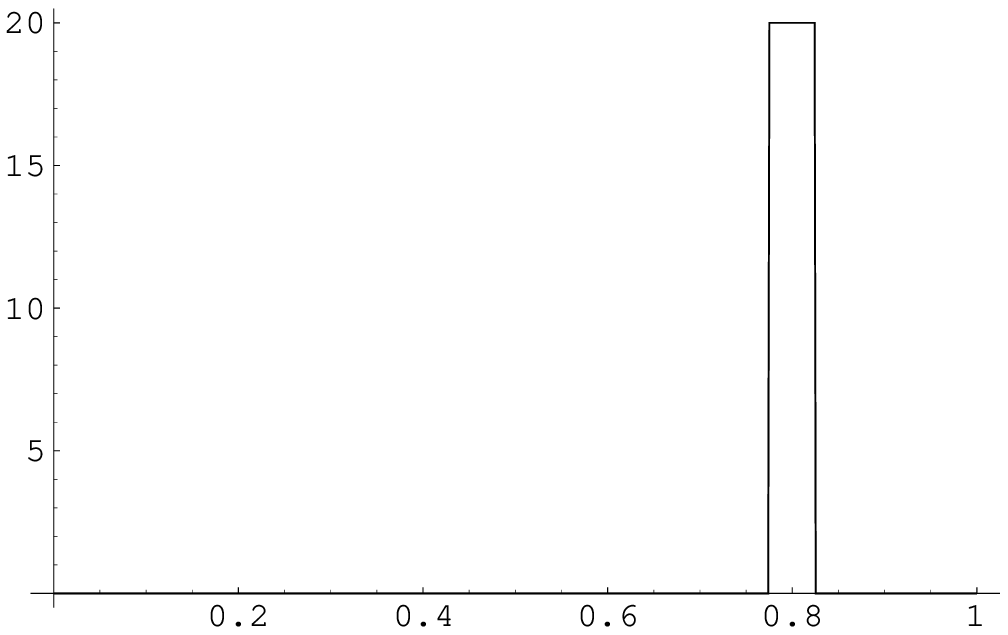} }
}
\def\imgfte{\raisebox{-5mm}{
\epsfxsize=2.5cm
\epsfysize=1.5cm
\epsfbox{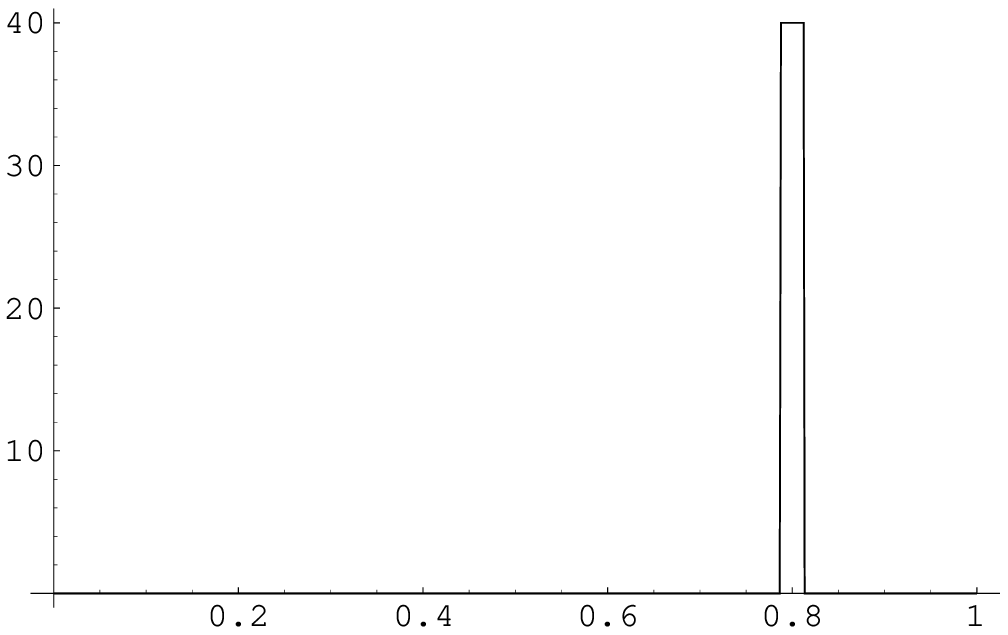} }
}

\begin{figure}
\noindent For $a=0.1$ ($K=10$):
\begin{eqnarray}
&&\int^{(a)} [ {\textrm d} \phi ] F[\phi,g] \stackrel{def}{=}
\int \text{d}\phi_0 ... \int \text{d}\phi_{9} F[\phi,g] = \label{main10} \\
&& F[\imgfaq] + F[\imgfbq] + ... + F[\imgftq] + ...\nonumber
\end{eqnarray}
For $a=0.05$ ($K=20$):
\begin{eqnarray}
&&\int^{(a)} [ {\textrm d} \phi ] F[\phi,g] \stackrel{def}{=}
\int \text{d}\phi_0 ... \int \text{d}\phi_{19} F[\phi,g] = \label{main11}\\
&& F[\imgfaw] + F[\imgfbw] + ... + F[\imgftw] + ...\nonumber
\end{eqnarray}
For $a=0.025$ ($K=40$):
\begin{eqnarray}
&&\int^{(a)} [ {\textrm d} \phi ] F[\phi,g] \stackrel{def}{=}
\int \text{d}\phi_0 ... \int \text{d}\phi_{39} F[\phi,g] = \label{main12}\\
&& F[\imgfae] + F[\imgfbe] + ... + F[\imgfte] + ...\nonumber
\end{eqnarray}
And at the continuum limit, $a\rightarrow 0$ ($K \rightarrow \infty$)
\begin{eqnarray}
&&\int [\text{d}\phi] F[\phi,g] \stackrel{def}{=}  \lim_{a\rightarrow 0} 
\underbrace{\int \text{d}\phi_0 ... \int \text{d}\phi_{K-1}}_{K \simeq L/a} F[\phi,g] = \label{main13}\\
&& F[\underbrace{\imgfar}_
{\int \phi^n(x) \text{d}x \text{ is finite}}] 
+ F[\underbrace{\imgfbr}_
{\int \phi^n(x) \text{d}x \text{ is finite}}] + ... 
+ F[\underbrace{\imgftr}_
{\int \phi^n(x) \text{d}x \rightarrow \infty}]+ ...\nonumber  
\end{eqnarray}
\caption{Example of lattice regularization of the Path Integral\label{main1}}
\end{figure}

\inbox{``Lattice regularization'' is the way to regularize the integral~(\ref{PI1}) 
by approximating the fields with sums of regularized distributions.
This is equivalent to discretize the space-time on which the fields are defined.} 

In practice one introduces a mimimum lenght scale, the ``lattice spacing'' $a$, and expresses the most general path (field configuration) $\phi(x)$ in terms of regularized delta functions (as shown in fig.~\ref{delta1}).
\begin{equation}
\phi (x)\simeq \phi ^{\text{latt}}(a,x)\stackrel{def}{=} \sum_{k=0}^{K-1}\phi_k\delta (a,x-ka)  \label{expansion0}
\end{equation}
where 
\begin{equation}
\phi _k\stackrel{def}{=} \frac1a \int_{ka}^{ka+a} \phi(x) \text{d}x 
\end{equation}
are ordinary integration variables and
$\delta (a,x)$ is a regularized delta function (it is normalized 
to one and peaked at zero within the length scale $a$, fig.~\ref{delta1}[left]).

With this prescription, for every finite $a=L/K$, one can define
\begin{equation}
\int^{(a)} [ {\textrm d} \phi ] F[\phi(x),g] \stackrel{def}{=}
\underbrace{\int \text{d}\phi_0
\int \text{d}\phi_1 ...
\int \text{d}\phi_{K-1}}_{K\simeq L/a} F[\phi ^{\text{latt}}(a,x),g] + O(a)
\label{reg1}
\end{equation}

Note the upper index $(a)$ which identifies the regularized Path Integral 
with lattice spacing set to $a$.

Due to discretization, the continuum variable $x$ has been replaced 
by the discrete index $k$.

For every finite $K$, the right-hand side of eq.~(\ref{reg1}) can be computed using
the Monte Carlo technique discussed in the preceding chapter. In this case
the integration variables $\phi_k$ replace the $x_k$ of the last chapter. 

\inbox{The regularized Path Integral, eq.~(\ref{reg1}), is well defined for any finite $a$. The limit $K \rightarrow \infty$ is infested by divergences and we must give a prescription to make sense to this limit.}

\inbox{Divergences associated with limit $a\rightarrow 0, K\rightarrow \infty$ (at $L=Ka=$constant) are called ultraviolet, while those associated with limit $K,L\rightarrow \infty $ (at $a=L/K=$constant) are called infrared.}

Fig.~\ref{main1} shows in a schematic way, how the path integral, eq.~(\ref{main13}) can can be approximated by finite multidimensional integrals, eqs.~(\ref{main10}-\ref{main12}), and the fields are defined on a lattice. For each finite $a$,
the ``infinite sum on the path'' (eqs.~(\ref{main10}-\ref{main12})) is well defined since all the divergences that may appear can be absorbed in the normalization of the integration measure and, for each path $\phi$, the functional $F[\phi,g]$ is finite. 
In the limit $a\rightarrow 0$ the regularized delta function $\delta(x,a)$ become more and more peaked and approaches a Dirac $\delta(x)$ function (eq.~(\ref{main13})). Since the integrand $F[\phi(x),g]$ is non-linear in the field and the product of delta functions is not defined, $F[\phi,g]$ diverges on those configurations $\phi(x) \simeq \delta(x)$. Therefore the Path Integral diverges.

These divergences are not physical since, in any practical experiment, one only has a finite resolution, $\bar a$, and one cannot discriminate a $\delta(x)$ form $\delta(x,a)$ if $a < \bar a$. This is why one is allowed to substract these divergences or ignore them by considering the regularized theory an ``effective theory''.

The way out is the following: 
\inbox{To have a well defined limit $a \rightarrow 0, K \rightarrow \infty$ (at $L=Ka=$constant) one must impose that the result of the regularized path integral is independent from $a$. The only way to do it is to make the field normalization and the coupling constant (the $g$ of eq.~(\ref{interactionterms})) dependent on $a$
\begin{equation} 
g \rightarrow g_R(a,\Lambda) 
\label{ga}
\end{equation} 
(the constant $\Lambda$ must be introduced because in general $a$ and $g$ 
do not have the same dimensions). 
This makes the physics independent by the lattice scale $a$.}

One does it by choosing a particular correlation function (identified by the functional integrand $F[\phi,g]$ and imposing the contraint
\begin{equation}
\frac{\text{d}}{\text{d}\,\log a} \left[ \underbrace{
\int \text{d}\phi_0 \int \text{d}\phi_1 ...
\int \text{d}\phi_{K-1}}_{K \simeq L/a} F[\phi(x), g_R(a,\Lambda)] \right] \simeq 0
\label{RGE3}
\end{equation}
This determines the behavior (the running) of $g_R(a)$.
Eq.~(\ref{RGE3}) is nothing else than a way to write the 
Renormalization Group Equation for a lattice regularized theory. 
The appearence of $\Lambda$ is called {\it dimensional trasmutation}.

The effects of very short distance physics (below the length scale $a$) 
only appear in the regularized correlation functions
through corrections which are proportional to $a$, or through the
renormalization of the coupling (and, in general, of the masses)~\cite{applequist}.

\inbox{It should be remarked that this procedure of defining the limit $a \rightarrow 0$ cannot be carried out for an arbitrary theory since there may be more sources of divergences than coupling constants. If this limit can be defined the theory is said to be renormalizable.}

Usually one distinguishes between the ``bare'' parameters that
appear in the regularized Lagrangian (for a finite value of the
cut-off, $a$) and the ``dressed'', ``physical'' or ``renormalized''
parameters that are defined, i.e. measured, by actual experiments. If one
takes the limit $a \rightarrow 0$, the bare parameters lose
any physical meaning and one must carefully define the renormalized
ones (one says to choose a prescription). 
If one is happy of keeping the cut-off small but finite one is
allowed to identify the renormalized and the bare parameters, because
these can now be measured. This is the approach one uses on the
lattice and it corresponds to the Kadanoff-Wilson approach to
renormalization.

One can write a RGE both for the bare parameters (as function of
the cut-off) or, equivalently, for the renormalized ones (as function of
the renormalization scale, i.e. the scale at which one performs 
the measurements). The two equations are formally identical at first 
order in pertubation theory%
\footnote{Only at second order in perturbation theory the RGE for the
renormalized 
parameters shows a dependence from their exact definition, the
renormalization prescription.}.
This is not surprising because one can always
define the renormalized coupling at a scale $\bar a$ to be equivalent
to the bare coupling with a cut-off $a=\bar a$.

In any practical lattice simulation one computes the regularized integrals, eq.~(\ref{reg1}), for a finite $a$, using the Monte Carlo technique. 
The coupling constant is an input parameter that fixes the scale of the problem. 
Relating the values of the coupling constants at different lattice spacings is, 
in general, a fine tuning problem.

\subsection{Improving the convergence}

It has been shown by Symanzik~\cite{symanzik} that it is possible to
improve the convergence of the correlation functions of a regularized 
theory to its continuum limit 
($a\rightarrow 0$) from $O(a)$ to $O(a^{n+1})$. In order to achieve
this, it is necessary \index{Symanzik improvement}
to add to the action ${\mathcal S}_{\text{E}}$ terms which are proportional to $%
a,a^2,...,a^n$ and adjust the corresponding coefficients. 
These tersm are called {\it irrelevant operators} since their contribution vanishes
in the limit $a \rightarrow 0$.
This {\it improvement} technique is heavily used in lattice simulations where the
minimum length scale, the lattice spacing $a$, cannot be reduced
arbitrarily, therefore one desires the dependence on $a$ of the correlation
functions to be as small as possible.

\subsection{Lattice regularization and momentum cut-off}  

We have shown as any given smooth field $\phi (x)$ defined in $[0,L]$ can be
approximated with $\delta (a,x)$ functions 
\begin{equation}
\phi (x)\simeq \phi ^{\text{\text{latt}}}(a,x)\stackrel{def}{=} \sum_{k=0}^{K-1}\phi
_k\delta (a,x-ka)  \label{expansion1}
\end{equation}
with $K = L/a$.

The same field $\phi$ expanded in Fourier components
\begin{equation}
\phi (x)=\sum_{n=0}^\infty b_ne^{ip_nx}  \label{expansion2}
\end{equation}
where 
\begin{equation}
p_n\stackrel{def}{=} \frac{2 \pi n}L; \qquad
b_n\stackrel{def}{=} \frac 1{2\pi }\int_{0}^L\phi (x)e^{-ip_nx}dx
\end{equation}

\begin{figure}
\begin{center} 
\begin{tabular}{ccc}
\epsfxsize=4cm
\epsfysize=4cm
\epsfbox{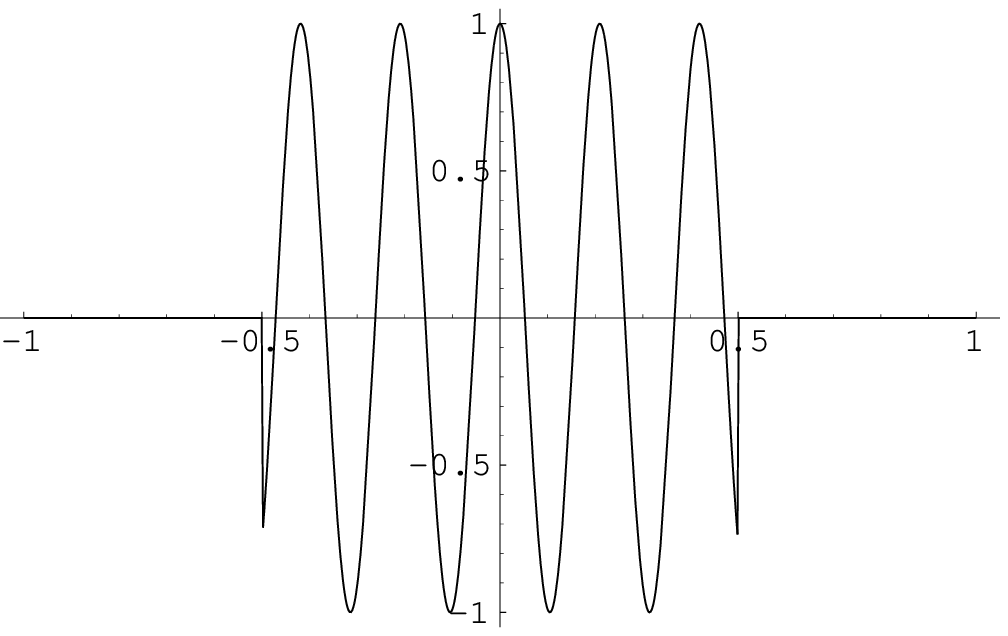} &
\epsfxsize=4cm
\epsfysize=4cm
\epsfbox{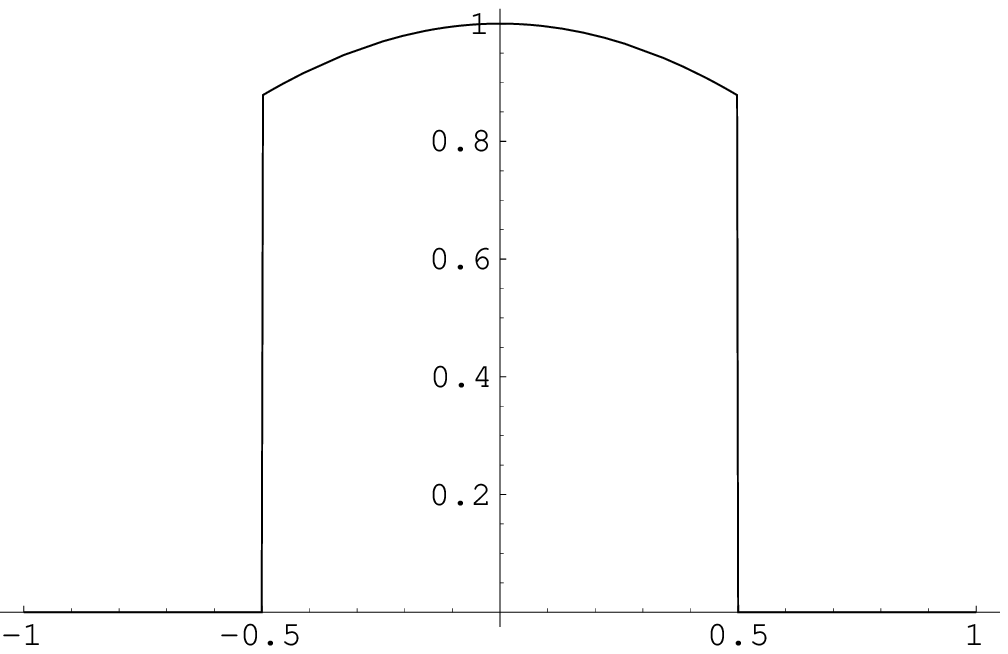}
\end{tabular}
\end{center}
\caption{Different behavior of the integrand of $b_n'$ for high
frequency modes (left) and low frequency modes (right) respectively. 
The $x$ axis is in units of $x/L$. \label{waves}}
\end{figure}

Also the right hand side of eq.(\ref{expansion1}) can be 
expanded in Fourier components and this can be written as 
\begin{equation}
\phi ^{\text{\text{latt}}}(a,x)=\sum_{n=0}^\infty b_n^{\prime }e^{ip_nx}
\end{equation}
where 
\begin{equation}
b_n^{\prime }=\frac 1{2\pi }\sum_{k=0}^{K-1}\left[ \phi
_k\int_{0}^L\delta (a,ka-x)e^{-ip_nx}dx\right] 
\end{equation}
It becomes evident that for $p_n>1/a$ the integrand oscillates fast and
the corresponding integral, $b_n^{\prime }$, has to be small; while for $%
p_n<1/a$ the integral is almost constant and approximately equal to $%
e^{-ip_nka}$, therefore $b_n^{\prime }\simeq b_n$. 
The different behavior of the integrand is shown in figure \ref{waves}.
This proves that eq.(\ref{expansion1}) can be written as 
\begin{equation}
\phi (x)\simeq \phi ^{\text{\text{latt}}}(a,x)\simeq \phi ^{\text{co}}(a,x)\stackrel{def}{=}
\sum_{n=0}^\infty \theta \left(\frac{1}{a}-p_n\right) b_ne^{ip_nx}
\end{equation}

If $\phi (x)$ describes some physical quantity and one has a finite
resolution in space, $\bar a$, then $\phi (x)$ can be replaced by its Fourier
expansion with a cut-off in momentum space $p_{\text{max}}<\bar a^{-1}$. 
Therefore a momentum cut-off can be regarded as an alternative procedure to regularize 
the path integral.

Note that $b_0$ is the mean value of $\phi(x)$. 
Moreover $p_1=2\pi/L$ represents the minimum energy/momentum mode that can
propagate on a finite unidimensional volume of length $L$.

\index{regularization!lattice}
\index{regularization!momentum cut-off}
\index{regularization!Pauly-Villars}
\index{regularization!dimensional}
The superscripts ``latt'' and ``co'', used to identify the two different
regularizations, are abbreviations of {\it lattice} and {\it cut-off}
respectively.

\subsection{Remarks on the physics of Effective Theories}

We have seen how, in QFT, one is forced to introduce three different
length scales:

\begin{itemize}
\item $\bar a$ the higher spatial (temporal) resolution of current experiments. 

\item  $a$: the scale that corresponds to the precision of the
mathematical description, the cut-off in the Kadanoff-Wilson approach to
Renormalization Group.
The theory is said to be renormalizable if this scale can be
arbitrarily small. This does not mean that one can trust the 
predictions of the theory with arbitrary precision.

\item  $1/\Lambda $: the typical lenght scale of the physics being studied. 
This scale is in nature and there is no freedom to fix it. 
For QCD it is $\Lambda_{\text{QCD}}\simeq 250$MeV 
($1/\Lambda_{\text{QCD}}\simeq 1$fm).
\end{itemize}

It is usually possible to model the same phenomenon using different QFTs,
which differ in the regularization-renormalization prescription and/or
the renormalization scale. The predictions of these different theories
must be compatible with each other apart from order $O(a)$ corrections.

Some QFTs have a finite number of coupling 
constants and it is possible to give a well
defined meaning to the limit $a\rightarrow 0$ because all the possible
divergencies can be absorbed in the renormalized constants. These QFTs are
said to be renormalizable. Other QFTs are not renormalizable because
it is not possible to absorb all the divergencies in a
finite set of constants.  The possible correlation functions, at different orders
in perturbation theory, exhibit an infinite variety of divergent behaviour.
Originally it was believed that only renormalizable QFT made sense%
\footnote{%
This requirement led Weinberg, Glashow and Salam to formulate the Standard
Model.}. \index{effective Lagrangian}
The modern picture is different: if the world is
described by a continuous QFT, it must be a renormalizable one. But the
world could have a minimum length scale and the renormalization requirement
is no longer a fundamental one. Perhaps the most important modern
interpretation of these results is that, if one wants to formulate a
QFT to describe physics down to a finite resolution, $\bar a$, and one
does not pretend it to be the theory of everything, it does not have to be
renormalizable (because one does not pretend to send the scale $a$ to
zero)~\cite{weinberg2}. 
These particular kind of theories are called {\it effective theories}%
\footnote{E.g. 
the Fermi theory of electroweak interactions is not renormalizable but it is
able to describe with good accuracy weak interactions at energy scales below 
$m_W$}.

In the real world, there might be new supersymmetric interactions or
superstring, or electrons and muons may have internal stucture, none of
which is incorporated in the Standard Model. Nevertheless the Standard Model, 
as it is today, explains the results of our experiments with a
typical accuracy of 10\%. Renormalization saves us by saying that one
does not need to know what happens at short distance (high momenta) in order to
understand low energy experiments. 

Since we will never be able to probe the physical world at every length scale,
every quantum field theory should be considered an effective theory.

 \clearpage\newpage

\section{Lattice QCD}

\outbox{He who can properly define and divide is to be considered a god}{Plato}

\index{effective Lagrangian!Lattice QCD}

At the typical hadronic scale, while electroweak effects can be computed in the
standard perturbative way, QCD effects are not. Therefore
it becomes necessary to formulate QCD in such a way to make it possible to
perform a numerical computation of the correlation functions.

To reach this goal we need to regularize the theory on a lattice by
introducing a finite lattice spacing $a$ and write down the QCD action 
in terms of the discretized fields.

In this section we will present one possible discretization of the action 
(due to Wilson) and the problems connected with the definition of chiral (massless) 
fermionic fields on the lattice. 

In principle one can compute the correlation functions of QCD with arbitrary 
precision by reducing the lattice spacing $a$.
In practice one only has a finite computational power therefore
one cannot arbitrarily reduce $a$. 
Hence we will show how to ``correct'' the action to improve 
the correction of the correlation functions from order $O(a)$ to order $O(a^2)$.

We will then discussed the errors associated with the typical lattice 
approximations.

In the end we will present, as an example, a full program to compute 
$f_B \sqrt{m_B}$ and some recent lattice results for this quantity.

\subsection{Basic degrees of freedom and action}

The Aharonov-Bohm experiment revealed that the \index{Ahronov-Bohm} 
gauge field $A_\mu=t^a A^a_\mu$ is not 
observable, because it is gauge dependent, but the phase of a
particle in a gauge field background, moving from $x$ to $y$, is an observable 
\begin{equation}
{\cal P}e^{ig\int_x^yA_\mu \text{d}x^\mu }
\end{equation}
(${\cal P}$ indicates a path-ordered exponential). \index{links}

For a finite lattice spacing $a$, it becomes convenient to write
the action of QCD in terms of the shortest paths on the lattice 
\begin{eqnarray}
U_\mu (x)  &\stackrel{def}{=}& e^{ig\int_x^{x+a\mu }A_\mu \text{d}x^\mu }
\simeq 1+ igaA_\mu (x+a\widehat{\mu }/2) \\
U_{-\mu }(x) &\stackrel{def}{=}& e^{ig\int_x^{x-a\mu }A_\mu \text{d}x^\mu }=U_\mu
^{\dagger }(x-a\widehat{\mu }) \simeq 1 - igaA_\mu (x-a\widehat{\mu }/2) 
\end{eqnarray}
(which are associated to the straight path connecting two consecutive lattice sites).
In QCD $U$ are $SU(3)$ matrices, called links.

The fermionic degrees of freedom instead, the quarks, can naively be defined
as $q_\alpha^i(x)$ fields associated to the lattice sites $x$ (where $\alpha$ 
is the spin index and $i$ is the color index).

The basic discretized operators which appear in the Lagrangian, are:

\begin{itemize}
\item  Ordinary derivative%
\footnote{We remark that a different choice could be made as long as in
the limit $a \rightarrow 0$ one recovers the ordinary derivative.}:
\begin{equation}
\partial _\mu q (x)\stackrel{def}{=} \frac 1{2a}\left[ q (x+a\widehat{\mu })-q (x-a%
\widehat{\mu })\right] 
\end{equation}

\item  Covariant derivative:
\begin{equation}
D_\mu q (x) \stackrel{def}{=} \frac 1{2a}\left[ U_\mu (x)q (x+a\widehat{\mu
})-U_{-\mu}(x)q (x-a\widehat{\mu })\right] 
\label{DiracOp}
\end{equation}
It is trivial to check that in the continuum limit it is equivalent to the
usual covariant derivative of QCD. In fact, up to order $a$
corrections,
\begin{eqnarray}
D_\mu q (x) &= & \frac 1{2a} 
\big[(1+igaA_\mu (x)+...) (1+a\partial _\mu +...) q(x) - 
\nonumber \\
&&\hskip 7mm (1-igaA_\mu (x)+...) ( 1-a\partial _\mu+...) q(x) \big] 
\nonumber \\
&=&(\partial _\mu +igA_\mu (x)+...)q (x)
\end{eqnarray}

\end{itemize}

For practical purposes, that will be examined below, one is usually
interested in the Euclidean formulation of Lattice QCD. 
\index{Wick rotation} This is achieved by performing the Wick
rotation%
\footnote{The way how different quantities transform under the Wick
rotation is reported in table~(\ref{wickrot})}  
\begin{equation}
x_0\rightarrow ix_0;\qquad x_i\rightarrow x_i
\end{equation}
Under this rotation the exponential term in the action becomes real
\begin{equation}
e^{i{\cal S}}=e^{ia^4\sum_x{\cal L}}\rightarrow 
e^{-{\cal S}_E}=e^{-a^4\sum_x{\cal L}_E}
\end{equation}
From now on all the quoted quantities (including $\gamma$ matrices)
will be Euclidean.
In terms of the links, any correlation function in QCD can be written as 
\begin{equation}
\langle 0 | {\cal O}(...) |0\rangle^{\text{latt}} \stackrel{def}{=} 
\int [\text{d}U][\text{d}q ][\text{d}\bar{q }%
]{\cal O}(...)e^{-{\cal S}_E[U,q,\bar{q}]}
\end{equation}
where~\cite{wilson}
\begin{equation}
{\cal S}_{\text{E}}={\cal S}_{\text{E}}^{\text{gauge}}+{\cal S}_{\text{E}}^{\text{quark}}
\end{equation}
and the two contributions are the gauge (pure Yang-Mills) and quark respectively. 
We write them in terms of the links as
\begin{eqnarray}
{\cal S}_{\text{E}}^{\text{gauge}} &\stackrel{def}{=}& \beta \sum_{x,\mu,\nu }
\left[ 1-\frac13 \text{Re tr}P_{\mu \nu }(x)\right] \nonumber \\
&=&a^4 \sum_x \frac14\text{tr}G^a_{\mu \nu }(x)G^{a\mu \nu }(x)+O(a^2) 
\label{llgauge} 
\end{eqnarray}
with
\begin{equation}
P_{\mu\nu}\stackrel{def}{=} U_\mu(x)U_\nu(x+a\hat\mu)
U_{-\mu}(x+a\hat\mu+a\hat\nu)U_{-\nu}(x+a\hat\nu) 
\end{equation}
and 
\begin{equation}
{\cal S}_{\text{E}}^{\text{quark}} \stackrel{def}{=} a^4\sum_x
\bar{q}(x)(\gamma ^\mu D_\mu +m)q(x) 
\label{llquarks}
\end{equation}

The variable $\beta=6/g^2(a)$ has been introduced to
conform to the standard notation of Lattice QCD.
In the gauge part of the action there
are no order $a$ corrections  and the first corrections arise at the
order $a^2$. On the other side,
in the quark part of the Lagrangian, order $a$
corrections play, in general,  a very important role for the
following two reasons:

\begin{itemize}
\item  If one neglects order $a$ corrections and naively uses the lattice
covariant derivative to implement ${\cal S}_E^{\text{quark}}$, 
one obtains a free
quark propagator of the form \index{Brillouin zone} \index{doubling problem}
\begin{equation}
S(p)=\frac{a}{i\gamma ^\mu \sin (p_\mu a) +am}
\end{equation}
which has 16 zeros in the Brillouin zone in the limit $m\rightarrow 0$, to
be confronted with the single zero of the continuum propagator
\begin{equation}
S(p)=\frac{1}{i\gamma ^\mu p_\mu +m}
\end{equation}
This problem is known as doubling.
To get rid of this proliferation of zero modes
Wilson proposed to add to the action a term of order $a$ of the form 
\begin{equation}
- a^5 \frac{r}{2} \sum_{x,\mu}
\bar{q}(x)\frac{1}{a^2}\left[U_\mu(x)q(x+a\widehat{\mu})
-2q(x)+U_{-\mu}(x)q(x-a\widehat{\mu})\right]
\label{mterm}
\end{equation}
In practical simulations $r$ is fixed to be
$1$, nevertheless it is convenient to show its dependence.
Note that eq.~(\ref{mterm}) is a mass term therefore it
explicitly breaks chiral symmery.

\item  When computing correlation functions one is always interested in the limit $%
a\rightarrow 0$ and one would like to improve the convergence of correlation
functions from order $a$ to order $a^2$. This can be done by adding to the
Action terms of order $a$ that compensate for the discretization errors
up to the same order~\cite{luscher98}. 
In particular one can choose a term of the form
\begin{equation}
-a^5\frac{c_{SW}}4\sum_{x,\mu>\nu }\bar{q }(x)\gamma ^\mu \gamma ^\nu
G_{\mu\nu}(x) q (x)
\label{clover_term}
\end{equation}
where $G_{\mu\nu}$ is a discretized version of the chromo-electro-magntic tensor 
and the constant $c_{SW}$ must be fixed somehow. 
\index{clover term} This term is usually referred to as {\it clover term}. 
It is important to observe that $c_{SW}$ is not a new free parameter of the theory, 
it is uniquely determined by the value of $\beta$ (i.e. by the lattice spacing).
\end{itemize}

Including these $O(a)$ corrections, the quark part of the action can be
re-written as
\begin{equation}
{\cal S}_E^{\text{quark}}[U,q,\bar{q}] 
=\frac{a^3}{2\kappa }\sum_y\bar{q }(x)Q_{xy}[U]q (y)
\label{actionsw}
\end{equation}
where \index{$\beta$} \index{$\kappa$} \index{$c_{SW}$}
\index{improvement!tree-level}  
\begin{eqnarray}
Q_{xy}[U] =&\delta _{xy} &-\kappa \sum_\mu \big[ (r-\gamma ^\mu )U_\mu
(x)\delta _{x,y-\mu }+(r+\gamma ^\mu )U_{-\mu }(x)\delta _{x,y+\mu }\big]
\nonumber \\
&& -\frac{\kappa c_{SW}}2\sum_{\mu >\nu }\gamma ^\mu \gamma ^\nu G_{\mu \nu
}(x)\delta _{xy} \label{qxy}
\end{eqnarray}
and, at tree-level%
\footnote{The fact that Lattice QCD with the action of eq.~(\ref{qxy}) is
$O(a)$ improved for 
on-shell quantities does not simply appear form a Taylor expansion. To show
the improvement it is necessary to list all dimension 5 operators, use the
equations of motion to reduce some of them and absorb the contribution of
the remaining two operators in the coupling constant and mass renormalization
\begin{equation}
g^2 \rightarrow g^2(1+b_gma); \qquad m \rightarrow m(1+b_mma)
\end{equation}}
\begin{equation}
\beta  =\frac 6{g^2(a)}; \qquad \kappa  =\frac 1{2ma+8r}; \qquad c_{SW}=1
\end{equation}
\index{Sheikoleslami-Wolhert action} The action
(\ref{actionsw}-\ref{qxy}) is referred to as {\it Sheikoleslami-Wolhert
action} (SW)~\cite{sw}. 

The coefficient $a^3/2\kappa$ can be absorbed in the definition of the
fermionic fields, $q$ and $\bar{q}$.

In practice, in lattice simulations, $a$ is not an input parameter, 
since it is dimensionfull. Therefore one fixes 
the lattice spacing, using dimensional transmutation, 
by setting a value for $\beta=6/g^2(a)$. 
One procedes in the following way:

One measures on the lattice some dimensionfull quantity, $m$
(for example the $K$ mass or the charmonium $1P-1S$ splitting), 
in adimensional units $a m$ 
and compares it with the experimental value $m^{\text{exp.}}$.
From the comparison one determines the value of $a$ corresponding to the
arbitrary input value of $\beta$.
So that one adjust $a$ by fine tuning the bare $\beta$.

It is not surprising that the only way to fix the physical parameters 
of the theory is by comparing them with physical experiments.

On the other size, the improvement coefficients, $c_{SW}$ for example, and
$\kappa_{crit}$ (the value of $\kappa$ associated to chiral fermions)
are uniquely associated with the value of $\beta$ and can be determined
theoretically.

\subsection{Simulation aspects and quenching}

First of all, one can consider correlation functions that do not depend
on quark fields and integrate out the quarks
\begin{eqnarray}
\langle 0 |{\cal O}(...) | 0\rangle^{\text{latt}} &=& 
\int [\text{d}U][\text{d}q ][\text{d}\bar{q}
]{\cal O}(...[U])e^{-{\cal S}_E[U,q,\bar{q}]} \nonumber \\
&=& \int [\text{d}U]{\cal O}(...[U])\det Q[U]e^{-{\cal
S}_E^{\text{gauge}}[U]} \nonumber \\
&=& \int [\text{d}U]{\cal O}(...[U]) P[U] \label{green3}
\end{eqnarray}
where
\begin{equation}
P[U]=e^{-{\cal S}_E^{\text{gauge}}[U]+\ln\det Q[U]} 
\label{probab}
\end{equation}

In practice one neglects \index{quenching}
the contribution of $\ln \det Q[U]$ in the probability $P[U]$,
eq.~(\ref{probab}). This is called the {\it quenched approximation}. 
This is a very crude approximation and it breaks the unitarity 
of the theory.
Its only motivation is the limitation in present computer power. 
It introduces a systematic error in the computations that
has to be quantified.

The probability $P[U]$ is real (this is why Lattice computations are
performed in Euclidean space) and resembles a Boltzman weight factor. 
Therefore standard statistical mechanics techniques can be applied.  

We have seen how, for a finite lattice spacing, the Path Integral reduces 
to a well defined $K$-dimensional integral which can be computed numerically
using Monte Carlo integration. 

In a typical Lattice QCD simulation one discretizes the space-time on 
grid of about $48$ sites in the temporal direction and $24$ sites in each spatial
direction. Therefore the computation of each correlation function correspond 
to the computation of a $K$-dimensional integral with $K=48\times24^3=663552$. 
This is quite a big integral!

Any standard lattice simulation begins with the creation of an ensemble of
gauge configurations $\{U^{[i]}\}$. It is created through a Markov
process~\cite{rothe}, i.e. each configuration $U^{[i]}$ is generated from
the preceding one,  $U^{[i-1]}$, using a Monte Carlo algorithm satisfying
the condition
\begin{equation}
P(U^{[i-1]} \rightarrow U^{[i]}) P[U^{[i-1]}]=
P(U^{[i]} \rightarrow U^{[i-1]}) P[U^{[i]}]
\end{equation}
where $P(U \rightarrow U')$ is the probability of generating the
configurations $U'$ from the configuration $U$. An example of such an 
algorithm is the {\it Metropolis} Algorithm.
A more sofisticate one is the {\it heatbath} algorithm.

Note that $P[U]$ depends
on  $\beta=6/g^2(a)$ which is the parameter that fixes the lattice scale. 
The initial configuration $U^{[0]}$ is usually chosen to be ``cold'',
i.e. when all its links are the identity, or ``hot'', when each link
is a random $SU(3)$ matrix~\cite{metropolis, cm}. 

The computation of any correlation function, as defined in eq.~(\ref{green3}), 
can be approximated, in analogy with eq.~(\ref{iii}), as an average over 
the ensemble of gauge configurations
\begin{equation} 
\langle 0 |{\cal O}(...) | 0\rangle^{\text{latt}} \simeq
\frac 1{N}\sum_{i} {\cal O}(...[U^{[i]}])
\end{equation}
where $N$ is the number of generated configurations.

\begin{table}
\begin{center}
\begin{tabular*}{\hsize}{@{}@{\extracolsep{\fill}}|llll|} \hline
State & $I^G$ & $J^{PC}$ & Operator $J$ \\ \hline
scalar & $1^{-}$ & $0^{++}$ & $\bar q q^{\prime }$ \\ 
& $1^{-}$ & $0^{++}$ & $\bar q \gamma ^0q^{\prime }$ \\ 
pseudoscalar & $1^{-}$ & $0^{-+}$ & $\bar q \gamma ^5q^{\prime }$ \\ 
& $1^{-}$ & $0^{-+}$ & $\bar q \gamma ^0\gamma ^5q^{\prime }$ \\ 
vector & $1^{+}$ & $1^{--}$ & $\bar q \gamma ^\mu q^{\prime }$ \\ 
& $1^{+}$ & $1^{--}$ & $\bar q \gamma ^\mu\gamma ^0q^{\prime }$ \\ 
axial & $1^{-}$ & $1^{++}$ & $\bar q \gamma ^\mu\gamma ^5q^{\prime }$ \\ 
tensor & $1^{+}$ & $1^{+-}$ & $\bar q \gamma ^\mu\gamma ^jq^{\prime }$ \\ 
octet & $\frac 12$ & $\frac 12^{-}$ & $(q^{Ti}\gamma ^2\gamma
^0q^{\prime j})(\gamma ^5q^{\prime \prime k }) \varepsilon _{ijk}$ \\ 
& $\frac 12$ & $\frac 12^{-}$ & $(q^{Ti}\gamma ^2\gamma ^0\gamma
^5q^{\prime j})(q^{\prime \prime k}) \varepsilon _{ijk}$ \\ 
decuplet & $\frac 32$ & $\frac 32^{+}$ & $(q^{Ti}\gamma ^2\gamma ^0\gamma
^iq^{\prime j})(q^{\prime \prime k}) \varepsilon _{ijk}$ \\ \hline
\end{tabular*}
\end{center}
\caption{Example of currents used on lattice and their relative
quantum numbers. $q,q^{\prime }$ and $q^{\prime \prime }$ are
different flavours. The superscripts $i$, $j$ and $k$ are color
labels. \label{lattice_currents}} 
\end{table}

\subsection{Correlation functions and fermions}

The typical quantities that are measured on the lattice are the
two and three-point correlation functions between currents and their
(discrete) Fourier transforms at zero momentum, eqs.(\ref{c2}-\ref{c3})~\cite{chris},
\index{correlation function!2-point}
\index{correlation function!3-point}
\begin{eqnarray}
C_2(t_x) &=& \sum_{\mathbf x} \langle 0 | J^0(x) J^{0\dagger}(0) 
| 0 \rangle \label{2cp} \\
C_{3 \cal O}(t_x,t_y) &=& \sum_{{\mathbf x}, {\mathbf y}}
\langle 0 | J^0(x) {\cal O}(0) J^{0\dagger}(-y)| 0 \rangle
\label{3cp}
\end{eqnarray}

Since the lattice metric is Euclidean, the asymptotic behaviour of the
spatial Fourier transform of the two point correlation function, in
the limit $t_x \rightarrow \infty$, is given by
\begin{equation}
	C_2(t_x) \stackunder{t_x \rightarrow \infty}{\simeq} Z_J^2 e^{-m_J t_x}
\label{fit}
\end{equation}
where $m_J$ is the mass of the lightest state $| 1_J \rangle$ 
created by the current $J^\dagger$ and
\begin{equation}
	Z_J=\frac{\left| \langle 0 \right| J^0(0) \left| 1_J \rangle  \right|}{\sqrt{2m_J}}
\end{equation}
From the measurement of $C_2(t_x)$ and its fit to
(\ref{fit}), it is possible to extract masses of particles, $m_J$.
In the same fashion from the asymptotic behaviour of the ratio between
the three and two-point correlation functions it is possible to
extract matrix elements~\cite{chris}
\begin{equation}
\frac{C_{3 \cal O}(t_x,t_y)}{C_2(t_x) C_2(t_x)} \stackunder{t_x,t_y \rightarrow \infty}{\simeq} 
\frac1{Z_J^2} 
\frac{\langle 1_J | {\cal O} | 1_J \rangle }{2m_J}
\end{equation}

The most general current $J^\mu(x)$ is expressed in terms of fundamental 
fermionic fields $q_\alpha ^i (x)$ (the quark fields). 
A list of some interesting currents $J$ is reported in
table~\ref{lattice_currents}. 
In expressions like eq.(\ref{2cp}) and (\ref{3cp}) these fields 
are Wick contracted
\begin{equation}
	S_{ab} ^{ij}(x,y) \stackrel{def}{=}
	\langle 0 |\{ q_a ^i(x), \bar q _b ^j(y) \} |0 \rangle
\label{light}
\end{equation}
Despite the fact that fermions are neglected when gauge configurations
are created, they are re-introduced at a later stage as particles
propagating in the gluonic background field.
Therefore the two and three point correlation functions can 
be written as appropriate traces of propagators, $S_{ab}
^{ij}(x,y,[U])$, in the backgroud gluonic field $U$.
For example the propagator of a heavy-light pseudoscalar meson (associated to the
current $J^\mu=\bar h^i \gamma^\mu \gamma^5 q^i$, where $h$, $q$ are the heavy 
and the light quark respectively), from $0$ to $x$ can be computed as
\begin{eqnarray}
\langle 0 | J^0(x)J^{0\dagger}(0) | 0\rangle &=& 
\langle 0 |\bar h_c^i(x) (\gamma^0 \gamma^5)^{ca} q_a^i(x) \bar q_b^j(0) (\gamma^0\gamma^5)^{bd}
h_d^j(0)| 0\rangle \label{piprop} \\
&=& \langle 0 |S_{q\,ab}^{ij}(x,0) (\gamma^0\gamma^5)^{ca} S_{h\,dc}^{ji}(0,x)
(\gamma^0\gamma^5)^{bd} | 0\rangle \nonumber \\
&=& \frac{1}{N} \sum_{\{U\}} S_{q\,ab}^{ij}(x,0,[U]) (\gamma^0\gamma^5)^{ca}
S_{h\,dc}^{ji}(0,x,[U]) (\gamma^0\gamma^5)^{bd} \nonumber
\end{eqnarray}
($i$,$j$ are color indices and $a$, $b$, $c$, $d$ are spin indices).
Then by making use of the H discrete symmetry (see Appendix A) and properties 
of the $\gamma$ matrices one obtains
\begin{equation}
\langle 0 | J^0(x)J^{0\dagger}(0) | 0\rangle=
 \frac{1}{N} \sum_{\{U\}} \text{tr} \left\{S_{q\,ab}(x,0,[U])
S_{h\,ab}^\dagger(x,0,[U])\right\} \label{piprop2}
\end{equation}
(the trace is only on the color indices $i$ and $j$).

On each gauge configuration $U$, the fermion propagator $S$ is
computed by inverting numericaly the fermionic matrix \index{fermion propagator}
\begin{equation}
S(x,y,[U])=(Q[U])^{-1}_{xy}
\label{inversion}
\end{equation}
This is the most expensive part of any lattice calculations.
In the computation of the propagator 
$\kappa$ and $c_{SW}$ are input parameters. 
\index{$\kappa_{crit}$} \index{quark!mass (lattice)} 
The former is in one to one correspondence with the fermion mass 
(the pole-mass)
\begin{equation}
m=\frac1a \ln \left( 1 + \frac12\left(\frac1{\kappa}-\frac1{\kappa_{crit}}\right) \right)
\end{equation}
and $\kappa_{crit}$ is a parameter depending on $\beta$. The chiral
limit corresponds to the limit $\kappa \rightarrow \kappa_{crit}$, when
the quark becames massless. In practice any
inversion algorithm for eq.(\ref{inversion}) converges slower and
slower as the chiral limit is approached and this can never be reached. 

There are two standard ways of computing the fermion propagator: exact
and stochastic. The former is very time expensive therefore it 
is usual normal practice to compute propagators ending in one single 
point of the lattice. 
The latter is less precise but allows one to compute propagators
from each point to each point of the lattice in a feasible time.

\subsection{Lattice errors}

Numerical simulations of Lattice QCD are characterized by 
a number of statistical and systematical errors which will have 
to be taken into account when quoting lattice results. 
What follows is a list of the most common errors one has to consider, 
possibly reduce and, hopefully, quantify:

\begin{itemize}
\item  {\bf Discretization errors} $a$. Physical results are extracted
from lattice in the limit of $a\rightarrow 0$. This limit cannot be reached
in real lattice simulation and in practice one performs simulations with a
finite (as small as possible) lattice spacing. The discrepancy between the
computed correlation functions and their continuum limit is usually of the order
of $a$ (or $a^2$ for improved lattice actions). In typical simulations $%
1/a=1\div 3$GeV.

\item  {\bf Statistical errors}. All Monte Carlo simulations are based on
statistical sampling therefore they introduce a statistical error that
is expected to decrease with $1/\sqrt{N}$ where $N$ is the number of
independent measurements (in case of one measurement for gauge configuration, 
$N$ is the number of uncorrelated gauge configurations). 
Since the gauge configurations
are created using a Markov chain based on small changes to link variables,
one may be worried about correlations among the different configurations.
Because of modern day computational power it is possible to generate
reasonable statistical samples and this is no more a major problem. 

\item  {\bf Finite volume}. Because of the finite volume, periodic or
anti-periodic boundary conditions are imposed for the field. 
Therefore every observable which is computed on the lattice suffers
from an unphysical contribution of mirror states. In any case,
these finite-volume contribution to the correlation functions falls off
exponentially with the lattice length and they are usually negligible for
a lattice size $L$ bigger than $5/m_\pi$.
On the other side the effects of mirror states are crucial in
preventing a direct determination of scattering 
phases from lattice~\cite{maiani}. 

\item  {\bf Quenching}. This approximation is the hardest to
justify. Its only reason is the present limitation in computational power. As
a consolation one can argue that present exploratory unquenched simulations
suggest that the effects of the quark loops in the mass spectrum are
small, but the error introduced by quenching in the determination of
$a^{-1}$ can be as big as 10\%-20\%.

\item  {\bf Chiral extrapolation}. It has been shown that because of the
finite volume effects, an infrared cut-off is naturally associated with the
lattice, therefore the $u$ and $d$ quarks are too light to be simulated, even by modern
day computers. Therefore one usually performs lattice simulations for values
of $m_u=m_d$ much bigger of the physical values (typically bigger than 
$50$ MeV), then performs an extrapolation of the results to the limit $%
m_u=m_d=0$. This extrapolation is called the \index{chiral
extrapolation} {\it chiral extrapolation}. It
corresponds to the limit $\kappa \rightarrow \kappa_{crit}$ For the
masses of light particles (the pseudo-Goldstone boson) this extrapolation is
guided by predictions of the Chiral Lagrangian such as the Gell-Mann-Okubo
formula.

\item  {\bf Heavy quarks}. The $c$ and $b$ quarks are very heavy therefore not all of their modes can propagate on a typical lattice. To solve the problem there are three common approaches.
One possibility is to simulate these quarks with a mass smaller than the
physical one and then to perform an extrapolation to the physical mass (guided by the Heavy Quark Effective Theory). 
The second possibility is to implement the HQET on lattice. This implies that
one considers the heavy quark as static (non-relativistic) 
and, in principle, 
systematically computes corrections to this approximation in 
the $1/m_h$ expansion. The third approach is also based on the HQET and 
is explained in ref.~\cite{fermilab}.

\item  {\bf Matching between lattice and continuum scheme}. 
\index{matching} Experimental
data are analyzed using some continuum renormalization scheme, usually
dimensional regularization with the $\overline{\text{MS}}$
prescription. 
To confront Lattice QCD results with phenomenology it is therefore
necessary to match the matrix elements between the two different
schemes. In general
\begin{eqnarray}
\langle 0 | {\cal O}_i (...) | 0\rangle ^{\overline{\text{MS}}}&=&
Z_{ij}\langle 0 | {\cal O}_j (...) | 0\rangle ^{\text{latt}} \\
&=&\left(\delta_{ij}+O(\alpha_s)\right) \langle 0 | {\cal O}_j
(...) | 0 \rangle^{\text{latt}}
\end{eqnarray}
where $Z_{ij}$ 
are called matching coefficients and have a perturbative
origin. The matching coefficients can be computed in
perturbation theory and usually they are known only at 1-loop. 
Since $\alpha_s$ is big at the typical lattice energy scale, corrections of higher order
in $\alpha _s$ can contribute to an error in the matching of as much as
$10\%$. Moreover in the matching procedure it is common that matrix elements
of some continuum operator mix with the corresponding matrix elements of
new operators that appear on lattice, because $Z_{ij}$ is, in general,
non diagonal. The contribution of these operators 
can be big and must be taken into account.
\end{itemize}

\subsection{One full application, $f_B\sqrt{m_B}$}

We have seen in section 1 how one could extract $f_B\sqrt{m_B}$ by computing the 
Path Integral on the right-hand side of eq.~(\ref{c2}), the 2-point correlation function,
and fitting the result with 
\begin{equation}
C_2(t) \stackunder{t \rightarrow \infty}{=} \frac{f_B^2 m_B}{2} e^{-m_B t}
\end{equation}

Program {\tt C2.C} in the Appendix is an example of a real QCD program 
(written in {\tt C++}) that computes $C_2(t)$. It is based on the mathematical
library described in ref.~\cite{mdp}.

This program is organized in the following way: It opens the libraries and declares the variable containing the parameters of the lattice simulation:
\begin{itemize}
\item {\tt beta}$\equiv\beta$, that fixes the size of the lattice spacing;
\item {\tt mq}$\equiv m_q$, the pole mass of the light quark;
\item {\tt mh}$\equiv m_h$, the pole mass of the heavy quark;
\item {\tt grid\_size}, the lattice size (for example $16 \times 6^3$);
\item {\tt lattice}, the object that contains the grid and its properties 
(including the functions to move on the lattice, a local random number generator 
and functions necessary for the parallelization).
\end{itemize}
It then defines the basic fields:
\begin{itemize}
\item {\tt U(x,mu)(ij)}$\equiv U_\mu^{ij}(x)$, the gauge field configuration;
\item {\tt Sq(x,a,b)(ij)}$\equiv \langle 0 |  \{q_a^i(x), \bar q_b^j(0) \}|0 \rangle$, 
the light quark propagator;
\item {\tt Sh(x,a,b)(ij)}$\equiv \langle 0 |  \{h_a^i(x), \bar h_b^j(0) \}|0 \rangle$,
the heavy quark propagator;
\end{itemize}
(they are fields of $3 \times 3$ color matrices and $a$ and $b$ are spin indices).

The program starts with a hot configuration ({\tt set\_hot(U)}) then computes 
and discards the first 100 gauge configurations of the Markov chain. The
Heatbath ({\tt heatbath(U)}) algorithm is used to generate the Markov chain. It is 
equivalent but more sofisticate than the Metropolis algorithm.

Then, each 10 gauge configurations of the chain, the program 
computes the light and heavy propagator and measures the 2-point 
correlation function $C_2(t)$, eqs.~(\ref{c2})(\ref{2cp}) and (\ref{piprop2}),
\begin{equation}
\sum_{\mathbf x} J^0(x)J^{0\dagger}(0) =
\sum_{\mathbf x} \text{tr} \left\{ S_{q\,ab}(x,0,[U])
S_{h\,ab}^\dagger(x,0,[U]) \right\} \label{piprop3}
\end{equation}
($t=x_0$ is kept fixed in the sum).

This is done in the following piece of code:
\begin{verbatim}
           forallsites(x) ... 
              t=x(0);
              ...	
              F(config,t)+=real(trace(Sq(x,a,b)*
                           hermitian(Sh(x,a,b))));
\end{verbatim}

Note that the average of the trace in question is always real by virtue 
of the theorem of Appendix A (applied to the case $\vec p = 0$). 
This is in agreement with naive expectations from eq.~(\ref{c2}).

Finally the program computes the average over the gauge configurations of $C_2(t)$ 
(with its the Bootstrap error) and prints out the results.

The output of the program is plotted in fig.~\ref{c2a} for some arbitrary input 
values of the parameters. In this example, in fact, we did not attempt to 
tune the parameters properly since our main concern was to have a fast running 
program for didactic purposes. 

\begin{figure}[here]
\begin{center} 
\begin{tabular}{cc}
\epsfxsize=6cm
\epsfysize=6cm
\epsfbox{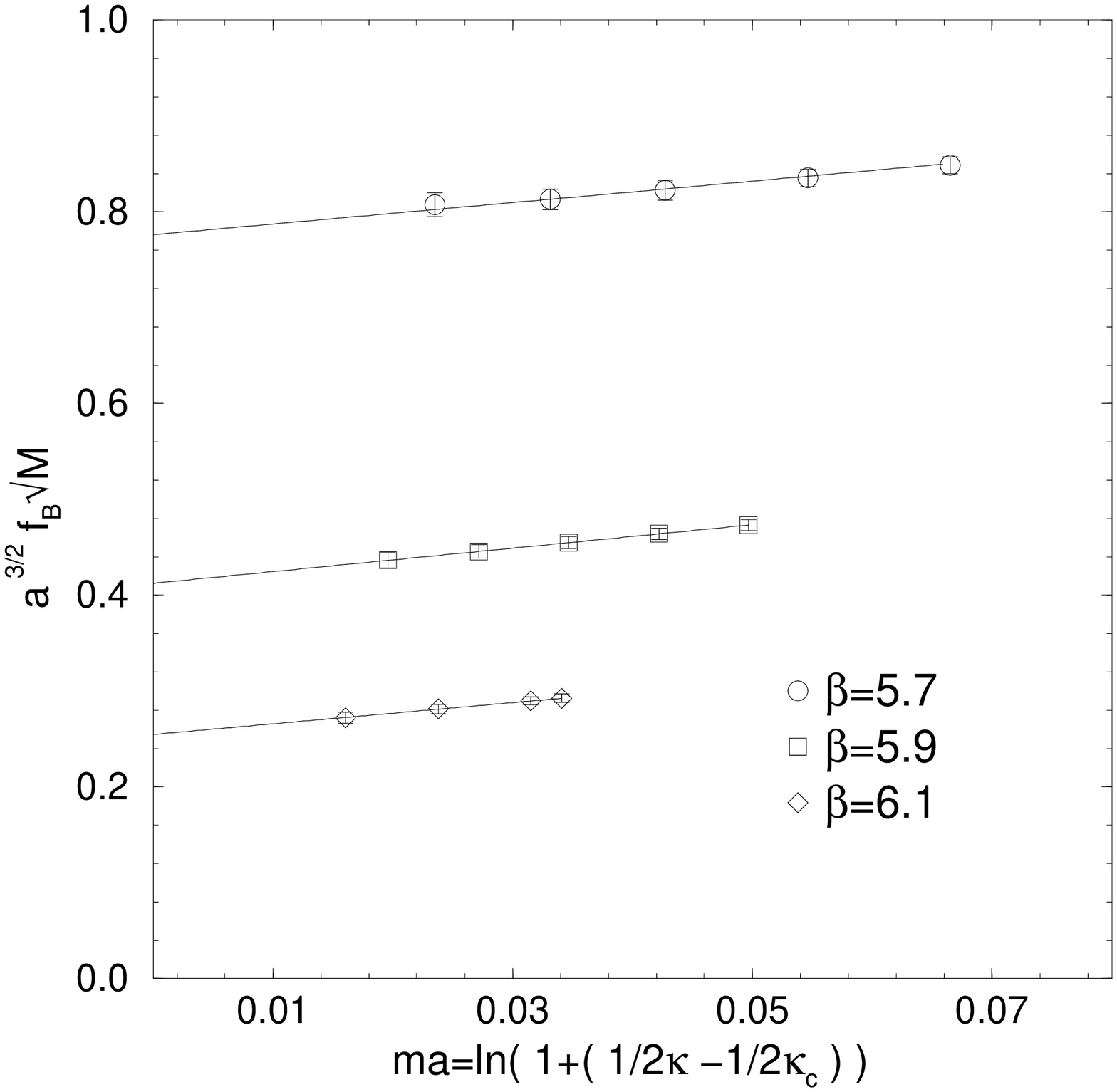} &
\epsfxsize=6cm
\epsfysize=6cm
\epsfbox{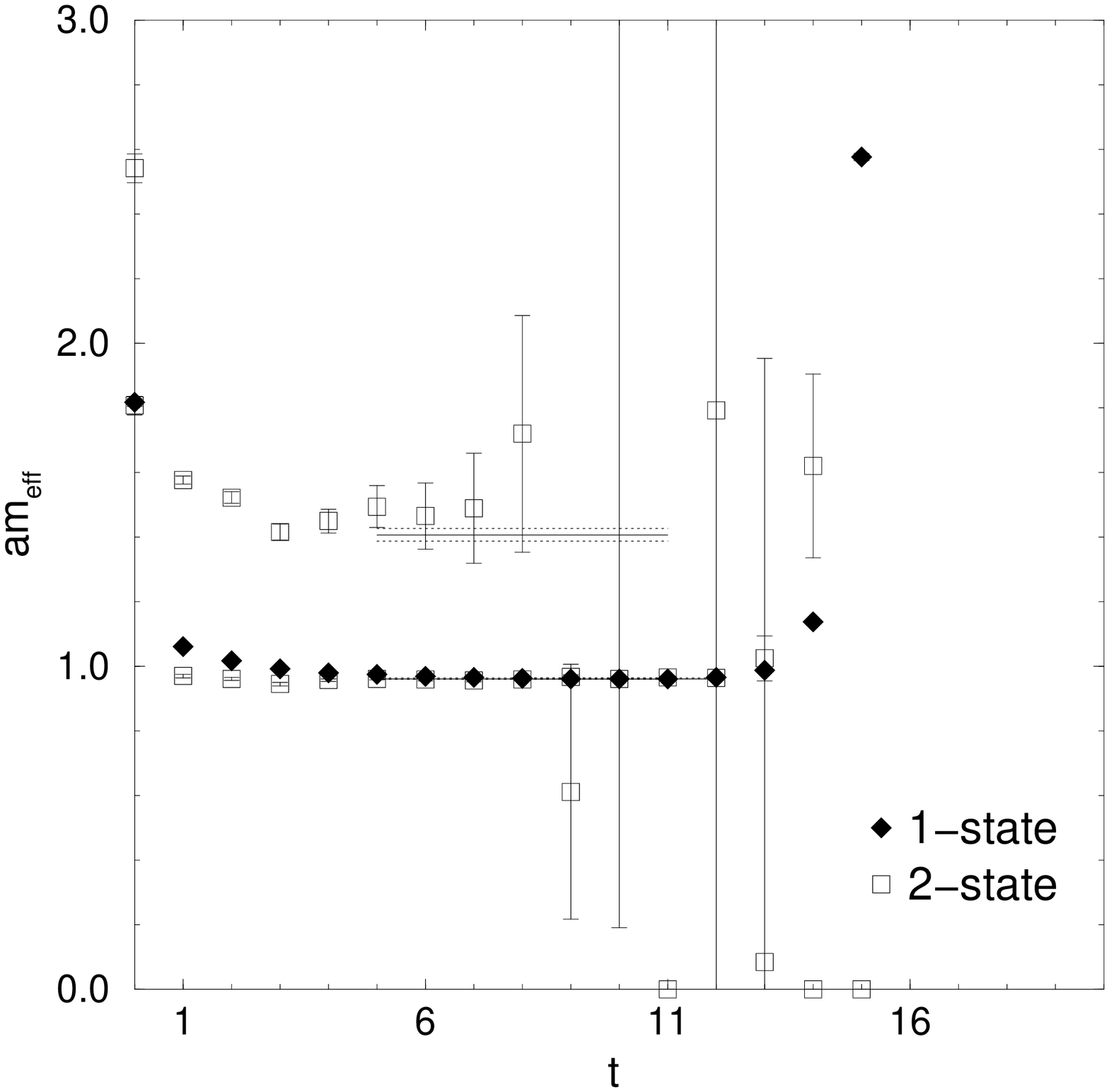} \\
\epsfxsize=6cm
\epsfysize=6cm
\epsfbox{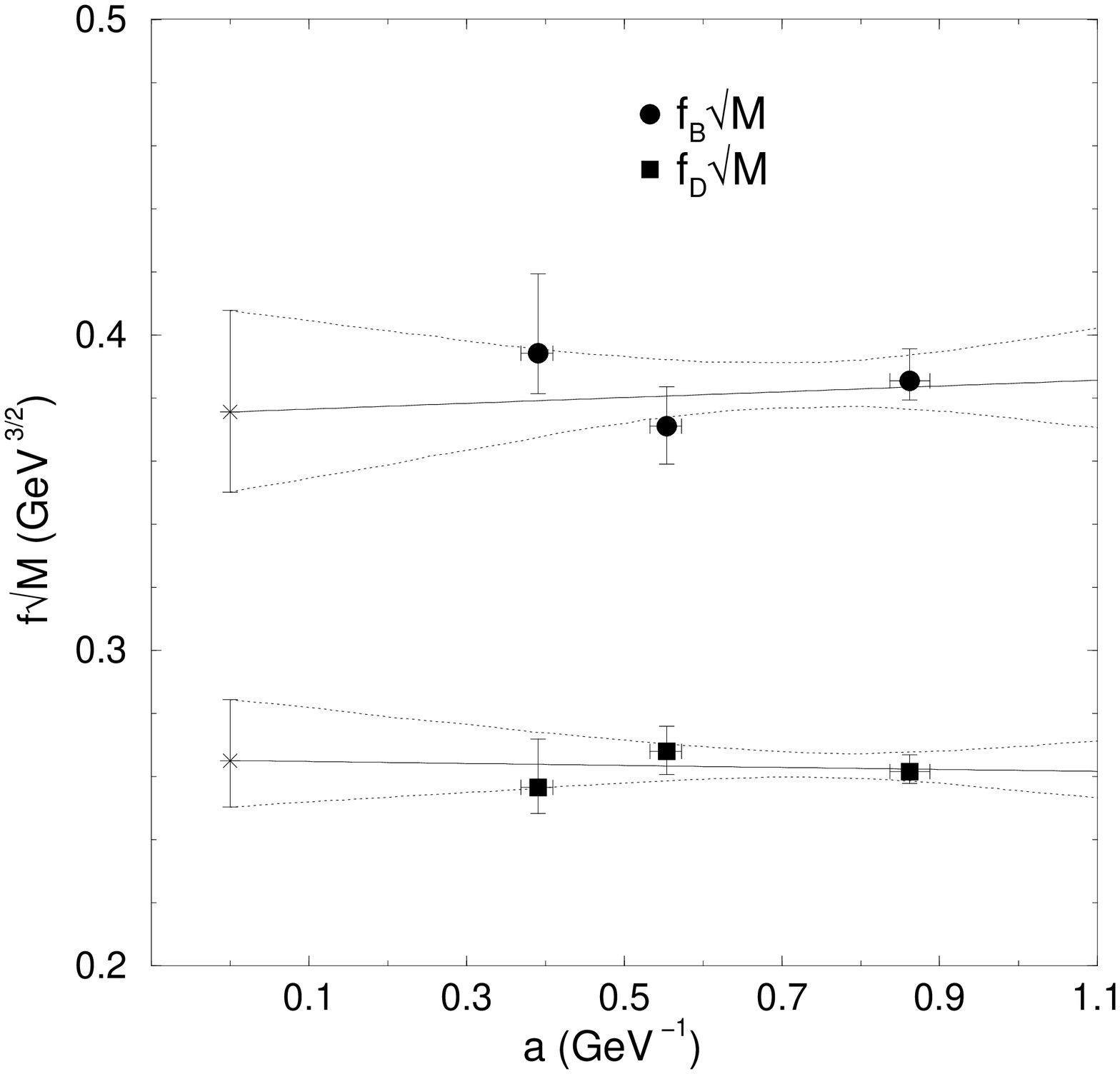} &
\epsfxsize=6cm
\epsfysize=6cm
\epsfbox{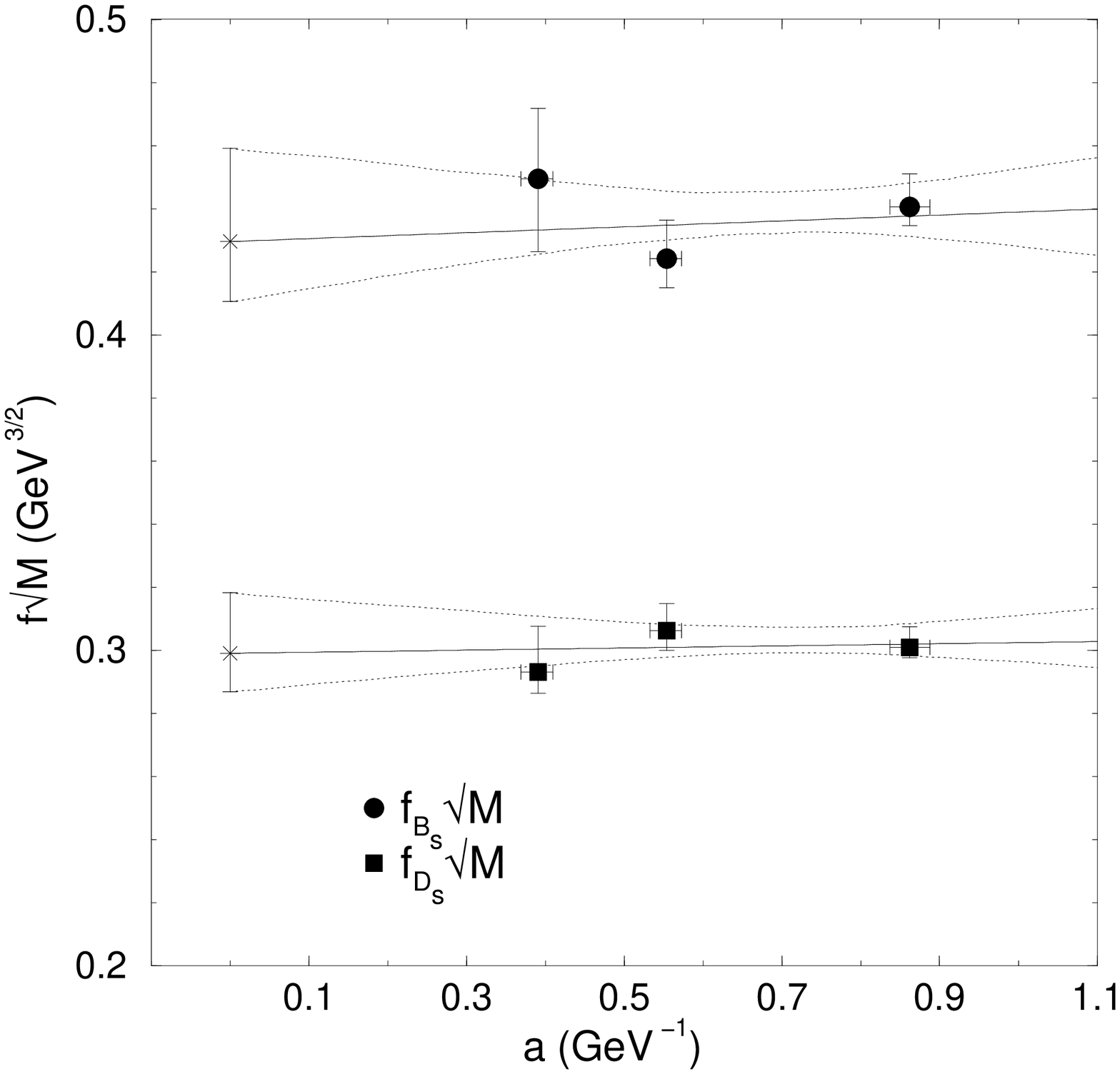}\end{tabular}
\end{center}
\caption{Determination of 
$f_B\sqrt{m_B}$, $f_D\sqrt{m_D}$,$f_{B_s}\sqrt{m_{B_s}}$ and $f_{D_s}\sqrt{m_{D_s}}$ (Fermilab~\cite{andreas}). The two point correlation functions, eq.~(\ref{c2}) are measured for different values of the lattice spacing $a$ and different masses for the light quarks $m=1/a \ln(1 + 1/(2\kappa) - 1/(2\kappa_c))$ and fitted to extract $f_B \sqrt{m_B}$. The results are extrapolated to the chiral limit $m \rightarrow 0$ (top-left). Then the lattice spacing is determined by measuring the 1P-1S mass splitting in the charmonium spectrum (top-right).
The chirally extrapolated values for $f_B \sqrt{m_B}$,  $f_D\sqrt{m_D}$,$f_{B_s}\sqrt{m_{B_s}}$ and $f_{D_s}\sqrt{m_{D_s}}$ are extrapolated to the continuum limit $a \rightarrow 0$ (bottom-left and bottom-right).\label{fbmb}}
\end{figure}

For a real state-of-the-art computation of $f_B \sqrt{m_B}$ we refer 
now to ref.~\cite{andreas}. The program used in that simulation is very 
similar to the one discussed here. The only operative differences are 
in the choice of the parameters and in the size of the lattice.
Fig.~\ref{fbmb} shows the results from ref.~\cite{andreas}

The values of $f_B\sqrt{m_B}$ are computed by fitting $C_2(t)$ for 
different sets of input parameters $m_q$ and $\beta$, while $m_h$ is 
tuned to the $B$ quark pole-mass. 
These results are first extrapolated to $m_q \rightarrow 0$ and
then extrapolated to $a \rightarrow 0$ ($\beta \rightarrow \infty$).

The results obtained in this way are renormalized in the lattice scheme at 
the lattice energy scale $a^{-1}$. To obtain the numbers renormalized
in the $\overline{\text{MS}}$ scheme at the $m_B$ energy scale they need 
to be corrected by a matching factor.

In the paper in exam the lattice spacing $a$ is measured (as function of $\beta$) 
by confronting the numerical result for the 1P-1S mass splitting in the charmonium 
spectrum with experimental results.

 \clearpage\newpage

\appendix

\section{Euclidean Space-Time in $d=4$ dimension}

\subsection{Wick rotation}

The Euclidean action is obtained from the Minkowskian one by performing a Wick rotation.
Under this rotation the basic vectors of the theory transform according with the following table
(E for Euclidean, M for Minkowski) 
\index{Wick rotation}
\begin{equation}
\begin{tabular}{|cc|cc|} \hline
E & M & E & M \\ \hline
$x^0$ & $ix^0$ & $x^i$ & $x^i$ \\
$\partial^0$ & $-i\partial _0$ & $\partial^i$ & $\partial _i$ \\
$A^4$ & $-iA_0$ & $A^i$ & $A_i$ \\ 
$F^0i$ & $-iF_{0i}$ & $F^{ij}$ & $F_{ij}$ \\
$\gamma^0 $ & $\gamma ^0$ & $\gamma^i$ & $-i\gamma ^i$ \\
$\gamma^5$ & $\gamma ^5$ & &  \\ \hline
\end{tabular}
\label{wickrot}
\end{equation}
and the integration measure transforms as follow 
\begin{equation}
\exp (-{\cal S}_E)=\exp (i{\cal S}_M)
\end{equation}
where 
\begin{equation}
{\cal S}_E=\int \text{d}^4 x_E {\cal L}_E[...]=-i\int 
\text{d}^4x_M{\cal L}_M[...]
\end{equation}
The choice $\text{d}^4x_E=i\text{d}^4x_M$ can be made, hence 
${\cal L}_E[...]=-{\cal L}_M[...]$

The Euclidean metric tensor is defined as
\begin{equation}
g_E^{\mu \nu }=-\delta ^{\mu \nu }=\text{diag}(-1,-1,-1,-1)
\end{equation}

\subsection{Spin matrices}

\begin{itemize}
\item  Dirac matrices (Dirac representation) \index{Dirac matrices!Euclidean}
\begin{equation}
\begin{tabular}{lll}
$\gamma^0=\left( 
\begin{array}{ll}
{\bf 1} & 0 \\ 
0 & -{\bf 1}
\end{array}
\right)$,\,\, & $\gamma^i=\left( 
\begin{array}{ll}
0 & -i\sigma _i \\ 
i\sigma _i & 0
\end{array}
\right)$,\,\, & $\gamma^5=\left( 
\begin{array}{ll}
0 & 1 \\ 
1 & 0
\end{array}
\right) $ 
\end{tabular}
\end{equation}

\item Dirac matrices (Chiral representation)
\begin{equation}
\begin{tabular}{lll} 
$\gamma^0=\left( 
\begin{array}{ll}
0 & {\bf 1} \\ 
{\bf 1} & 0
\end{array}
\right)$,\,\, & $\gamma^i=\left( 
\begin{array}{ll}
0 & -i\sigma _i \\ 
i\sigma _i & 0
\end{array}
\right)$,\,\, & $\gamma^5=\left( 
\begin{array}{ll}
-{\bf 1} & 0 \\ 
0 & {\bf 1}
\end{array}
\right) $%
\end{tabular}
\end{equation}
All the Euclidean Dirac matrices are hermitian. The following relations hold 
\begin{eqnarray}
g^{\mu \nu } &=&\frac 12\{\gamma^\mu ,\gamma^\nu \} = \delta^{\mu\nu} \\
\sigma^{\mu \nu } &=&\frac i2[\gamma^\mu ,\gamma^\nu ] \\
\gamma^5 &=&\gamma^0\gamma^1\gamma^2\gamma^3
\end{eqnarray}
and all the $\sigma ^{\mu \nu }$ are hermitian.

\item  Projectors 
\begin{equation}
L=\frac{1-\gamma^5}2\qquad R=\frac{1+\gamma^5}2
\end{equation}

\item  Traces 
\begin{eqnarray}
\text{tr}(\gamma^\mu \gamma^\nu ) &=& 4\delta ^{\mu \nu } \\
\text{tr}(\gamma^\mu \gamma^\nu \gamma^\rho ) &=& 0 \\
\text{tr}(\gamma^\mu \gamma^\nu \gamma^\rho 
\gamma^\sigma ) &=& 4(\delta^{\mu \nu }\delta^{\rho
\sigma }-\delta^{\mu \rho }\delta^{\nu \sigma }+\delta^{\mu
\sigma }\delta^{\rho \nu }) \\
\text{tr}(\gamma^5\gamma^\mu\gamma^\nu\gamma^\rho \gamma^\sigma ) 
&=&4\epsilon _E^{\mu \nu \rho\sigma }
\end{eqnarray}
where $\epsilon _E^{0123}=-1.$
\end{itemize}

\subsection{Lattice discrete symmetries}

The lattice formulation of QCD is invariant under the following
discrete symmetries of the quark propagator~\cite{bernard}
\index{CPT symmetries}\index{H symmetry}
\begin{itemize}
\item {\bf Parity, $P$}:
\begin{equation}
S_{\alpha \beta}^{ij}(x,y,[U])=
\gamma^0_{\alpha\alpha'} S_{\alpha' \beta'}^{ij}(x^P,y^P, [U^P])
\gamma^0_{\beta'\beta} 
\end{equation}

\item {\bf Charge conjugation, $C$}: 
\begin{equation}
S_{\alpha \beta}^{ij}(x,y,[U])=
(\gamma^0\gamma^2)_{\alpha\alpha'} S_{\alpha' \beta'}^{ji}(y,x,[U^C])
(\gamma^2\gamma^0)_{\beta'\beta}
\end{equation}

\item {\bf Time reversal, $T$}: 
\begin{equation}
S_{\alpha \beta}^{ij}(x,y,[U])=
(\gamma^0\gamma^5)_{\alpha\alpha'} S_{\alpha' \beta'}^{ij}(x^T,y^T, [U^T])
(\gamma^5\gamma^0)_{\beta'\beta}
\end{equation}

\item {\bf $H$ symmetry}: 
\begin{equation}
S_{\alpha \beta}^{ij}(x,y, [U])=
\gamma^5_{\beta\alpha'} S_{\alpha' \beta'}^{ji}(y,x,[U])
\gamma^5_{\beta'\alpha} 
\end{equation}
\end{itemize}

$U^P,U^C,U^T$ are the parity reversed, charge conjugate, time reversed
gauge configurations respectively.

\subsection{Theorems about correlation functions}

These discrete symmetries play a very important role because they put
contraints on the Euclidean correlation functions. In particular one can
consider correlation function of the form
\begin{equation}
G^{(n,m)}(\vec p_1,...,\vec p_n)=\int
\text{d}^4x_1e^{ip_1x_1}...\text{d}^4x_n e^{ip_n x_n}
\left<0\right|\text{tr}\{S\gamma^{\mu_1}S\gamma^{\mu_2}...S\gamma^{\mu_m}\}
\left|0\right> 
\end{equation}
where the trace is in spin and color, and $S$ are quark propagators
connecting an arbitrary couple of points in the ensemble $\{x_1,...,x_n\}$.
Imposing invariance under $P$ (parity), one obtains that
\begin{equation}
G^{(n,m)}(\vec p_1,...,\vec p_n)=(-1)^N G^{(n,m)}(-\vec p_1,...,-\vec p_n)
\label{using_p}
\end{equation}
where $N$ is the number of indices $\mu_1$,...,$\mu_m$ that differ
from 0. Eq.~(\ref{using_p}) is true also in the Minkowski space.

Imposing invariance under $PCH$ one obtains that
\begin{equation}
G^{(n,m)}(\vec p_1,...,\vec p_n) 
= (-1)^N \left[G^{(n,m)}(\vec p_1,...,\vec p_n)\right]^\ast
\label{using_pch}
\end{equation}
This tells whether any correlation function is real or imaginary.

Eq.~(\ref{using_pch}) is not true in Minkowski space. It is
replaced by an equivalent expression where $N$ counts the total
number of indices $\mu_1$,...,$\mu_m$ that are equal to 5.

As an example we consider
\begin{equation}
G^{(3,2)}(\vec p) = \int \text{d}^4 x e^{ipx} \left<0\right| \text{tr} \left\{
S(x,0)\gamma^1\gamma^5S(0,x)\gamma^2 \right\} \left|0\right>
\end{equation}
Since $N=3$, eq.~(\ref{using_p}) tells that it is odd under
parity and eq.~(\ref{using_pch}) tells that it is imaginary. Hence
for $p=0$ it must be zero (because it is odd under parity).

 \clearpage\newpage

\section{Example programs}

\subsection{Basic integration: {\tt program1.c}}
{\footnotesize
\begin{verbatim}
// program1.c
#include "random.c"

double f(double x) {
  return sin(Pi*x);
};

int main() {
  double Sum, I1, I2, I3, x0, x1, x2;
  double alpha=0;
  double beta=1;
  long N=20,i;

  // //////////////////////
  // Basic method
  // ////////////////////// 
 
  for(N=2; N<10000; N*=2) {
  Sum=0;
  for(i=0; i<N-1; i++) {
    x0=alpha+(beta-alpha)*i/N;
    Sum=Sum+f(x0);
  };
  I1=Sum*(beta-alpha)/N;

  // //////////////////////
  // Newton-Cotes method
  // //////////////////////  
  Sum=0;
  for(i=0; i<N-1; i++) {
    x0=alpha+(beta-alpha)*i/N;
    x1=x0+(beta-alpha)/N;
    Sum=Sum+(f(x0)+f(x1))/2.0;
  };
  I2=Sum*(beta-alpha)/N;

  // //////////////////////
  // Simpson's method
  // //////////////////////  
  Sum=0;
  for(i=0; i<N-2; i++) {
    x0=alpha+(beta-alpha)*i/N;
    x1=x0+(beta-alpha)/N;
    x2=x1+(beta-alpha)/N;
    Sum=Sum+(f(x0)+4.0*f(x1)+f(x2))/6.0;
  };
  I3=Sum*(beta-alpha)/N;
  
  printf("%i\t%f\t%f\t%f\n",N, I1, I2, I3);
  };
};
\end{verbatim}}

\subsection{Monte Carlo integration: {\tt program2.c}}
{\footnotesize
\begin{verbatim}
// program2.c
#include "random.c"
 
double f(double x) {
    return sin(Pi*x);
};
 
int main() {
    double I, Sum,x;
    long   N;
    Sum=0;
    for(N=1;; N++) {
      x=Random();
      Sum=Sum+f(x);
       I=Sum/N;
       printf("N = %i, I(N) = %f\n",N,I);
    };
};
\end{verbatim}}

\subsection{Monte Carlo integration in 3D: {\tt program3.c}}
{\footnotesize
\begin{verbatim}
// program3.c
#include "random.c"
 
double f(double *x) {
  return 3.0*x[0]*x[0]*x[1]+2.0*x[2]*x[2]*x[2];
};

int main() {
  double I, Sum, x[3];
  long N;
  Sum=0;
  for(N=1;; N++) {
    x[0]=Random();
    x[1]=Random();
    x[2]=Random();
    Sum=Sum+f(x);
    I=Sum/N;
    printf("N = %i, I(N) = %f\n",N,I);
  };
};
\end{verbatim}}

\subsection{Metropolis Monte Carlo integration: {\tt program4.c}}
{\footnotesize
\begin{verbatim}
// program4.c
#include "random.c"

double P(double *x) {
  return exp(-(x[0]*x[0]+x[1]*x[1]+x[2]*x[2]));
};
 
double f(double *x) {
  return 128.0*x[0]*x[0]*x[0]*x[1]*x[1]*x[2];
};

int main() {
  double I, Sum, x[3], y[3],Py,Px;
  long N,j,step, Nstep=1000;
  
  Sum=0;
  
  x[0]=Random();
  x[1]=Random();
  x[2]=Random();
  
  for(N=1;; N++) {
    for(step=0; step<Nstep; step++) {
      y[0]=Random();
      y[1]=Random();
      y[2]=Random();
      if(P(y)/P(x) > Random()) { 
        x[0]=y[0]; x[1]=y[1]; x[2]=y[2];
      };    
    };

    Sum=Sum+f(x);
    I=Sum/N;
    printf("N = %i, I(N) = %f\n",N,I);
  };
};
\end{verbatim}}

\subsection{Metropolis and Bootstrap: {\tt program5.c}}
{\footnotesize
\begin{verbatim}
// program5.c
#include "random.c"

long D=3;        // number of dimensions
long N=8192;     // number of Montecarlo samples
long M=100;      // number of Bootstrap samples
long Nstep=1000; // number of interations per config.

double P(double *x) {
  return exp(-(x[0]*x[0]+x[1]*x[1]+x[2]*x[2]));
};

double f(double *x) {
  return 128.0*x[0]*x[0]*x[0]*x[1]*x[1]*x[2];
};

void swap(double &a, double &b) {
  double c;
  c=a; a=b; b=c;
};

int main() {
  double I, Sum;
  double x[N][D]; 
  double y[D];
  double Ibar[M];
  long  i,j,d,k[N][M], step;
  
  // /////////////////////////////////////////
  // generate confifgurations using Metropolis
  // /////////////////////////////////////////
  
  for(d=0; d<D; d++) x[0][d]=Random();
  for(i=1; i<N; i++) {
    for(d=0; d<D; d++) x[i][d]=x[i-1][d];
    for(step=0; step<Nstep; step++) {
      for(d=0; d<D; d++) y[d]=Random();
      if(P(y)/P(x[i]) > Random())
        for(d=0; d<D; d++) x[i][d]=y[d];
    };
  };

  // //////////////////////////////////////////
  // compute the numerical integral I(N)
  // //////////////////////////////////////////
  
  Sum=0;
  for(i=0; i<N; i++)Sum=Sum+f(x[i]);
  I=Sum/N;
  printf("N = %i, I(N) = %f\t", N, I);
  
  // //////////////////////////////////////////
  // create Bootstrap configurations
  // //////////////////////////////////////////
  
  for(i=0; i<N; i++)
    for(j=0; j<M; j++)
      k[i][j]=(long) ((double) N*Random());
  
  // //////////////////////////////////////////
  // create Bootstrap integrals Ibar[j]
  // //////////////////////////////////////////
  
  for(j=0; j<M; j++) {
    Sum=0;
    for(i=0; i<N; i++) Sum=Sum+f(x[k[i][j]]);
    Ibar[j]=Sum/N;
  };
  
  // //////////////////////////////////////////
  // sort the Ibar[j]
  // //////////////////////////////////////////
  
  for(j=0; j<M-1; j++)
    for(i=j+1; i<M; i++)
      if(Ibar[j]>Ibar[i]) 
        swap(Ibar[i],Ibar[j]);
  
  // //////////////////////////////////////////
  // print the confidence interval
  // //////////////////////////////////////////
  
  printf(" and %f < I < %f\n", Ibar[16], Ibar[84]);
};
\end{verbatim}}

\subsection{Marsaglia random number generator: {\tt random.c}}
{\footnotesize
\begin{verbatim}
/* random.c */
// ////////////////////////////////////////////
// Remastered version of the generator used by 
// the UKQCD Collaboration.
// It is based on the Marsaglia generator
// ////////////////////////////////////////////
#include <stdio.h>
#include <math.h>
#include <complex.h>
#include <sys/time.h>
#include <unistd.h>

#define and            &&
#define or             ||
#define Pi             3.14159265359
#define PRECISION      1e-16
#define TRUE  1
#define FALSE 0

float Random(long ijkl=0) {
  static float u[98];
  static float c;
  static float cd;
  static float cm;
  static int ui;
  static int uj;
  static int first=0;
  int i, j, k, l, ij, kl;

  if((first==0) || (ijkl!=0)) {
    printf("initializing the Random generator\n");
    if( (ijkl < 0) || (ijkl > 900000000) )
      exit(1);
    ij = ijkl/30082;
    kl = ijkl - (30082 * ij);
    i = ((ij/177) % 177) + 2;
    j = (ij % 177) + 2;
    k = ((kl/169) % 178) + 1;
    l = kl % 169;
    if( (i <= 0) || (i > 178) )
      exit(1);
    if( (j <= 0) || (j > 178) )
      exit(1);
    if( (k <= 0) || (k > 178) )
      exit(1);
    if( (l < 0) || (l > 168) )
      exit(1);
    if (i == 1 && j == 1 && k == 1)
      exit(1);
    int ii, jj, m;
    float s, t;
    for (ii = 1; ii <= 97; ii++) {
      s = 0.0;
      t = 0.5;
      for (jj = 1; jj <= 24; jj++) {
        m = ((i*j % 179) * k) % 179;
        i = j;
        j = k;
        k = m;
        l = (53*l+1) % 169;
        if (l*m % 64 >= 32) s += t;
        t *= 0.5;
      };
      u[ii] = s;
    };
    c  = 362436.0   / 16777216.0;
    cd = 7654321.0  / 16777216.0;
    cm = 16777213.0 / 16777216.0;
    ui = 97;
    uj = 33;


    first=1;
  };

  float luni;			/* local variable for Float */      
  luni = u[ui] - u[uj];
  if (luni < 0.0)
    luni += 1.0;
  u[ui] = luni;
  if (--ui == 0)
    ui = 97;
  if (--uj == 0)
    uj = 97;
  if ((c -= cd) < 0.0)
    c += cm;
  if ((luni -= c) < 0.0)
    luni += 1.0;
  return ((float) luni);
};
\end{verbatim}}

\subsection{$f_B$ and $m_B$. Program: {\tt C2.C}}

{\footnotesize \begin{verbatim}
// ////////////////////////////////////////////////////////
// Program C2.C written by Massimo Di Pierro @ July 2000
// ////////////////////////////////////////////////////////
// WORKING EXAMPLE of a Lattice QCD program to compute the 
// Euclidean propagator of an Heavy-Light Meson, C2(t)
// To extract m_B and f_B fit output with
//
// C2(t) = 1/2 f_B^2 m_B exp(-m_B t) + ... 
//
// and extrapolate to 
// mq -> 0    (GeV)
// mh -> mb   (GeV) (the b quark pole mass)
// a  -> 0    (GeV^(-1))
// ////////////////////////////////////////////////////////

// #define PARALLEL
// ////////////////////////////////////////////////////////
// open the libraries: Matrix Distributed Processing 1.0
// (for a description read: hep-lat/0004007)
// ////////////////////////////////////////////////////////

#include "MDP_Lib2.h"
#include "MDP_MPI.h"

// ////////////////////////////////////////////////////////
// open the FermiQCD libraries
// ////////////////////////////////////////////////////////
#include "MDP_Gauge.h"
#include "MDP_Fermi.h"
#define GeV 1

// ////////////////////////////////////////////////////////
// main program
// ////////////////////////////////////////////////////////

int main(int argc, char **argv) {
  mpi.open_wormholes(argc, argv); // open communications

  // //////////////////////////////////////////////////////
  // declare parameters of the simulation
  // //////////////////////////////////////////////////////
  int          Nt=16, Nx=6, Ny=6, Nz=6; // lattice size
  int          Nc=3;          // set colors, SU(Nc)
  float        beta=5.7;      // set lattice spacing
  float        mq=0.2*GeV;    // set light quark pole-mass
  float        mh=0.7*GeV;    // set heavy quark pole-mass
  
  // //////////////////////////////////////////////////////
  // additional parameters (they only depend on beta!)
  // //////////////////////////////////////////////////////
  float        a      =0.91/GeV; // lattice spacing  
  float        cSW    =1.57;   
  float        kappa_c=0.14315;


  // //////////////////////////////////////////////////////
  // define gamma matrices in auclidean space
  // //////////////////////////////////////////////////////
  define_base_matrices("FERMILAB");

  // //////////////////////////////////////////////////////
  // define the grid size on which the lattice is defined
  // //////////////////////////////////////////////////////
  int                grid_size[]={Nt,Nx,Ny,Nz};

  // //////////////////////////////////////////////////////
  // associate the lattice to the grid
  // //////////////////////////////////////////////////////
  generic_lattice    lattice(4, grid_size);

  // //////////////////////////////////////////////////////
  // define a gauge field 
  // U(x,mu) = exp(i g A(x,mu) )
  // to the sites of the lattice
  // //////////////////////////////////////////////////////
  gauge_field        U(lattice,Nc);

  // //////////////////////////////////////////////////////
  // define a light and a heavy propagator
  // S = <0| q(x) \bar q(0) |0>
  // to the sites of the lattice
  // //////////////////////////////////////////////////////
  fermi_propagator   Sq(lattice,Nc);
  fermi_propagator   Sh(lattice,Nc);

  // //////////////////////////////////////////////////////
  // define a variable site 
  // to move on the lattice
  // //////////////////////////////////////////////////////
  site               x(lattice);

  // //////////////////////////////////////////////////////
  // define the number of gauge configurations to be used
  // //////////////////////////////////////////////////////
  int                Nconfig=100;

  // //////////////////////////////////////////////////////
  // define some more auxiliary variables
  // //////////////////////////////////////////////////////
  int                i,j,t, config;

  // //////////////////////////////////////////////////////
  // define an object for bootstrap error
  // //////////////////////////////////////////////////////
  JackBoot           C2(Nconfig,Nt);

  // //////////////////////////////////////////////////////
  // creating initial gauge configuratino
  // //////////////////////////////////////////////////////
  set_hot(U);

  // //////////////////////////////////////////////////////
  // create and skip 100 gauge configuration
  // //////////////////////////////////////////////////////
  U.param.beta=beta;  
  heatbath(U,100);

  // //////////////////////////////////////////////////////
  // setting the parameters for the light and heavy quarks
  // //////////////////////////////////////////////////////
  Sq.param.kappa=1.0/((exp(mq*a)-1.0)*2.0+1.0/kappa_c);
  Sh.param.kappa=1.0/((exp(mh*a)-1.0)*2.0+1.0/kappa_c);
  Sq.param.cSW=Sh.param.cSW=cSW;
  Sq.precision=Sh.precision=1e-7; 

  // //////////////////////////////////////////////////////
  // loop over the gauge configurations
  // //////////////////////////////////////////////////////
  for(config=0; config<Nconfig; config++) {

    heatbath(U,10);      // each 10 gauge configurations

    compute_em_field(U); // compute electromagnetic-field

    generate(Sq,U);      // compute light propagator 

    generate(Sh,U);      // compute heavy propagator
    
    // ////////////////////////////////////////////////////
    // compute the pion propagator by
    // wick contracting 
    //
    // C_2(t_x) = 
    //   \sum_{x} \bar h(x) Gamma5 q(x) \bar q(0) Gamma5 h(0)
    //
    // as function of t = t_x
    // ////////////////////////////////////////////////////
    for(t=0; t<Nt; t++) C2(0,t)=0;
    forallsites(x) {
      t=x(0); 
      for(i=0; i<4; i++)
        for(j=0; j<4; j++)
          C2(config,t)+=
            real(trace(Sq(x,i,j)*hermitian(Sh(x,i,j))));
    };
    // ////////////////////////////////////////////////////
    // for each t print out C_2(t) with the Bootstrap error 
    // ////////////////////////////////////////////////////
    printf("\nRESULT FOR C2(t) (@ gauge = %i)\n", config);
    printf("==================================\n");
    printf("t\tC2\t\t(error)\n");
    printf("==================================\n");
    for(t=0; t<Nt; t++) {
      C2.plain(t);
      printf("%i\t%e\t%e\n", t, C2.mean(), C2.b_err());    
    };    
    printf("==================================\n\n");
    fflush(stdout); 
  };

  mpi.close_wormholes(); // close communications
  return 0;
};
\end{verbatim}}
 \clearpage\newpage

\section{Status of Lattice QCD}

In this appendix we will briefly report a very small subset of 
recent Lattice QCD results which have 
direct phenomenological interest and which provide an example of the state-of-the-art 
in Lattice QCD simulations%
\footnote{The papers quoted below are chosen as examples and we do not aim to provide 
a complete list of references on any of the subjects.}.

A huge amount of work has also been dedicated 
by the lattice community to the study of some theoretical and numerical aspects of field 
theories and the properties of their lattice regulated versions. They include the study 
of different possible discretizations for the Dirac operator ($/\!\!\!\! D$), 
study of low energy eigenvalues of these operators, the confining properties of 
different Yang-Mills theories and chiral symmetry breaking. These studies 
have given some important insights in the understanding of QCD and constitute the 
foundations on which any simulation of phenomenological interest relies on.

\begin{figure}[t]
\centerline{\epsfxsize=10cm \epsfbox{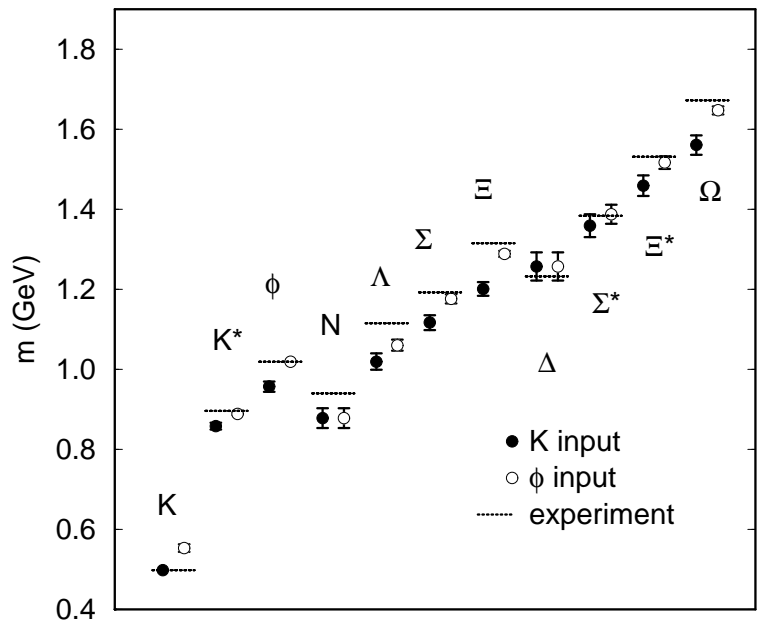}}
\caption{Light hadrons spectrum as computed by the CP-PACS collaboration~\cite{cppacs}}
\label{cppacs_spectrum}
\end{figure}

\begin{figure}[t]
\centerline{\epsfxsize=9cm \epsfbox{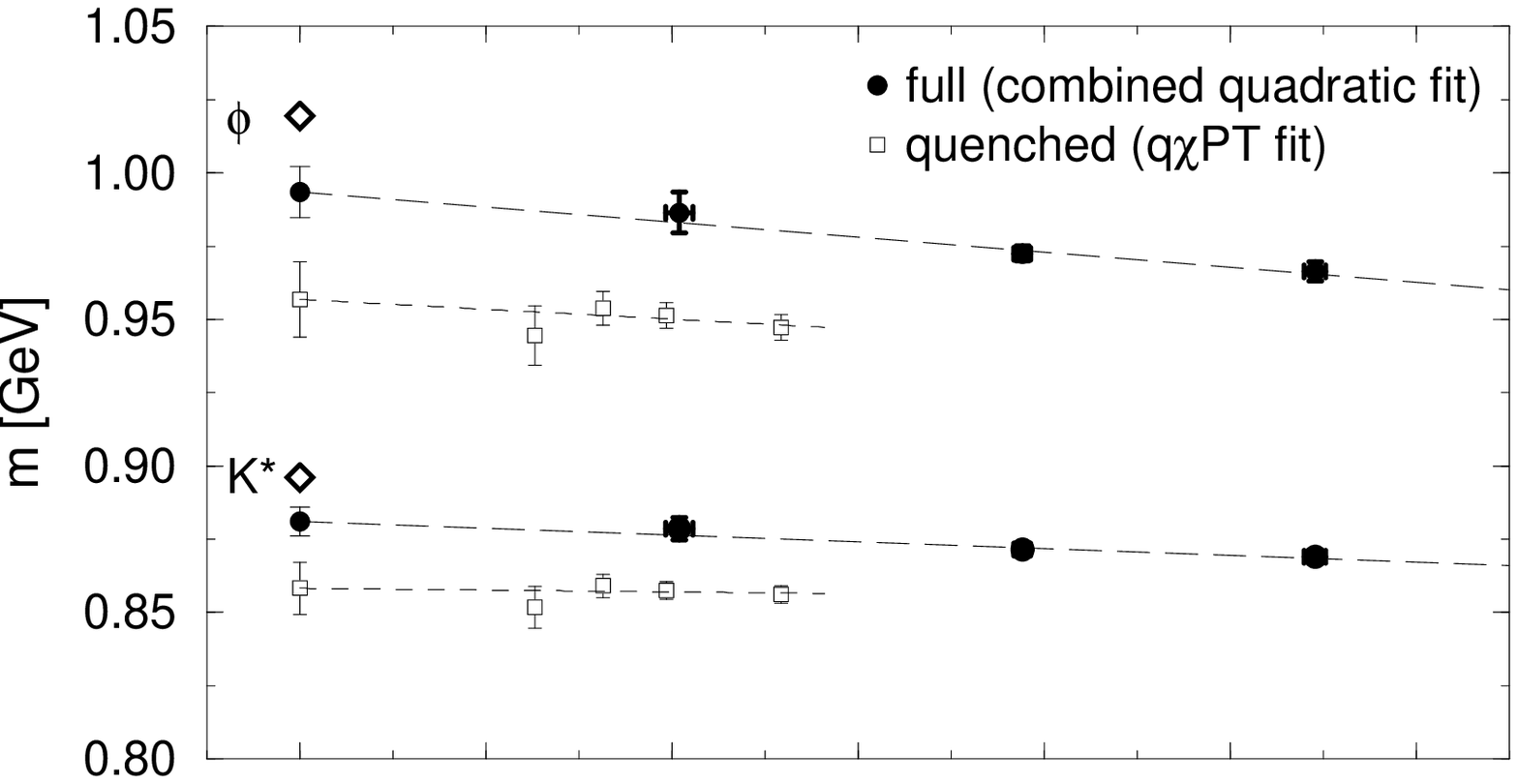}}\vskip -7mm
\centerline{\epsfxsize=9cm \epsfbox{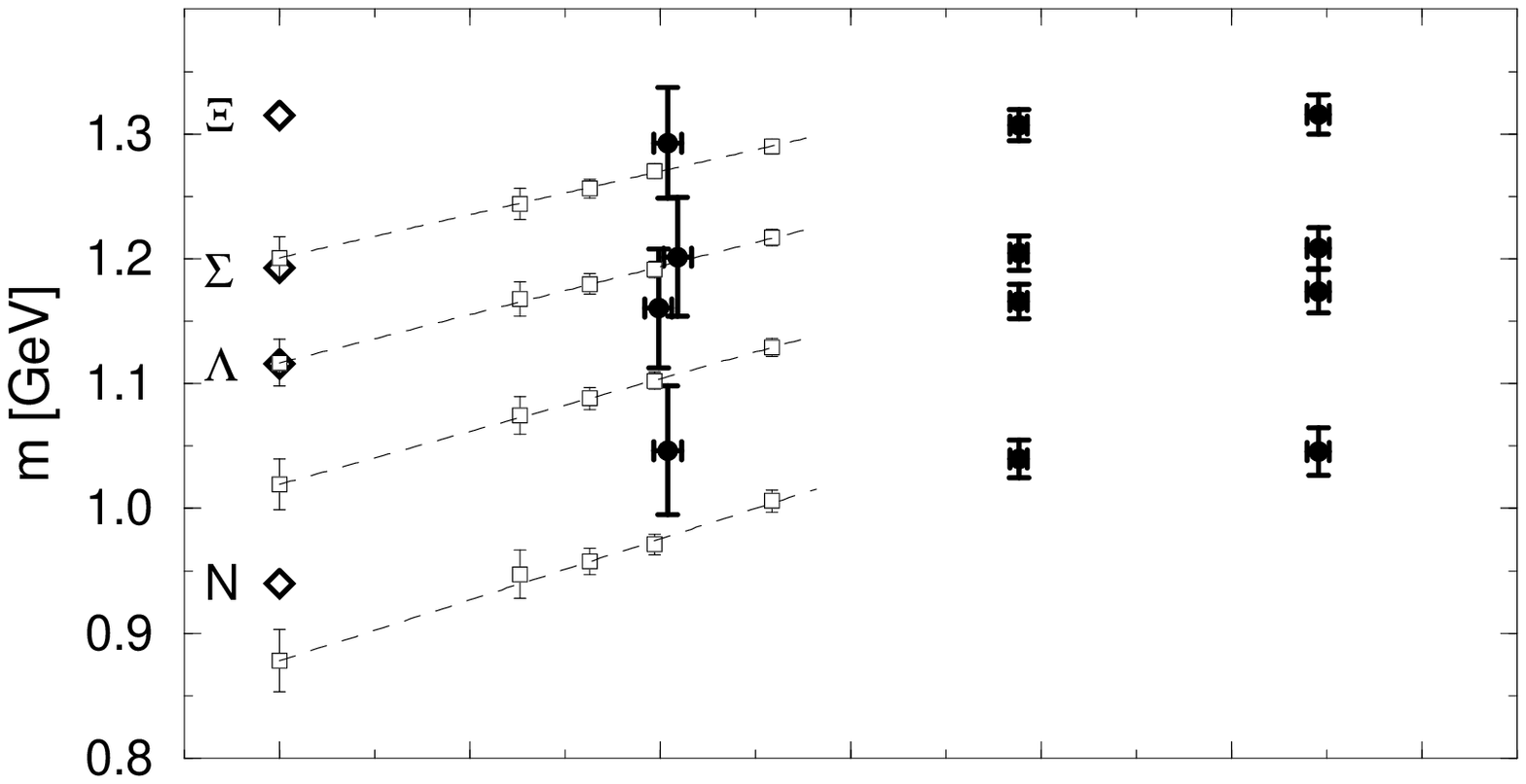}}\vskip -7mm
\centerline{\epsfxsize=9cm \epsfbox{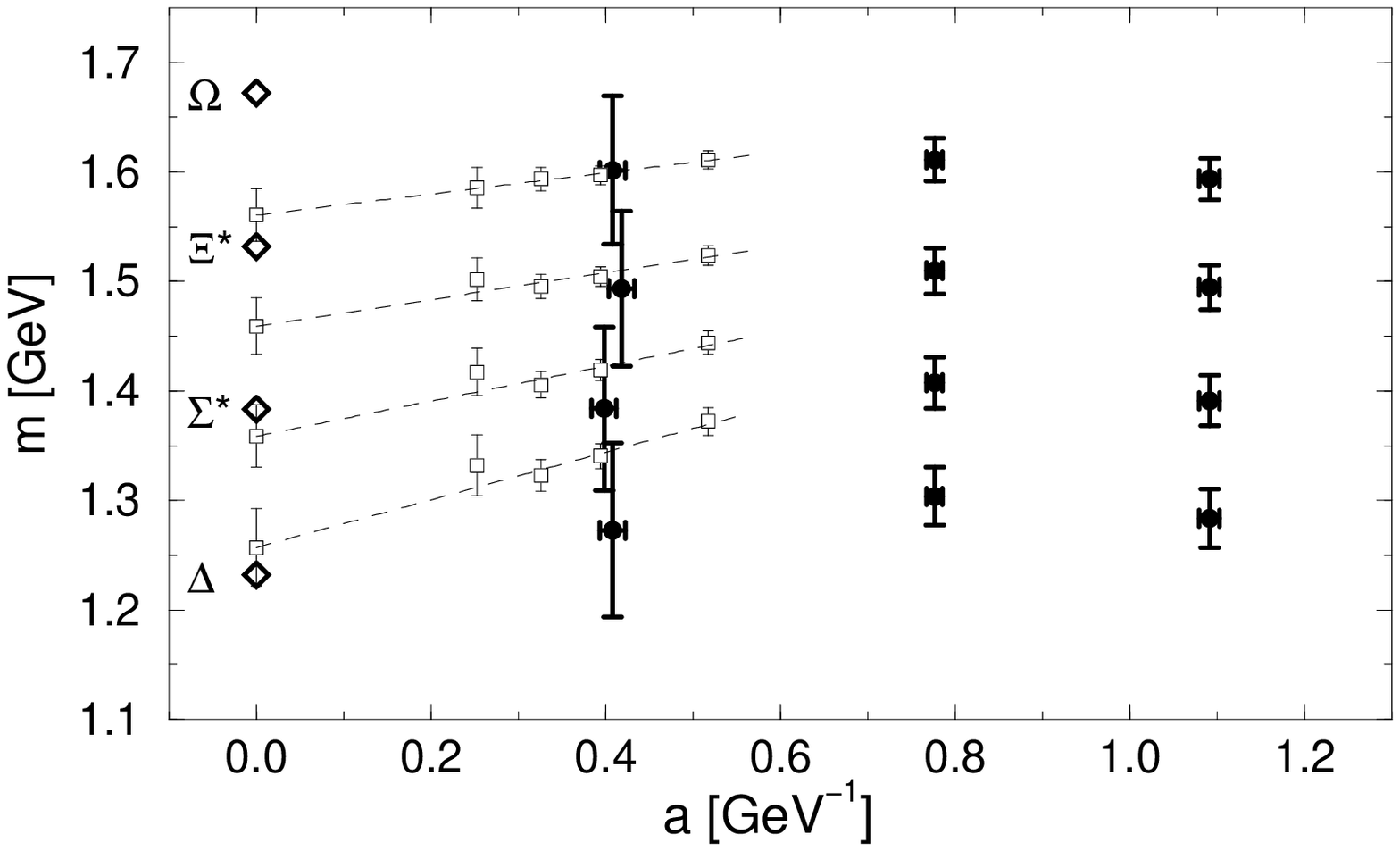}} 
\caption{Hadron masses in full QCD as function of the lattice spacing 
compared with quenched results (with Wilson fermions). $m_K$ is used as input.
Results form the CP-PACS collaboration~\cite{cppacs}} 
\label{quenching_effects}
\end{figure}

\begin{figure}[t]
\centerline{\epsfxsize=10cm \epsfbox{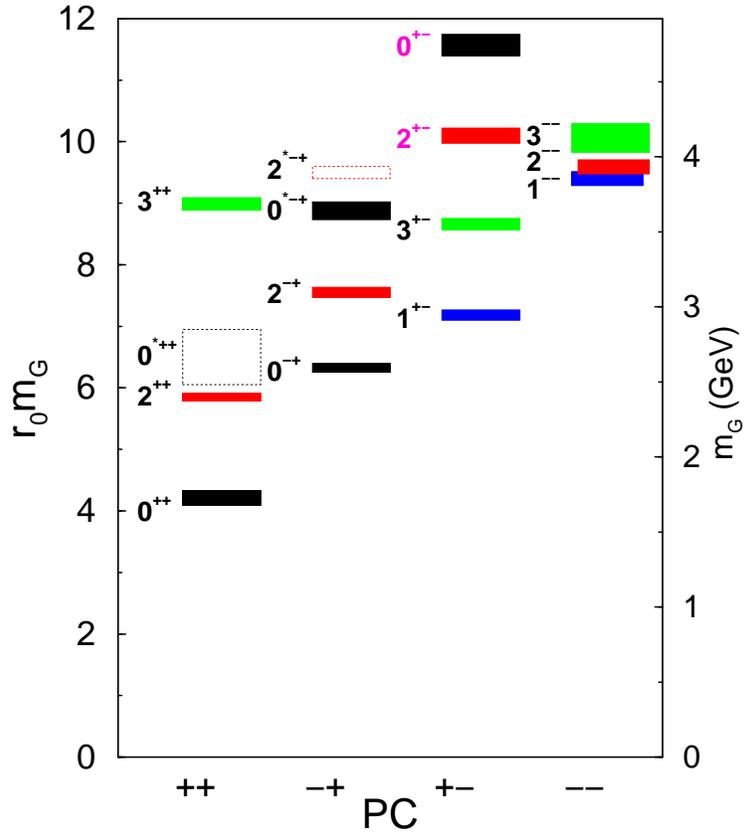}}
\caption{Mass spectrum in the SU(3) quenched theory as computed in 
ref.~\cite{glueballs}. The scale in the right vertical axis is based 
on the assumption that $r_0=410$MeV.}
\label{glueball_spectrum}
\end{figure}

\begin{figure}[t]
\centerline{\epsfxsize=10cm \epsfbox{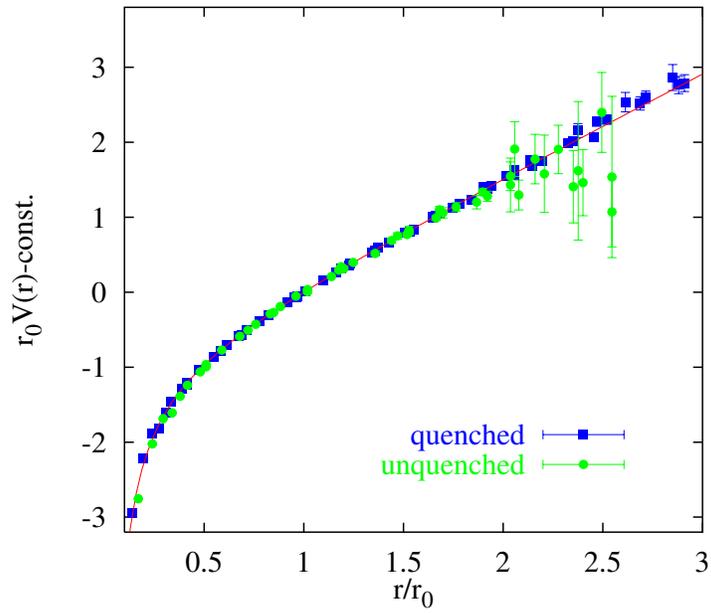}}
\caption{String tension in quenched and unquenched (two dynamical 
light flavours at $\kappa =0.1575$), extracted from ref.~\cite{bali}}
\label{string_tension}
\end{figure}

\begin{figure}[t]
\centerline{\epsfxsize=8cm \epsfbox{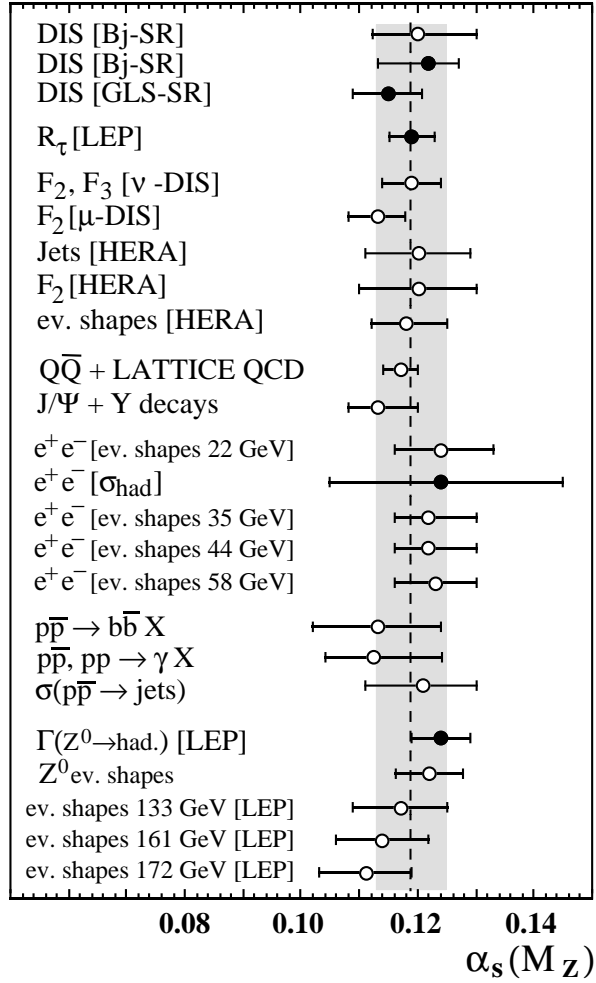}}
\caption{Comparison among different determinations for $\alpha_s$ at the $Z$ pole mass.}
\label{alphas}
\end{figure}

\begin{itemize}
\item {\bf Chiral symmetry on the Lattice}.
The light spectrum of QCD is dominated by spontaneous chiral symmetry breaking 
(in fact the pion is both a bound state and a Goldstone boson). At a 
classical level the continuum Dirac operator, $/\!\!\!\! D$, 
preserves the chiral symmetry, which we rewite as the Ginsparg-Wilson relation
\begin{equation}
	\gamma^5 /\!\!\!\! D + /\!\!\!\! D \gamma^5 =0
\label{chiral1}
\end{equation} 
The chiral symmetry is broken, at the quantum level, by the chiral anomaly.

The lattice regularized Dirac operator, eq.~(\ref{DiracOp}), explicitely breaks this 
symmetry and therefore does not provide a satisfatory description of chiral physics.

Since today no-one succeded in writing down a discretized versions of chiral fermions because
of the famous Nielsen-Ninomiya no-go theorem~\cite{nielsen}. Recently two solutions 
have been found to this problem and they are both equivalent to modify the chirality 
condition, eq~(\ref{chiral1}), into
\begin{equation}
	\gamma^5 /\!\!\!\! D + /\!\!\!\! D \gamma^5 = a /\!\!\!\! D \gamma^5 /\!\!\!\! D
\label{chiral2}
\end{equation}
where the right-hand side vanishes in the limit $a \rightarrow 0$. 

The two approaches are known as domain wall fermions~\cite{domainwall} and Neuberger 
fermions~\cite{neuberger}. They, from a theoretical point of view,
are equivalent~\cite{shamir1,shamir2}.

The work in this subject has still to be considered exploratory but seems very promising. 
Moreover it provides a theoretical testing ground for other well-estabilished 
lattice discretizations of fermions, such as Sheikoleslami-Wolhert (also known as Clover) 
and Kogut-Susskind (also known as Staggered) fermions. 

At present the most significant phenomenological lattice results are computed using Clover 
(and Staggered) fermions. Here ``significant'' means that simulations with Clover (and Staggered) 
fermions have been performed in a wide range of lattice spacings 
($1\text{GeV} < a^{-1} < 3 \text{GeV}$), on relatively large lattices 
(up to $48 \times 24^3$ and even bigger) and for many different values of the light quark 
masses. Large statistical Monte Carlo samples (of the order of hundreds of configurations) 
have been created and scrutinized.

Moreover the most important results of phenomenological interest have been reproduced 
indipendently by different international collaborations and agree with each other within 
the statistical errors.


\item {\bf Light hadronic spectrum}.
Fig.~\ref{cppacs_spectrum} shows some of the most recent lattice results
for the mass specrtum of light mesons and baryons, as computed by the CP-PACS 
collaborations. These results are obtained 
using Clover fermions, extrapolated to the chiral limit, in the quenched approximation. 
The error includes the effect of the chiral extrapolation but does not include the unknown 
effect of quenching. 

Fig.~\ref{quenching_effects} shows a comparison between quenched and unquenched 
results for the mass of light hadrons as function of the lattice spacing.
Quenched results are easier to compute therefore have a much smaller 
statistical error.  At present their error is dominated by the systematic one. 
The comparison with unquenched results can be used to quantify this systematic error.

\item {\bf Glueball spectrum}.
One sector in which lattice calculations have absolutely no competitors (because no models 
are available) is the computation of the glueball spectrum. Fig.~\ref{glueball_spectrum} 
shows some lattice results obtained in ref.~\cite{glueballs}. Some candidate resonances 
have been observed in experiments.

\item {\bf Confinement}.
Lattice has been the first theoretical tool to give direct evidence of the 
phenomenon of confinement. This is done through the computation of the Wilson loop, 
from which one obtains the chromo-electro-magnetic potential 
between a couple of static quarks-antiquark.
This potential (both for quenched and unqueched simulations) 
is shown in fig.~\ref{string_tension}.
For large distances the behavior of this potential, in the quenched approximation, 
grows linearly because no new particles can appear in 
the vacuum to screen the two static particles. The linearity is explained with the 
dual-superducting property of the QCD vacuum~\cite{digiacomo}. It forces the 
chromoelectric field 
of the system of two particles to be squeezed into a flux tube connecting them: the string.
In the unquenched (full) theory, instead, this potential only grows up to the point when the 
string contains enough potential energy to create in the QCD vacuum a new quark-antiquark pair
that breaks the string itself. 
Hence the potential reaches a plateau in correspondence to the treshold energy.

The deviation between the quenched and the unquenched potential has not been observed so far 
because, for numerical reasons, dynamical quark masses are still too heavy and lattice 
volumes too small.

The computation of the chromo-electromagnetic potential, combined with experimental measurements,
provides the best present determination of $\alpha_s$ (fig.~\ref{alphas}), which is the expansion 
parameter of any perturbative QCD calculation.

\begin{table}
\begin{center}
\begin{tabular}{|l|l|l|} \hline
Matrix element & Process & Parameters\\ \hline
$\langle 0 | \bar b \gamma^\mu q | B \rangle \equiv f_B p^\mu$ & $B \rightarrow \text{leptons}$ 
& $|V_{td}|$, $|V_{ts}|$\\
$\langle 0 | \bar c \gamma^\mu q | D \rangle \equiv f_D p^\mu$ & $D \rightarrow \text{leptons}$ 
&\\
$\langle D | \bar c \Gamma s | K \rangle$  & $D \rightarrow K + \text{leptons}$ 
& $|V_{cs}|$\\
$\langle K | \bar s \Gamma u | M_{u} \rangle$ & $K \rightarrow M_{u} + \text{leptons}$
&$|V_{us}|$\\
$\langle D | \bar c \Gamma d | M_{d} \rangle$ & $D \rightarrow M_{d} + \text{leptons}$
&$|V_{cd}|$\\
$\langle B | \bar b \Gamma u | M_{u} \rangle$ & $B \rightarrow M_{u} + \text{leptons}$
&$|V_{ub}|$\\
$\langle \bar B | \bar b \Gamma c | D^\ast \rangle$ & $B \rightarrow D^\ast + \text{leptons}$
&$|V_{cb}|$ \\
$\langle B | \bar q \Gamma q | B^\ast \rangle$ & $B \rightarrow \pi + \text{leptons}$ 
& $|V_{ub}|$ \\
$\langle K | \bar s \Gamma q \, \bar s \Gamma q | \bar K \rangle\propto B_K$ & 
($K$-$\bar K$ mixing) & $|V_{td}|$, $|V_{ts}|$\\
$\langle D | \bar c \Gamma q \, \bar c \Gamma q | \bar D \rangle\propto B_D$ & 
($D$-$\bar D$ mixing) & \\
$\langle B | \bar b \Gamma q \, \bar b \Gamma q | \bar B \rangle\propto B_B$ & 
($B$-$\bar B$ mixing) & $|V_{td}|$, $|V_{ts}|$\\
$\langle B | \bar b \Gamma b | B \rangle$ & $B$ kinetic and magnetic energy 
&\\
$\langle B | \bar b \Gamma q \, \bar q \Gamma b | B \rangle$ & inclusive $B$ decay &
\\
$\langle \Lambda_b | \bar b \Gamma q \, \bar q \Gamma b | \Lambda_b \rangle$ & 
inclusive $\Lambda_b$ decay & \\ \hline
\end{tabular}
\end{center}
\caption{Examples of matrix elements usually computed on the lattice and related processes
($M_{q}$ stands for the most general $(qq')$ meson, for example $\pi$, $\rho$ or $K$; $\Gamma$ is the most general spin$\otimes$color matrix). 
The table also shows the $V_{CKM}$ matrix elements that are associated
to the processes.}
\label{table_heavy}
\end{table}

\item {\bf Matrix elements and decays}. Almost all decays of hadrons
can be parametrized in terms of matrix elements that encode the 
non-perturbative contribution of QCD. For many of these matrix elements lattice 
computations have been able to produce satisfactory results. 
In table~\ref{table_heavy}
we list, as an example, some of those matrix elements.
Some other results are not conclusive and occasionally very controversial 
(for example attempts to compute $\epsilon' / \epsilon$). 

Matrix elements that include a contribution of final state interactions 
(for example $B \rightarrow \pi \pi$) have so far been outside the reach 
of lattice computations because of the Maiani-Testa no-go theorem~\cite{maiani}.

\item {\bf Lattice, effective field theories and models}.
Since lattice can be used to compute matrix elements and these can be confronted 
with predictions from effective theories or models, it becomes possible to use the
lattice results to extract their effective parameters. As an example in ref.~\cite{giulia}
the effective coupling of the Heavy  Meson Chiral Lagrangian is measured on the lattice
using a numerical computation for the following matrix element
\begin{equation}
	E (r) = 
	\frac13 \sum_{\mu=1,2,3} \int \langle B | A_\mu({\mathbf x}) | B^\ast \rangle \text{d}
	{\Omega_{\mathbf x}}
\end{equation}
where $A_\mu(x)$ is the axial current and $\Omega_\mathbf x$ is the solid angle associated to 
the 3D vector $\mathbf{x}$. Fig.~\ref{quark_model} shows a comparison between the lattice result 
for $E(r)$ and a prediction of the Chiral Quark Model. 
The parameters of the model have been adjusted to
fit the experimental mass spectrum of heavy mesons. See ref.~\cite{me_last} for further details.
The agreement between the lattice (a first principle simulation) and the model (consequence 
of experimental observations and some theoretical assumptions) is remarkable.

\item {\bf Heavy quarks and CKM Matrix}.
As we have shown in Section 1 the Cabibbo-Kobayashi-Maskawa matrix elements have to be 
extracted from a comparison between experimental and theoretical predictions. 
The latter, at present, are computed on the lattice with a non-negligible uncertainty.
Table~\ref{table_heavy} includes a list of those processes that mainly contribute to the 
determination of the CKM matrix elements and have to be computed using lattice simulations. 

Fig.~\ref{fig_now}(top) shows the present constraints on the CKM mixing angle in the ($\bar \rho$, 
$\bar \eta$) plane (where $\rho$ and $\eta$ are parameters of the Wolfenstein parametrization 
of the CKM matrix) and should be compared 
with fig.~\ref{fig_now}(bottom) which is obtained using the same experimental and 
theoretical input but assuming a possible future uncertainty instead of the 
``real'' present one for the lattice parameters. The future uncertainty is based on 
being able to generate 1000 gauge configurations with a lattice spacing of 
$a = 0.08$fm and with $m_\pi / m_\rho = 0.4$~\cite{sach3} (the latter constraint measures how well 
one is approaching the chiral limit). This estimate also assumes that systematic 
error due to quenching are under control.

The comparison shows how important it is to invest in lattice simulations while, 
contemporary, investing in experimental facilities. In fact for many fundamental 
quantities the theoretical uncertainty is as significant (if not more)
than the experimental one.

\end{itemize}

\clearpage

\begin{figure}[t]
\begin{center} 
\epsfxsize=10cm
\epsfbox{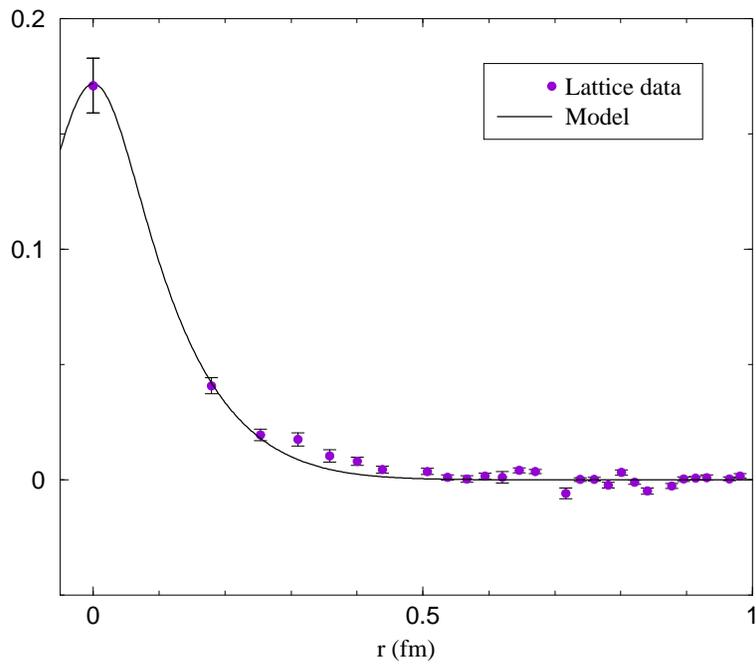} 
\end{center}
\caption{Comparison of a Lattice QCD matrix element, $E(r)$, and the same matrix element 
evaluated in the Chiral Quark Model.}\label{quark_model}
\end{figure}

\begin{figure}[t]
\begin{center}
\centerline{\epsfxsize=8cm\epsfbox{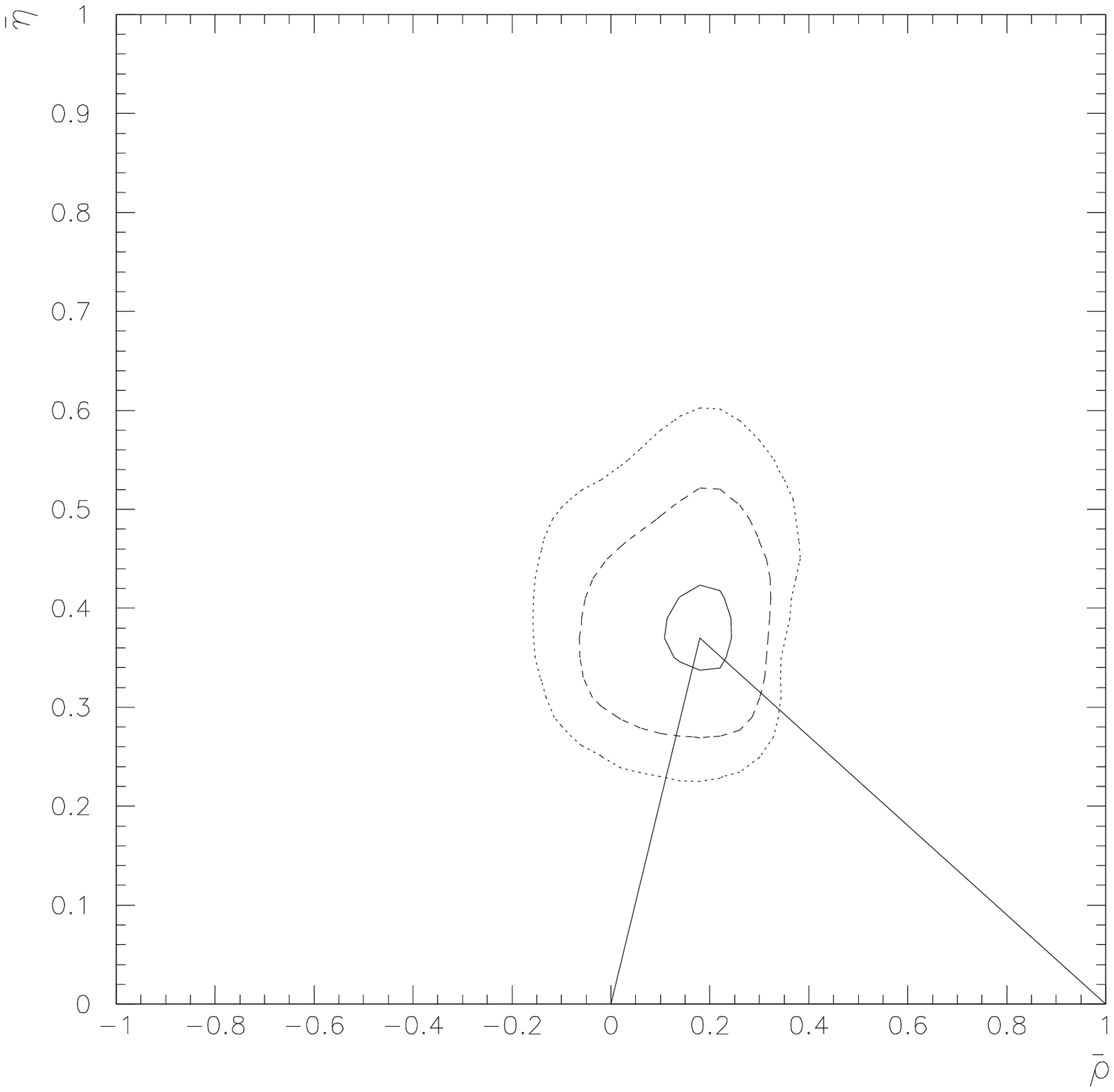}}
\vskip -5mm
\centerline{\epsfxsize=8cm\epsfbox{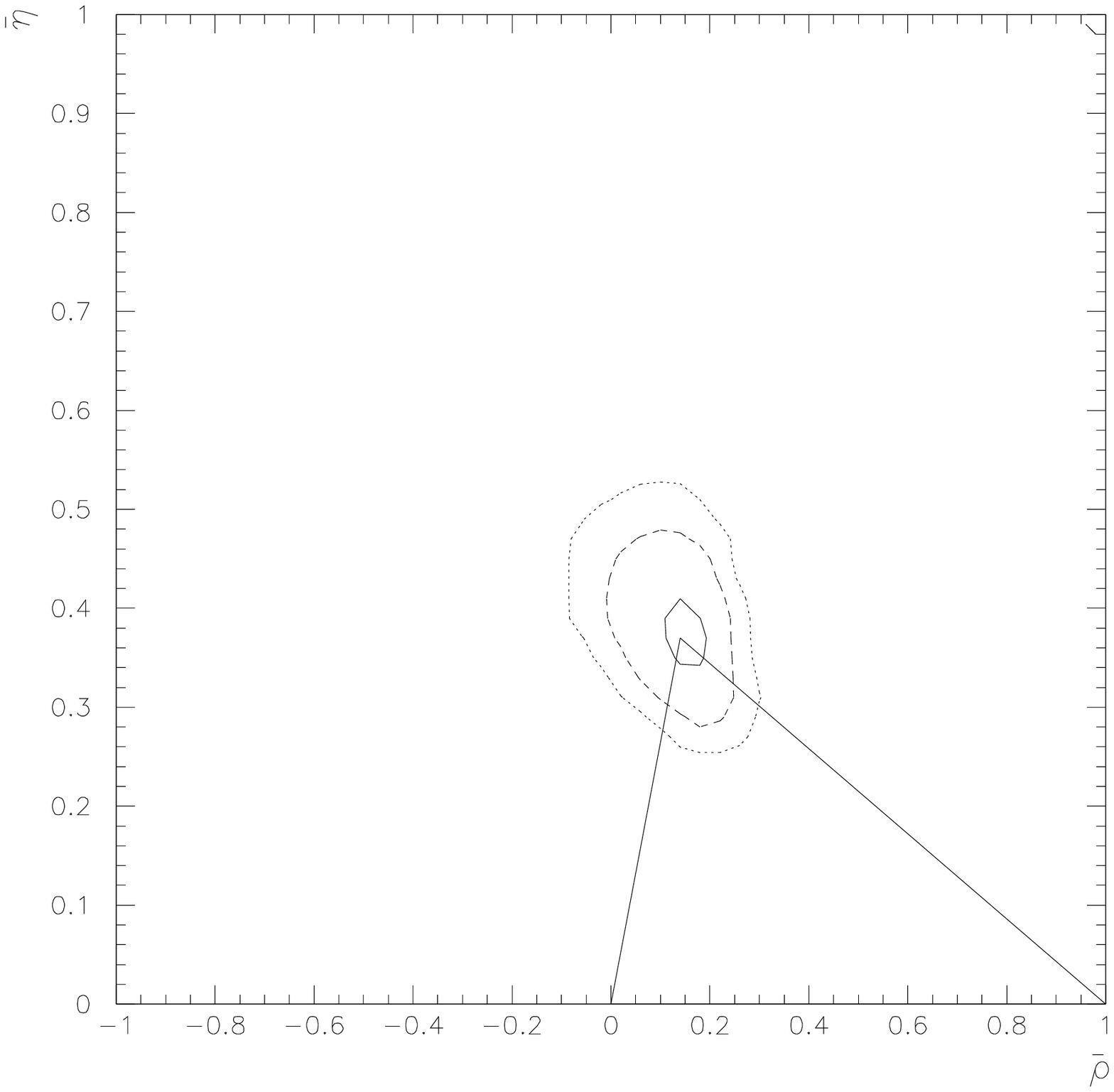}}
\caption{Current (top) and future (bottom) allowed regions for the point ($\bar \rho$, $\bar \eta$), 
the vertex of the unitarity triangle. The three regions correspond to 5\%, 68\% and 95\% 
confidence levels. The future estimate is based on the assumptions 
discussed in ref.~\cite{sach3}}
\label{fig_now}
\end{center}
\end{figure}

\newpage

\end{document}